\documentclass[12pt]{article}
\usepackage{amsmath}
\usepackage{graphicx}
\usepackage{enumerate}
\usepackage{natbib}
\usepackage{url} % not crucial - just used below for the URL 
\usepackage[toc,page]{appendix}
\usepackage[utf8]{inputenc} % Cracteres con acentos.
\usepackage{latexsym} % Simbolos
\usepackage{graphicx} % Inclusian de graficos. Soporte para \figura (mas abajo)
\usepackage[pdftex=true,colorlinks=true,linkcolor=blue]{hyperref} % Soporte hipertexto
\usepackage{amsthm}
\usepackage{xcolor}
\usepackage{totcount}
\usepackage{ragged2e}
\regtotcounter{section}
\usepackage{multido}
\usepackage{color,verbatim}
\usepackage{amsmath,bibentry}
\usepackage{mathtools}
\usepackage{amsfonts,rotating}
\usepackage{float}
\usepackage{colortbl}
\usepackage{subfig}

\newcommand{\blind}{1}

\usepackage{natbib}
\theoremstyle{definition}
\newtheorem{definition}{Definition}[section]
\begin{document}

\def\spacingset#1{\renewcommand{\baselinestretch}%
{#1}\small\normalsize} \spacingset{1}

%%%%%%%%%%%%%%%%%%%%%%%%%%%%%%%%%%%%%%%%%%%%%%%%%%%%%%%%%%%%%%%%%%%%%%%%%%%%%%

\if1\blind
{
\title{Improved $q$-values for discrete uniform and homogeneous tests: a comparative study} 
  	\author{ Marta Cousido-Rocha\\
  		\footnotesize{Department of Statistics and Operations Research $\&$ SiDOR Research Group, University of Vigo},\\
  		Jacobo de U\~na-\'Alvarez\\
  		\footnotesize{Department of Statistics and Operations Research $\&$ SiDOR Research Group, University of Vigo}\\
  		and\\
  		Sebastian D\"ohler\\
  		\footnotesize{CCSOR and Departament of Mathematics, University of Applied Sciences Darmstadt.}}
  \maketitle}
\fi

\if0\blind
{
  \bigskip
  \bigskip
  \bigskip
  \begin{center}
    {\LARGE\bf Improved $q$-values for discrete uniform and homogeneous tests: a comparative study}
\end{center}
  \medskip
} \fi

\bigskip 
\begin{abstract}
{Large scale discrete uniform and homogeneous $P$-values often arise in applications with multiple testing. For example, this occurs in genome wide association studies whenever a nonparametric one-sample (or two-sample) test is applied throughout the gene loci. In this paper we consider $q$-values for such scenarios based on several {existing} estimators for the proportion of true null hypothesis, $\pi_0$, {which take} the discreteness of the $P$-values into account. The theoretical guarantees of the {several} approaches with respect to the estimation of $\pi_0$ and the false discovery rate control are reviewed. The performance of the {discrete} $q$-values is investigated through intensive Monte Carlo simulations, including location, scale and omnibus nonparametric tests, {and possibly dependent $P$-values}. The methods are applied to genetic and financial data for illustration purposes too. Since the particular estimator of $\pi_0$ used to compute the $q$-values may influence the power, relative advantages and disadvantages of the reviewed procedures are discussed. Practical recommendations are given.

}
\end{abstract}

\noindent%
{\it Keywords:}  Multiple testing procedures; Discrete $P$-values; High-dimensional data; Homogeneous $P$-values.
\vfill

\newpage
\spacingset{1.5} % DON'T change the spacing!

\section{Introduction}\label{se:intro}

In many modern applications a large number of hypotheses are simultaneously tested leading to large scale $P$-values. Classical
approaches to deal with the multiplicity problem focus on the control of the number of false positives. Two well-known error rates which multiple comparison procedures (MCP) aim to control are 
the familywise error rate
(FWER), which is the probability of having at least one false positive, and the false discovery rate (FDR), which is the expected proportion of true null hypotheses rejected out of all rejected hypotheses \citep[see][]{BH}. Research on FDR-controlling procedures has been booming; see \cite{BH10} for existing proposals up to that date. The majority of these procedures have been developed in the setting of continuously distributed test statistics; such procedures can be overly conservative when the $P$-values follow
a discrete distribution. For example, for continuous $P$-values the FDR of Benjamini and Hochberg (1995) procedure, henceforth
referred to the BH method, is $({m_0}/{m})\alpha$ when applied at nominal level $\alpha$. Here $m$ and $m_0$ denote the number of hypotheses and the number of true null hypotheses, respectively. For discrete $P$-values, the FDR of the BH
method may be much smaller than $({m_0}/{m})\alpha$ \citep[see][Section 1]{Heller2012}, thus yielding a conservative decision rule and, consequently, a loss in power. This can be prevented, however, by developing procedures that appropriately incorporate the discreteness of the $P$-values. Indeed, by exploiting the discrete nature of the $P$-values dramatic improvements in power can be achieved, especially when the $P$-values are highly discrete. 

Even though discrete $P$-values arise in many applications, few papers explicitly deal with this aspect of multiple testing. \cite{hey} introduced a discrete BH procedure, which takes advantage of the discrete distribution of the $P$-values. However, Heyse's method
may be anti-conservative, i.e., the actual FDR level may be larger than nominal. \cite{sebas} constructed similar BH-type procedures
that incorporate the discrete and heterogeneous structure of the data and
guarantee FDR-control, filling the gap of \cite{hey}. On the other hand, \cite{Heller2012} proposed a step-down procedure that exploits the discreteness of the $P$-values and obtains
FDR levels closer in magnitude to the nominal level. Their method can be considered as a discrete version of the classical method of \cite{BL} which controls the FDR for continuous $P$-values under independence or positive dependence. {Recently, \cite{new1} investigated the BH procedure when applied to mid $p$-values, providing in this way a correction of the BH method for discrete $P$-values. More precisely, they proved the FDR control of the BH procedure applied to two-sided mid $P$-values of Binomial tests and Fisher’s exact tests.
	 In the same line of research, \cite{new2} proposed a new BH procedure which controls the FDR when
	applied to mid-$P$-values and to $P$-values with general distributions.}

%%% Correci�n Sebastan notas 22/05
In this article we investigate a particular type of discrete $P$-values, which are homogeneous (that is, identically distributed) and which we term discrete uniform in the sense of Definition \ref{def:1.1} below. To formalize things, suppose that one tests a large number of null hypotheses, $m$, and that the resulting $P$-values $\{pv_1,
\dots, pv_m\}$ are observations of the random variables $PV_i, i=1, \dots, m$. Assume that all the $P$-values are identically distributed {under the null hypothesis} sharing a common support ${A}=\{t_1, \dots, t_s, t_{s+1}\}$ with $t_0 \equiv 0<t_1 <\dots <t_s<t_{s+1} \equiv 1$. Furthermore, throughout the paper it is assumed that the $P$-values follow the cumulative distribution function (cdf) introduced in the following definition.
	
	\begin{definition}\label{def:1.1} (\textit{Discrete uniform cdf}). Given ${A}=\{t_1, \dots, t_s, t_{s+1}\}$ with $t_0 \equiv 0<t_1 <\dots <t_s<t_{s+1} \equiv 1$ (the support set of the distribution of the $P$-values), the discrete uniform cdf with support $A$, $H_{A} \equiv H_{\{t_1, \dots, t_s, t_{s+1}\}}$, is defined as
		
		$$H_{\{t_1, \dots, t_s, t_{s+1}\}} (x) = \left \{ \begin{matrix} 0 & \;\text{for}\; x<t_1
		\\ t_j &\;\;\;\;\;\;\;\;\;\;\;\text{for}\; x \in [t_j,t_{j+1})\\
		1 & \text{for} \; x\geq 1\end{matrix}\right. $$
	\end{definition}
	
	Note that $H_{A}$ is a step function that jumps up by $t_j-t_{j-1}$ at $t_j$ for $j=1, \dots, s+1$. The classical discrete uniform cdf is $H_{A}$ where $A$ contains equally spaced points, i.e., $A=\{1/N, 2/N, \dots, (N-1)/N, 1\}$, $N \in \mathbb{N}$. Therefore, Definition \ref{def:1.1} generalizes this concept to possibly non-equidistant support points. Summarising, we refer to any member of the class $\mathcal{H}=\{H_A| A \subset (0,1], A \;\; \text{countable}\}$ as discrete uniform distribution.
	
	$P$-values whose cdf belongs to the class $\cal{H}$ are often found in practice. These include nonparametric one sample or two-sample tests such as Kolmogorov-Smirnov test, Wilcoxon location test or Siegel-Tukey test for scale. For example, the two-sample Kolmogorov-Smirnov test with samples sizes $n_1=n_2=4$ leads to $P$-values following $H_A$ where $A=\{{1}/{35}, {8}/{35}, {27}/{35}, 1\}$. As another example, the two-sample absolute group mean difference test in \cite{Lia} is a permutation test which draws $P$-values from $H_A$ where $A=\{1/N, 2/N, \dots, (N-1)/N, 1\}$, $N$ being the number of permutations that lead to different values of the statistic (for example $N=35$ for sample sizes $n_1=n_2=4$). See Section 3 for other examples and further illustration.
%	Figure \ref{fi:exA} reports a plot of the step function $H_A$ for $A=\{{1}/{35}, {8}/{35}, {27}/{35}, 1\}$ (Kolmogorov-Smirnov test).
	
%	\begin{figure}[h!]
%		\begin{center}
%			\captionsetup{width=1\textwidth}
%			\includegraphics[width=0.55\linewidth]{1.pdf}
%			\caption{{Plot of the function $H_A$ for $A=\{{1}/{35}, {8}/{35}, {27}/{35}, 1\}$. Cumulative distribution function of the $P$-values derived from the  two-sample Kolmogorov-Smirnov test with samples sizes $n_1=n_2=4$, and, in blue, {cumulative distribution of a continuous uniform variable in $(0,1)$}}} 
%			\label{fi:exA}
%		\end{center}
%	\end{figure}

Discrete corrections of MCP like those in \cite{sebas} and \cite{Heller2012} are irrelevant for homogeneous discrete uniform (hdu) $P$-values, which are special to this regard. Indeed, the adjusted discrete $P$-values of \cite{Heller2012} and \cite{hey} reduce to the ones for continuous $P$-values in \cite{BH} and \cite{BY}, respectively, when applied to \textit{any type} of homogeneous $P$-values, leaving the results unchanged. { The same holds true the method of \cite{new2}. Therefore, we decide to focus our research on the $q$-value approach proposed by \cite{Storeyq} {based on estimators of the proportion of true null hypothesis, $\pi_0$, which take the discreteness of the $P$-values into account. The estimators of $\pi_0$ we consider are well-suited for hdu $P$-values and generally lead to a power increase when compared to standard estimators for continuous $P$-values; see Section \ref{se:simul3} for more on this.}

The paper is organized as follows. In Section \ref{se:MCP} we review the $q$-value method and {several} corrections of such approach for hdu $P$-values. The theoretical guarantees of the proposed methods with respect to the estimation of the proportion of true null hypotheses, the estimation of the FDR and the FDR control are summarised too. In Section \ref{se:tests} we enumerate and briefly describe several two-sample nonparametric tests, including location, scale and omnibus tests, which lead to hdu $P$-values. The performance of the proposed discrete $q$-values in such two-sample settings is investigated through intensive Monte Carlo simulations in Section \ref{se:simul3}. Both settings with independent and dependent tests are considered. The performance of the standard $q$-value approach for continuous $P$-values is studied for comparison purposes too. In Section \ref{se:real3} we illustrate the behaviour of the proposed methods through two real data examples. Finally, in Section \ref{se:conclu} we give the
main conclusions of {our comparative study} and we provide some practical recommendations. Tables with simulation results and additional simulations for the one-sample problem are provided in the online Supplementary Material. The methods investigated in this paper have been implemented in the user-friendly \texttt{DiscreteQvalue} package \cite{R3} of the free software
\texttt{R}.

\section{Multiple comparison procedures: $q$-value method}\label{se:MCP}

In this section we review the $q$-value method and we {several} ways of estimating $q$-values when the $P$-values are hdu. Consider a family of $m$ null hypotheses $H_{0i}$, $i=1, \dots, m$, with associated $P$-values $pv_i$, $i=1, \dots, m$, which are observations of the random variables $PV_{i}, i=1, \dots, m$. The number of true null hypotheses is denoted by $m_0$; 
$R_m$ is the number of rejected null hypotheses, while $V_m$ the number of true null hypotheses
which are rejected (Type I errors). The most popular error rates to control the Type I errors in a simultaneous
way are the FWER
and the FDR. The $q$-value method aims at controlling the latter, which is defined as the the expected value of the proportion of Type
I errors among the rejected hypotheses, i.e.,
{$\operatorname{FDR}=E\left[{V_m}/{R_m}\right].$ The $q$-value method decides whether each one of the $H_{0i}$, $i=1, \dots, m$, should be rejected or not based on a measure of each feature's significance (referred to as its $q$-value) which
	automatically takes multiplicity into account. The $q$-value of a feature $i$ is defined as the minimum FDR that can be attained when declaring that feature significant:
	
	\begin{equation}\label{ec:qvalue}
	q(pv_{i})=\min_{t \geq pv_i} \operatorname{FDR}(t),
	\end{equation}
	where FDR$(t)$ denotes the FDR when one rejects the hypotheses with $P$-values smaller than or equal to $t$.}

{Note that the FDR is undefined if $R_m=0$; actually, the formal definition of th FDR is given by 
$\operatorname{FDR}=E\left[({V_m}/{R_m})|R_m>0\right]\operatorname{P}(R_m>0)$. However, since the $q$-value is interpreted under the assumption that the feature is called significant, the inclusion of the term $\operatorname{P}(R_m>0)$ in the definition of the FDR is strange. Hence, the $q$-value is most technically defined as the minimum positive false discovery rate, pFDR$=E\left[({V_m}/{R_m})|R_m>0\right]$, at which the feature can be called significant. In our framework $m$ is large, implying that $\operatorname{P}(R_m>0)\approx 1$, which leads to FDR $\approx$ pFDR. Hence, the distinction between both error rates is not relevant for our aim (see Appendix A in \cite{S2} for more details).}

%where $R_p$ is the number of rejected hypotheses, $V_p$ the number of true null hypotheses

 In practice, FDR$(t)$ is unknown and must be estimated.
Hence, one can estimate the $q$-value of a feature $i$ by plugging a FDR estimator in (\ref{ec:qvalue}).  We consider the FDR estimator employed in  \cite{S3} which is 

\begin{equation}\label{ec:FDR}
\widehat{\operatorname{FDR}}(t)= \dfrac{m \widehat{\pi}_0 t}{\#\{i|p_{vi} \leq t\}},
\end{equation}
where $\widehat{\pi}_0$ is an estimator of the proportion of true null hypotheses $\pi_0=m_0/m$. Once the estimated $q$-values are computed, the $q$-value method rejects the null hypotheses whose $q$-values
are less than or equal to the nominal level $\alpha$. This is equivalent to applying the \cite{BH} method at level $\alpha/\widehat{\pi}_0$, this method is known as adaptive Benjamini and Hochberg (adaptive BH). Hence, for a given nominal level $\alpha$, the $q$-value method is more powerful than the \cite{BH} method except when $\widehat{\pi}_0=1$ (they are equivalent in this case), or when the estimator of $\pi_0$ is unacceptable because it reports values greater than 1. 

%{\color{red} Although the $q$-value method and the adaptive BH are equivalent, we decide to present the method through the first approach since in the majority of areas of application where the hdu $P$-values appears, for example genomewide studies, \cite{S2} is well known and wide applied.}

Different versions of the $q$-value method can be defined depending on which $\pi_0$ estimator is plugged in (\ref{ec:FDR}). In Section \ref{se:qcont} two versions of the $q$-value method for continuous $P$-values are reviewed. Furthermore we consider in Section \ref{se:qd} three versions of the $q$-value method for hdu $P$-values. One of them is an adaptive BH method introduced in \cite{chen} for discrete and possibly heterogeneous null distributions, for which a simplified version is proposed for the case of hdu $P$-values.}

{In the setting of multiple testing it is important to distinguish three different issues: (a) conservativeness of the $\pi_0$ estimator; (b) conservativeness of the FDR estimator (\ref{ec:FDR}); and (c) FDR control of the $q$-value method based on (\ref{ec:qvalue}) and (\ref{ec:FDR}). Below we discuss these issues for each of the $q$-value methods.
}

\subsection{$q$-value method for continuous $P$-values}\label{se:qcont}

The classical $\pi_0$ estimator proposed in \cite{S1} is 

\begin{equation}\label{ec:pi0}
\widehat{\pi}_0(\lambda)=\dfrac{\#\{pv_i>\lambda ; i=1, \dots,m\}+1}{m(1-\lambda)},
\end{equation}
where $\lambda \in \left[0,1\right]$ is well-chosen according to some procedure. A standard choice for $\lambda$, for continuous $P$-values, is $1/2$ \citep{S1}. Henceforth, we refer to the $\pi_0$ estimator given by (\ref{ec:pi0}) and $\lambda=1/2$ as standard Storey estimator (abbr. $\widehat{\pi}_0^{SS}$), and to the corresponding $q$-value method as standard Storey (SS) $q$-value method.  \cite{Blanchard} recommend $\lambda$ equal to the nominal level $\alpha$ instead of $\lambda=1/2$ since it leads to a more robust procedure under positive dependence, but at the price of being more conservative.

Additionally  \cite{S2} proposed an automatic method to estimate $\pi_0$ which avoids the selection of the $\lambda$ parameter in (\ref{ec:pi0}). Specifically they suggested $\widehat{\pi}_0^{ST}=\widehat{f}(1)$, where $\widehat{f}$ is the natural cubic spline with 3 degrees of freedom of $\widehat{\pi}_0(\lambda)$ on $\lambda$, with
$\lambda=0, 0.01,0.02, \dots, 0.95$ (or another sequence of $\lambda$ values between 0 and 1) and  $\widehat{\pi}_0(\lambda)$ is the estimator in (\ref{ec:pi0}). Henceforth, we refer to this estimator and the corresponding $q$-value method as ST estimator and ST $q$-value method, respectively.

When the null (continuous) $P$-values are uniformly distributed in $(0,1)$, it is easy to see that $E(\widehat{\pi}_0(\lambda))\geq \pi_0$, i.e., the estimator in (\ref{ec:pi0}) is conservative. \cite{S3} proved in their Theorem 1 that, for a fixed $\lambda$ and under certain conditions, the estimator in (\ref{ec:FDR}) is conservative too, in the sense that $E(\widehat{\operatorname{FDR}}(t))\geq \operatorname{FDR}(t)$. A flaw in the proof of such result was corrected by \cite{LiangNettleton}, who required (besides the uniform distribution of the null $P$-values) the \textit{null independence condition}: the null $P$-values are independent among themselves, and they are independent of the alternative $P$-values. These theoretical results are established for fixed $\lambda$ and do not include the situation with data-driven selection of this parameter, thus excluding the ST method. Extended theory for dynamic adaptive (i.e. data-driven) procedures was given by \cite{LiangNettleton}, who proved conservativeness for both $\widehat {\pi}_0(\lambda)$ and $\widehat{\operatorname{FDR}}(t)$ when the data-driven $\lambda$ is a stopping time with respect to the filtration $\mathcal{F}_s=\sigma\{I\{p_i\leq u\},0\leq u\leq s,1\leq i\leq m\}$, $0\leq s<1$. Unfortunatelly, ST method does not fulfill such condition and, hence, the development of formal theory for this procedure remains undone.

	Regarding the FDR control of the $q$-value method, \cite{S2} pointed out two interesting properties: (i) for large $m$ ($m \rightarrow \infty$), the
	FDR is $\leq \alpha$; and (ii) the estimated $q$-values
	are simultaneously conservative for the true $q$-values ($m \rightarrow \infty$). Indeed, \cite{S2} indicate that these properties can be formally proved
	from minor modifications to some of the main results in \cite{S3}. It should be noted, however, that these results are asymptotic, and that the proofs refer to the situation with a fixed $\lambda$.
	
	An important issue is the possible weak dependence among the large number of features or variables. We are not aware of any theoretical result on the conservativeness of the SS and ST estimators for $\pi_0$ and FDR in such a setting. However, the aforementioned results on the FDR control of the $q$-value method include the case of weakly dependent $P$-values. Theoretical guarantees for SS and ST methods with respect to the estimation of $\pi_0$ and FDR, as well for the FDR control of the corresponding $q$-value method, are summarized in Table \ref{ca:summary}. Information in Table \ref{ca:summary} refers to the special type of weak dependence considered by \cite{S2}.

The two $\pi_0$ estimators presented in this section are suitable for continuous $P$-values but can be overly conservative for discrete $P$-values. 
For this reason, in the next section we introduce three $\pi_0$ estimators which take into account the discrete distribution of the $P$-values.

\subsection{$q$-value method for discrete $P$-values}\label{se:qd}

{ In Section \ref{se:Liang} the $q$-value method based on the  $\pi_0$ estimator of \cite{Lia} is considered.}{ To the best of our knowledge,} {the performance of the $q$-value method based on such estimator is studied for the first time in this paper (Section \ref{se:simul3}). In Section  \ref{se:rand}  the $q$-value method based on a $\pi_0$ estimator based on randomized $P$-values is considered. On the other hand, the $q$-values which arise from the $\pi_0$ estimator in Section \ref{se:chen} can be regarded as a simplification of the adaptive FDR-procedure in \cite{chen} for hdu $P$-values.
}

\subsubsection{$q$-values based on Liang method}\label{se:Liang}
\cite{Lia} proposed a $\pi_0$ estimator for large scale hdu $P$-values. Let $B = \{b_1, \dots, b_{s+1}\}$ be the sample frequencies of
every element in $A$, i.e., $b_i = \#\{{pv}_j : {pv}_j=t_i\}$ for $i = 1, \dots, s +
1$. His procedure is based on finding the smallest support point such that the $b_i$'s to its right are roughly equal, i.e, it is a right-boundary procedure. The method finds the smallest $\lambda$ for which $\widehat{\pi}_0(\lambda)$ stops decreasing, where $\lambda$ is chosen from a subset of {$\{t_0, \dots, t_s\}=A\setminus t_{s+1}$ (see Definition \ref{def:1.1}).} %The reason for choosing $\lambda$ within the set $A\setminus t_{s+1}$ is that other choices generally lead to unnecessary conservativeness; see \cite{Lia} for furthert details.

Formally, Liang's $\pi_0$ estimator is $\widehat \pi_0(\lambda_L)$, where $\widehat{\pi}_0(\lambda)$ is the estimator in (\ref{ec:pi0}) and $\lambda_L$ is defined in Definition \ref{def:stoprule}.

\begin{definition}\label{def:stoprule}
	Let {$\Lambda=\{\lambda_1, \dots,\lambda_{\nu}\} \subseteq \{t_0, \dots, t_s\}=A \setminus t_{s+1}$, see Definition \ref{def:1.1},} be a candidate set for $\lambda$ such that $0 \equiv \lambda_0 < \lambda_1< \dots < \lambda_{\nu}<\lambda_{{\nu}+1}\equiv 1$. Then, the $\lambda$ chosen is $\lambda_L$ where $L=\min\{1 \leq i \leq {\nu}-1: \widehat{\pi}_0(\lambda_i)\geq \widehat{\pi}_0(\lambda_{i-1})\}$ if $\widehat{\pi}_0(\lambda_i)\geq \widehat{\pi}_0(\lambda_{i-1})$ for some $i=1, \dots,{\nu}-1$ and $\lambda_L=\lambda_{\nu}$ otherwise. 
\end{definition}

\begin{figure}[h!]
	\begin{center}
		\captionsetup{width=1\textwidth}
		\includegraphics[width=0.65\linewidth]{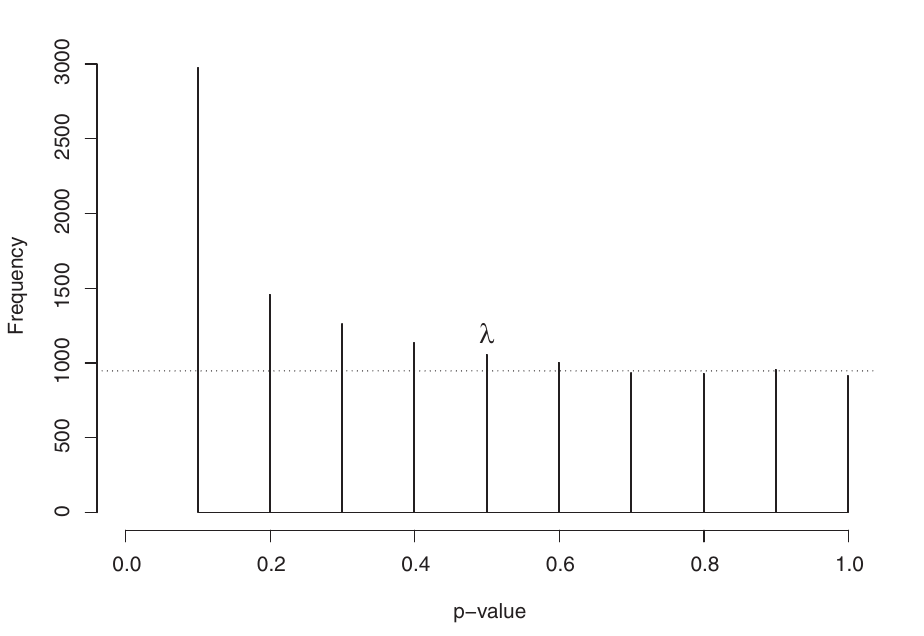}
		\caption{{The histogram of the $P$-values in the application in \cite{Lia}, Section 6. His method takes
				$\lambda_L=0.5$, and the dotted horizontal line is the expected number
				of true null $P$-values at every support point.}} 
		\label{fi:liang}
	\end{center}
\end{figure}

In order to illustrate Liang's method, we report in
Figure \ref{fi:liang} the histogram of the $P$-values in the application in \cite{Lia}, Section 6. In this example $A=\{0.1, \dots, 0.9, 1\}$, $\Lambda=\{0, 0.1, \dots, 0.5\}$, $\lambda_L=0.5$ and $\widehat{m}_0=9474$; the dotted horizontal line is the expected number
of true null $P$-values at every support point, 947.

{
	\cite{Lia} proves the conservativeness of his $\pi_0$ estimator, and that of the corresponding FDR estimator according to (\ref{ec:FDR}), for independent and hdu $P$-values. Furthermore, he also proves the conservativeness of the FDR estimator under a type of ``weak dependence'' of the $P$-values \citep[more details about this particular type of dependence in Section 3 of][]{Lia}. The type of weak dependence considered by \cite{Lia} matches the one in \cite{S3}.
	
	Since the $q$-value method is equivalent to the corresponding adaptive BH method, FDR control would follow from $E(1/\widehat{\pi}_0) \leq {1}/{\pi_0 }$ \citep{Blanchard}. However, such condition is stronger than $E(\widehat{\pi}_0) \geq {\pi_0 }$, which is what it is proved in \cite{Lia}, and hence FDR control for this method remains unclear.
	See Table \ref{ca:summary} for a summary of the properties of the estimators and $q$-value method of \cite{Lia}. Note that the validation of the FDR-control is performed for the first time in this paper, see Section \ref{se:simul3}.
}

\subsubsection{$q$-values based on Chen method}\label{se:chen}
\cite{chen}  proposed a $\pi_0$ estimator for $P$-values which follow discrete and possibly heterogeneous null distributions. We present a simplified version of Chen's algorithm for the case of hdu $P$-values. 

{\cite{chen} studied the bias of the $\pi_0$ estimator (\ref{ec:pi0}) in the discrete paradigm. In order to reduce this bias they followed an idea similar to that in \cite{Lia} but, instead of choosing a single $\lambda$ parameter, they suggested to consider several $\lambda$'s and then to average the resulting estimates for $\pi_0$.
	The steps of the Chen's algorithm are (with $A$ as in {Definition \ref{def:1.1}}):
	
	\begin{itemize}
		\item[1:] Set $q=\inf\{c: c\in A\}$. Pick a sequence of $B$ increasing, equally spaced ``guiding values'' $\{\tau_j\}_{j=1}^{B}$ such that $q=\tau_0 \leq \tau_1 \leq \dots \leq \tau_B <1$.
		\item[2:] For each $j \in \{1, \dots, B\}$, set
		$T_{j}=\{\lambda \in A: \lambda \leq \tau_j\}$
		and $\lambda_j=\sup\{\lambda: \lambda \in T_j\}$. For each $j \in \{1, \dots, B\}$, define the ``trial estimator'' 
		$ \beta(\tau_j)={1}/({(1-\tau_j)m})+ ({1}/{m}) \sum_{i=1}^{m} {I\{{pv}_{i}>\lambda_j\}}/({1-\lambda_j}).$
		Truncate $\beta(\tau_j)$ at 1 when it is greater than 1. 
		
		\item[3:]  Set 
		$\widehat{\pi}_0^{G}=({1}/{B}) \sum_{j=1}^{B} \beta(\tau_j)$
		as the estimate of $\pi_0$.
		
	\end{itemize}
	
		{
		\begin{table}
			\begin{center}
				\scalebox{0.75}[0.75]{ 
					\begin{tabular}{cccccccc}
						\hline
						& 	 \multicolumn{3}{c}{Independence} & &\multicolumn{3}{c}{Dependence}\\
						\cline{2-4}\cline{6-8} \\
						&	$\widehat{\pi}_0$	& FDR & $q$-value && $\widehat{\pi}_0$	& FDR & $q$-value \\	
						SS & T & T& S& & & & S\\
						ST &  &  & S &&  &  & S\\
						Liang & T & T & S &&  & T & S\\
						Chen &  T &  & T &&  & & S\\
						Rand &  &  &S && &  & S\\
						\hline
				\end{tabular}}	
			\end{center}
			\caption{Conservativeness of the several estimators for $\pi_0$ and FDR introduced in Section \ref{se:MCP}, and FDR-control of the corresponding $q$-values. {For each case, the table reports ``T'' if a theoretical proof is available in the literature, and ``S'' if so far the result is only supported by simulation studies. Empty cells correspond to missing theoretical or by-simulation validation.}}
			\label{ca:summary}
	\end{table}}
	
	{The first term in $\beta(\tau_j)$ is technical and only useful to prove theoretical properties of adaptive MCP's.}
	The sequence $\{\tau_j\}_{j=1}^{B}$ used in  \cite{chen} is $\tau_1=\tau_0+0.5\times\left(0.5-\tau_0\right)$, $B=100$ if $\tau_0<0.5$, otherwise set $\tau_1=\tau_{B}=0.5$ and $B=1$. 
	{An in depth study of the sensitivity of Chen method to the choice of $\{\tau_j\}_{j=1}^{B}$ may be of practical interest, but it is beyond the scope of the present work. However, it is worth to mention that we checked via simulation the behaviour of Chen $\widehat{\pi}_0$ based on different sequences of ``guiding values'' (results not shown). Firstly, we tried Chen $\widehat{\pi}_0$ with $ \{\tau_j\}_{j=1}^{B}=A$, and the mean squared error (MSE) was always larger than that obtained using the $\{\tau_j\}_{j=1}^{B}$ recommended by \cite{chen}. This is probably related to the fact that, for large values in $A$, the $\pi_0$ estimator is based on few $P$-values, leading to a poor performance. Secondly, we fixed $\{\tau_j\}_{j=1}^{B}$ to be the support points smaller than $1/2$, and the MSE was approximately equal to that attached to the sequence proposed by \cite{chen}. Further investigation is required before reaching solid conclusions to this regard.}

	{\cite{chen} proved that their $\pi_0$ estimator satisfies  $E(1/\widehat{\pi}_0) \leq {1}/{\pi_0 }$ for independent $P$-values. {From this condition using Jensen's inequality we obtain that} $E(\widehat{\pi}_0) \geq {\pi_0 }$, i.e., their estimator is conservative. The $q$-value method respects  the false discovery rate nominal level for independent $P$-values since $E(1/\widehat{\pi}_0) \leq {1}/{\pi_0 }$ \citep[see Theorem 11 of][]{Blanchard}. Regarding the conservativeness of  the FDR estimator defined by plugging their $\widehat{\pi}_0$ in (\ref{ec:FDR}) we are not aware of results describing its theoretical behaviour. A simulation study considering dependent $P$-values has been carried out in the referred paper and, according to the obtained results, it seems that the theoretical properties may hold under some general type of dependence too. This is supported by our simulations in Section \ref{se:simul3} too. Table \ref{ca:summary} summarizes the comments in this paragraph.}

	\subsubsection{Randomized $q$-values}\label{se:rand}
	
	Other approaches to take the discreteness into account
	have been suggested in the literature. \cite{ku} and \cite{ha}, among others, suggested procedures based on randomized $P$-values. {\cite{ha} extends to the multiple testing setting the randomized $P$-value, (non-randomized) mid $P$-value and abstract randomized $P$-value which are recommended when the test statistic has a discrete distribution. \cite{ku} introduce fuzzy MCP's as a solution to the problem of multiple comparisons for discrete test statistics.}
	The randomized $P$-values follow a continuous uniform distribution under the global null hypothesis, and therefore classical methods to estimate $\pi_0$ as (\ref{ec:pi0}) can be applied. The randomized procedure used here is a simple one described in the next steps. {It uses the definition of randomized $P$-values in \cite{Dickhaus}}. Suppose that we want to define the randomized version of ${pv}_{i}$ with $i \in \{1, \dots,m\}$. Remember that the support of the $P$-values is denoted by  ${A}=\{t_1, \dots, t_s, t_{s+1}\}$ with $t_0=0<t_1 <\dots <t_s<t_{s+1}=1$ {(see Definition \ref{def:1.1}).}
	
	\begin{itemize}
		\item[1.] Generate an observation $u$ from a $U(0,1)$.
		\item[2.] Suppose ${pv}_{i}=t_k$, $k \in \{1, \dots, s+1\}$; then, the randomized $P$-value is defined by
		$${pv}_{i}^{Rand}={pv}_{i}-u(t_k-t_{k-1}).$$
		
	\end{itemize}
	
	Applying this algorithm to each $P$-value we obtain a set of randomized $P$-values $\{{pv}_{i}^{Rand}, i=1, \dots,m\}$. The next step is to compute (\ref{ec:pi0}) using the randomized $P$-values {and $\lambda=0.5$}. This procedure can be repeated a large number of times $L$ reporting $L$ values of (\ref{ec:pi0}) which can be summarized using the average and reported it as our final estimator, i.e., {$\widehat{\pi}_0^{Rand}(\lambda)=({1}/{L})\sum_{j=1}^{L} \widehat{\pi}_{0,j}^{Rand}(\lambda)$ where $\widehat{\pi}_{0,j}^{Rand}(\lambda)$ is the estimator in (\ref{ec:pi0}) computed using the randomized $P$-values obtained in the $j$-th simulation run}.

	We refer to the $q$-value method which plug in this $\pi_0$ estimator as \emph{randomized $q$-value method} (abbr. Rand). {Unfortunally, we are not aware of theoretical results describing the performance of the randomized $\widehat{\pi}_0$, $\widehat{\text{FDR}}$ and $q$-value method. But their behaviour is studied via simulations in Section \ref{se:simul3} for both, independent and dependent data.}}

\section{Two-sample tests}\label{se:tests}

In the simulation study in Section \ref{se:simul3} we consider the two-sample problem with low sample size and a large number of variables. The particular two-sample tests that are used to generate the hdu $P$-values in the simulation study are reviewed in this Section.

The data at hand are represented by two random matrices  $X=\left[X_1, \dots, 
X_m\right]^T$ and $Y=\left[Y_1, \dots, Y_m\right]^T$ of respective 
dimensions $m \times n_1$ and $m \times n_2$, where
$X_i=(X_{i1}, \dots,$ $X_{i{n}_1})$ and $Y_i=\left(Y_{i1}, \dots,
Y_{i{n}_2}\right)$, $i=1, \dots,m$. Here, $n_1$ and $n_2$ are the sample sizes in each of the two groups, whereas $m$ is the number of variables. As mentioned above, we 
consider the setting $n_1 <<m$ and $n_2 <<m$, which is known as low sample size and large
dimension.
Given sequences of cumulative
distribution functions $\{F_1,F_2,\ldots\}$ and
$\{G_1,G_2,\ldots\}$, it is assumed that $X_{i1}, \dots, 
X_{i{n}_1}$ and $Y_{i1},\ldots,Y_{i{n}_1}$ are independent random samples from
$F_i$ and $G_i$, respectively, $i=1,\dots,m$. We are interested in testing the null hypotheses  $H_{0i}:F_{i}\equiv
G_{i}$, for $1 \leq i \leq m$. The distributions $F_i$ and $G_i$ may differ in location, scale or more generally in shape. In the three following subsections we group the different tests according to the departure they aim to detect.

{
 As mentioned above, our simulation study covers different two-sample tests for detecting differences in location, scale and shape. When the $P$-values are continuous  their null distribution does not depend on the particular test and, hence, considering different types of tests is not critical. The situation changes in the discrete setting, since different tests lead to different discrete uniform distributions, see Table \ref{ca:sp}, and the performance of the methods may vary depending on the null distribution of the $P$-values. Under such point of view, the simulation study in this paper brings relevant novelties over the existing literature, which has been traditionally focused on tests for location.}

\subsection{Two-sample tests for location}

The most popular parametric two-sample test for location is the Student's $t$ test for the equality of means. When the samples are independent there are two versions of this test, depending on whether the two population variances are assumed to be equal or not; in the latter case it is referred as Welch's test \citep[see Section 9.1 of][]{Gi}. The $t$-test assumes that both samples are normally distributed although it is robust, usually performing well even in cases where this assumption is violated. When nothing is assumed on the underlying distributions, one of the most popular nonparametric tests for testing the equality of locations is the Wilcoxon rank-sum test, also known as Wilcoxon-Mann-Whitney test. The Wilcoxon test is based on the ranks of the observations. It uses the idea that, if the null hypothesis is true, it is expected that the ranks corresponding to the combined sample are interspersed while, under the alternative, it is expected that the ranks of the observations of each sample
are separated in two groups \citep[see Section 9.2 of][]{Gi}. In our framework the sample size is small, hence the distribution of the Wilcoxon's statistic {is determined} using a permutation test. Finally, for testing the equality of two populations means we also consider the test used in \cite{Lia} whose statistic is defined as the absolute difference between the sample means, i.e., for each $i \in \{1, \dots, m\}$, $D_i=|\overline{X}_i -\overline{Y}_i|$ where $\overline{X}_i=(1/{n}_1)\sum_{j=1}^{{n}_1} X_{ij}$ and $\overline{Y}_i=(1/{n}_2)\sum_{j=1}^{{n}_2} Y_{ij}$. Its null distribution is also {determined} using a permutation approach. Henceforth, this test is referred as \textit{absolute value test} (abbr. abs). Note that the $P$-values derived from its application follow a classical discrete uniform distribution.

\subsection{Two-sample tests for scale}
When two distributions differ in their variances, the classical parametric test is the  $F$-test of equality of variances. This test assumes that both samples are normally distributed, and is very sensitive to the violation of the normality assumption \citep[see e.g. Section 10.1 of][]{Gi}. A more robust parametric test is the \emph{Levene} test proposed in \cite{le}. There exist nonparametric tests for scale too. The \emph{Siegel-Tukey} test \citep{sie} is a nonparametric test for detecting differences in scale between two samples. It is a rank-sum test which uses a simple ranking idea,
and the already known null distribution of the Wilcoxon test. For the Siegel-Tukey test there are two options available to rank the observations which can lead to different values of the statistic and may even lead to different final conclusions. Hence, \cite{an} proposed a rank test which avoids this inconvenience
by essentially  averaging the two
Siegel-Tukey schemes for ranking.

\subsection{General two-sample tests}
The tests introduced above are designed to detect only one specific type of difference between the distributions, i.e. location or scale. We also investigate the performance of two tests which can detect any type of differences. We consider the well-know Kolmogorov-Smirnov test (abbr. KS) which tests the equality of distributions by measuring the distance in the supremum norm between the two empirical distribution functions obtained from each of the two samples \citep[see Section 7.3 of][]{Gi}. In our framework, i.e. small sample sizes, the distribution of the Kolmogorov-Smirnov's statistic is obtained using a permutation  test as well.

{Finally, we consider the nonparametric test based on the $L_2$-distance between the two empirical characteristic functions; specifically, in order to test the null hypothesis $H_{0i}:F_{i}\equiv	G_{i}$, we consider the test statistic
	
	{\footnotesize
		\begin{eqnarray*}
			J_i&=& \dfrac{1}{{n}_1({n}_1-1)} \sum_{j=1}^{{n}_1} \sum_{l=1, l \not =
				j}^{{n}_1} \exp\left(-\dfrac{1}{2}\left(\dfrac{X_{ij}-X_{il}}{\sqrt{2}b}\right)^2\right)+\dfrac{1}{{n}_2({n}_2-1)} \sum_{j=1}^{{n}_2} \sum_{l=1, l \not =
				j}^{{n}_2} \exp\left(-\dfrac{1}{2}\left(\dfrac{Y_{ij}-Y_{il}}{\sqrt{2}b}\right)^2\right)\\&-&\dfrac{2}{n_1{n}_2}\sum_{j=1}^{{n}_1} \sum_{l=1}^{{n}_2} \exp\left(-\dfrac{1}{2}\left(\dfrac{X_{ij}-Y_{il}}{\sqrt{2}b}\right)^2\right),
		\end{eqnarray*}}

	 \noindent where $b>0$ is a smoothing parameter. The test statistic $J_i$ can be regarded as the $L_2$-norm of the difference between the kernel density estimators pertaining to the two samples. The average of the statistics $J_i$, $1\leq i \leq m$, was proposed and investigated in \cite{Marta2} in order to test for the global null hypothesis $H_0= \bigcap_{i=1}^{p}  H_{0i}$. However, here we investigate for the first time the performance of the individual tests (the $J_i$'s) in the multiple testing setting in which the aim is to identify which particular variables are differently distributed. We define the permutation test by determining the distribution of each $J_i$ under the permutation hypothesis, which yields a set of $P$-values following a discrete uniform distribution (in the classical sense) with support points $\{1/N,2/N, \dots, N/N\}$. Here, $N$ is the number of permutations that lead to different values of the statistic.

The null distribution of the $P$-values corresponding to the $J_i$ permutation test, absolute value test, Kolmogorov-Smirnov test, Wilcoxon test, Ansari-Bradley test and Siegel-Tukey test is discrete. More precisely, these $P$-values follow some discrete uniform distributions. In order to better understand the results reported in the next section, Table \ref{ca:sp} shows the corresponding support points of the distributions of the $P$-values for some sample sizes. Note that the discreteness of the $P$-values corresponding to the Kolmogorov-Smirnov test, Wilcoxon test, Ansari-Bradley test and Siegel-Tukey test is stronger than for the $J_i$-permutation test and absolute value test for which the support points are equally spaced. 

	\begin{table}[htb]
	\begin{center}
		\scalebox{0.6}[0.6]{ 
			\begin{tabular}{cllll}
	&{$J_i$ permutation test} & & abs test \\
	\hline		${n}_1,{n}_2$  & Support points && Support points \\
	4,4 & $\left\{\dfrac{1}{35}, \dfrac{2}{35}, \dots, \dfrac{35}{35}\right\}$ && $\left\{\dfrac{1}{35}, \dfrac{2}{35}, \dots, \dfrac{35}{35}\right\}$\\
	5,5 &  $\left\{\dfrac{1}{126}, \dfrac{2}{126}, \dots, \dfrac{126}{126}\right\} $ && $\left\{\dfrac{1}{126}, \dfrac{2}{126}, \dots, \dfrac{126}{126}\right\} $\\
					&{KS test} & & Wilcoxon test \\
	\hline		${n}_1,{n}_2$  & Support points && Support points \\
	4,4 & $\left\{\dfrac{1}{35}, \dfrac{8}{35}, \dfrac{27}{35}, \dfrac{35}{35}\right\} $ && $\left\{\dfrac{1}{35},\dfrac{2}{35},\dfrac{4}{35},\dfrac{7}{35},\dfrac{12}{35},\dfrac{17}{35},\dfrac{24}{35},\dfrac{31}{35},\dfrac{35}{35}\right\} $\\
	5,5 & $\left\{\dfrac{1}{126}, \dfrac{10}{126}, \dfrac{45}{126}, \dfrac{110}{126}, \dfrac{126}{126}\right\} $&&$\left\{\dfrac{1}{126},\dfrac{2}{126},\dfrac{4}{126},\dfrac{7}{126},
	\dfrac{12}{126},\dfrac{19}{126},\dfrac{28}{126},\dfrac{39}{126},\dfrac{53}{126},
	\dfrac{69}{126},\dfrac{87}{126},\dfrac{106}{126},\dfrac{126}{126} \right\} $ \\
					&{Siegel-Tukey test} & & Ansari-Bradley test \\
	\hline		${n}_1,{n}_2$  & Support points && Support points \\
	4,4 &  $\left\{\dfrac{1}{35},\dfrac{2}{35},\dfrac{4}{35},\dfrac{7}{35},\dfrac{12}{35},\dfrac{17}{35},\dfrac{24}{35},\dfrac{31}{35},\dfrac{35}{35}\right\} $ && $\left\{\dfrac{1}{35},\dfrac{5}{35},\dfrac{14}{35},\dfrac{26}{35},\dfrac{35}{35}\right\} $\\
	5,5 &  $\left\{\dfrac{1}{126},\dfrac{2}{126},\dfrac{4}{126},\dfrac{7}{126},
	\dfrac{12}{126},\dfrac{19}{126},\dfrac{28}{126},\dfrac{39}{126},\dfrac{53}{126},
	\dfrac{69}{126},\dfrac{87}{126},\dfrac{106}{126},\dfrac{126}{126} \right\} $ && $\left\{\dfrac{2}{126},\dfrac{6}{126},\dfrac{18}{126},\dfrac{38}{126},
	\dfrac{68}{126},\dfrac{104}{126},\dfrac{126}{126}\right\} $\\
			\end{tabular}
		}
	\end{center}
	\caption{Support points of the $P$-values derived from the $J_i$ permutation test, abs test, KS test, Wilcoxon test, Ansari-Bradley test and Siegel-Tukey test for different sample sizes.}
	\label{ca:sp}
\end{table}
\section{Simulation study}\label{se:simul3}
{In this section we consider the two-sample problem with low sample size, along a large number of variables. In the Suplementary Material additional simulations for
	the one-sample problem in the same low sample size and high dimensional setting are provided too.}
{
	The aims of the simulation study are the following:
	\begin{itemize}
		\item[(1)] to compare the performance of the different $q$-value methods in Section \ref{se:MCP};
		\item[(2)] to compare the performance of the $\pi_0$ estimators in Section \ref{se:MCP};
		\item[(3)] to study the behaviour of the different two-sample tests in Section \ref{se:tests}.
\end{itemize}}

%Conclusions are reported separately for location, scale and more general differences in distribution in Sections \ref{se:lo}, \ref{se:sd} and \ref{se:lss}, respectively.

We consider a vector autoregressive model of order 1 (or multivariate autoregressive model), VAR$(1)$, defined as 
$
W_{t}=A W_{t-1}+ \varepsilon_t,
$
where $W_t=(W_{t1}, \cdots, W_{t\eta})^T$, $A=(a_{ij})$ is an $\eta
\times \eta$ design matrix such that the proccess $(W_t)_{t \in \mathbb{N}}$
is stationary, $\eta$ is the sample size, and
$\varepsilon_t \in \mathbb{R}^{\eta}$ are i.i.d. 
random vectors (the innovations). We generate a time series of length $m$ from the vector autoregressive 
model with innovations $\varepsilon_t \sim N_{\eta}(0, I_{\eta})$ and initial
point $W_0 \sim
N_{\eta}(0, \Sigma)$ where $\Sigma$ is the stationary covariance matrix,
i.e,  $\Sigma= A^T \Sigma A+ I_{\eta}$ \citep[Lyapunov equation; see][]{hamilton}. The vectors $X_i$ (resp. $Y_i$, $i=1, \dots, m$) consist on i.i.d. observations $W_i$, $i=1, \dots, m$. Specifically, $X=[X_1, \dots, X_m]^T$ and $Y=[Y_1, \dots,Y_m]^T$ are based on a standarization of $W=[W_1, \dots, W_m]^T$.

Depending on the choice of the design matrix $A$ a particular degree of dependence is obtained.

%{\color{blue}Considering dependent data in our simulated study we can conclude if the dependence induced on the $P$-values, and hence its effects on the null distribution of the $P$-values has or not an effect on the performance of the considered methods. This provides a completed simulation study.}

In this study, we consider two possibilities for $A$,
each of which
is an $\eta \times \eta$ lower triangular matrix with elements $a_{ij}$
satisfying $a_{ij}=0$ for $i-j>1$ ($\eta={n}_1$ or $\eta={n}_2$ depending on
whether one is simulating $X$ or $Y$): 
\begin{itemize}
	\item {Independence} is simulated by setting
	\begin{equation}\label{ec:A13}
	a_{ii}=0, \;i=1, \dots,\eta,\; \text{and} \;a_{i,i-1}=0, \;i=2, \dots,\eta. 
	\end{equation} 
	\item {Medium dependence}
	of $X_{ij}$ and $X_{kj}$ for $i\not=k$ and {strong dependence} of $X_{ij}$ and $X_{lk}$ for $i \not= l$ and $j \not= k$ is  is simulated by setting
	\begin{equation}\label{ec:A23}
	a_{ii}=0.5, \;i=1, \dots,\eta,\; \text{and} \;a_{i,i-1}=0.4, \;i=2, \dots,\eta. 
	\end{equation}
\end{itemize}

%The smoothing parameter $b_i$ for the statistics $J_i$
%was selected 
%
%in \cite{Marta2} as $\widehat{b}=1.144 s_{pool} \left(({{n}_1+{n}_2})/{2}\right)^{-1/5},$
%where $s_{pool}^2$ is the average of
%$(({n}_1-1)s_{X_i}^2+({n}_2-1)$$s_{Y_i}^2))/({n}_1+{n}_2-2)$, $i=1, \dots,m$, and
%$s_{X_i}^2$ and $s_{Y_i}^2$ are the sample variances of $X_i$ and
%$Y_i$, respectively, $i=1, \dots, m$. This proposal is reasonable when the aim is to test the global null hypothesis. However, in the current paper we focus on the individual hypotheses $H_{0i}$. For this reason, instead of using a common value $\widehat{b}$ for all the statistics $J_i$, $i=1, \dots,m$, we propose to use
%$\widehat{b}_i=1.144 \sqrt{\left(\left(({n}_1-1)s_{X_i}^2+({n}_2-1)s_{Y_i}^2\right)/({n}_1+{n}_2-2)\right)}\left(({{n}_1+{n}_2})/{2}\right)^{-1/5},\; i=1, \dots, m,$
%which takes into account the population-specific smoothing degree.

%Firstly, we focused on the setting in which the two populations differ only in their location. We start by describing the procedure which generates the data sets  $X=\left[X_1, \dots, X_m\right]^T$ and $Y=\left[Y_1, \dots, Y_m\right]^T$ with$X_i=(X_{i1}, \dots,$ $X_{i{n}_1})$ and $Y_i=\left(Y_{i1}, \dots,Y_{i{n}_2}\right)$, $i=1, \dots,m$.

In order to simulate $X_i$ we first define $X_i^{(0)}=\Sigma^{-1/2} W_i$, where $W_i$ are the vectors generated from the VAR$(1)$ model
with stationary covariance matrix $\Sigma$. Let $\{f_1,f_2, f_3, f_4\}$ be a collection of four densities, and let $I = \{I_j : j \in \{1, \dots,m\}\}$ be a
sequence of i.i.d. random variables such that $P(I_1=j)=\omega_j$,
with $\omega_j=1/4$ for $j=1, \dots,4$. Then, we take  
${X}_i=F_{I_i}^{-1}(\Phi(X_i^{(0)}))$, where $F_i$ is
the cdf corresponding to the density
$f_i, i=1, \ldots,4$, and $\Phi$ stands for the cdf of the standard normal. On the other hand, the data set $Y$ is generated as ${Y}_i=F_{L_i}^{-1}(\Phi(Y_i^{(0)}))$, where $Y_i^{(0)}=\Sigma^{-1/2} W_i$ and where $L = \{L_j : j \in \{1,
\dots,m\}\}$ is a sequence of i.i.d random variables defined in the following way: given $I=i$, $L$ takes the same value with probability $P(L_1=i| I_1=i) = r_{ii}=1-\delta$, and a different value with probabilities 
$P(L_1=j | I_1=i) = r_{ij}$, where $r_{31}=r_{42}=r_{13}=r_{24}=\delta$ and $r_{ij}=0$ otherwise; here we take 
$\delta=0, 0.3, 0.5$. Note that the proportion of null hypotheses in these settings is $\pi_0=1-\delta$. 

%Table \ref{table:parameters1} summarizes all the different parameters and their respective ranges.

%PROPONGO SUPRIMIR LA TABLA 3.

 %	The procedure of generating the data sets $X$ and $Y$ in the setting in which exist scale differences between the two distributions rather than differences in location is analogous to the one described above. The unique difference is that the collection of densities from which the samples are generated changes to $\{f_1,f_2,f_3,f_4\}=\{N(0,1/4),N(3,1/4), N(0,4),N(3,9)\}$. The parameters $\delta$ and $m$ take the same values as in the previous sections, whereas the sample sizes $n_1,n_2$ change to ${n}_1={n}_2=8$ (for smaller sample sizes the methods report no power).

%Finally, following the same procedure of generating the data sets $X$ and $Y$ we designed a setting where the populations differ in location (although only slightly), scale and shape.
%The collection of densities from which the samples are generated changes to $\{f_1,f_2,f_3,f_4\}=\{N(2.5,1/4),N(3.5,1/4), Exp(1/2),$
%$Exp(1/3)\}$. The values of $\delta$, $n_1,n_2$ and $m$ are the same as in the scale differences setting.

%The simulated settings can be classified in three groups, depending on the type of differences between the two distributions being compared. The first group  of settings corresponds to location differences, whereas in the second group the distributions differ due to their scale parameters. Finally, we consider settings where the distributions differ in location, scale and shape.

The family of densities $\{f_1,f_2, f_3, f_4\}$ is chosen in order to simulate differences in location, scale or shape. Specifically,

\begin{itemize}
\item $\{f_1,f_2, f_3, f_4\}=\{N(0,1),N(0,1/4),N(\mu,1),N(\mu,1/4)\}$ with $\mu=2$ or $\mu=3$ for location;
\item $\{f_1,f_2,f_3,f_4\}=\{N(0,1/4),N(3,1/4), N(0,4),N(3,9)\}$ for scale;
\item $\{f_1,f_2,f_3,f_4\}=\{N(2.5,1/4),N(3.5,1/4),Exp(1/2),Exp(1/3)\}$ for shape.
\end{itemize}

The third scenario involves differences in scale too, location differences being minor otherwise. The dimension is $m=100$ or $m=1000$. The proportion of true null hypothesis $\pi_0=1-\delta$ is $1$, $0.7$ or $0.5$. The sample sizes are $n_1=n_2=4$ and $n_1=n_2=5$ for location differences, and are increased to $n_1=n_2=8$ for scale and shape differences in order to get some statistical power. The number of Monte Carlo replicates is $1000$. 
%(CORRECTO? NO HE ENCONTRADO ESTA INFO EN EL PAPER NI EN EL SUPPL MAT)

%{
%	\begin{table}[H]
%		\begin{center}
%			\scalebox{0.8}[0.8]{ 
%				\begin{tabular}{|c|c|c|c|l|}
%					\hline
%					Sample sizes $n_1$, $n_2$ & Dimension $m$ & Location parameter $\mu$ & $\delta=1-\pi_0$ & Design matrix $A$\\
%					4,4 & 100& 2& 0 & (\ref{ec:A13}) Independence\\
%					5,5 & 1000 & 3 & 0.3 & (\ref{ec:A23}) Dependence\\
%					& & &0.5& \\
%					\hline
%			\end{tabular}}	
%		\end{center}
%		\caption{The different parameters considered in the simulation settings with location differences and their respective ranges.}
%		\label{table:parameters1}
%	\end{table}

Under the global null hypothesis ($\pi_0=1$), all the tests control the FDR at the nominal level (results not shown). The FDR is approximately zero for the nonparametric tests, whereas for the parametric ones the FDR is about 0.03. These results suggest that the tests are overly conservative. The full set of simulation results for $\pi_0<1$ (i.e. $\delta>0$) is provided along seventeen Tables in the Supplementary Material. In general, it is seen that the statistical power increases with the proportion of non-true nulls. The same holds true for the effect $\mu$ in the case of location differences. However, the power remains roughly the same when moving from the scenario with $m=100$ hypotheses to that with $m=1000$. In Figures \ref{fig:Table8} and \ref{fig:Table10} (location differences), Figure \ref{fig:Table16} (scale differences) and  Figure \ref{fig:Table22} (shape differences) we graphically display results on the FDR and power for selected scenarios. The Monte Carlo bias and standard deviation of the several estimators of $\pi_0$ in one of the location scenarios are given in Table \ref{ca:pi0}.

%, although a piece of these results is shown graphically in Figures \ref{fig:Table8} and \ref{fig:Table10} (Location setting), Figure \ref{fig:Table16} (Scale setting) and  Figure \ref{fig:Table22} (Location, scale and shape setting). Table \ref{ca:pi0} also shows results above the performance of the $\pi_0$ estimators in the Location setting.

{

	\begin{figure}[H]
		\begin{center}$
			\begin{array}{ccc}
			\includegraphics[width=0.32\linewidth]{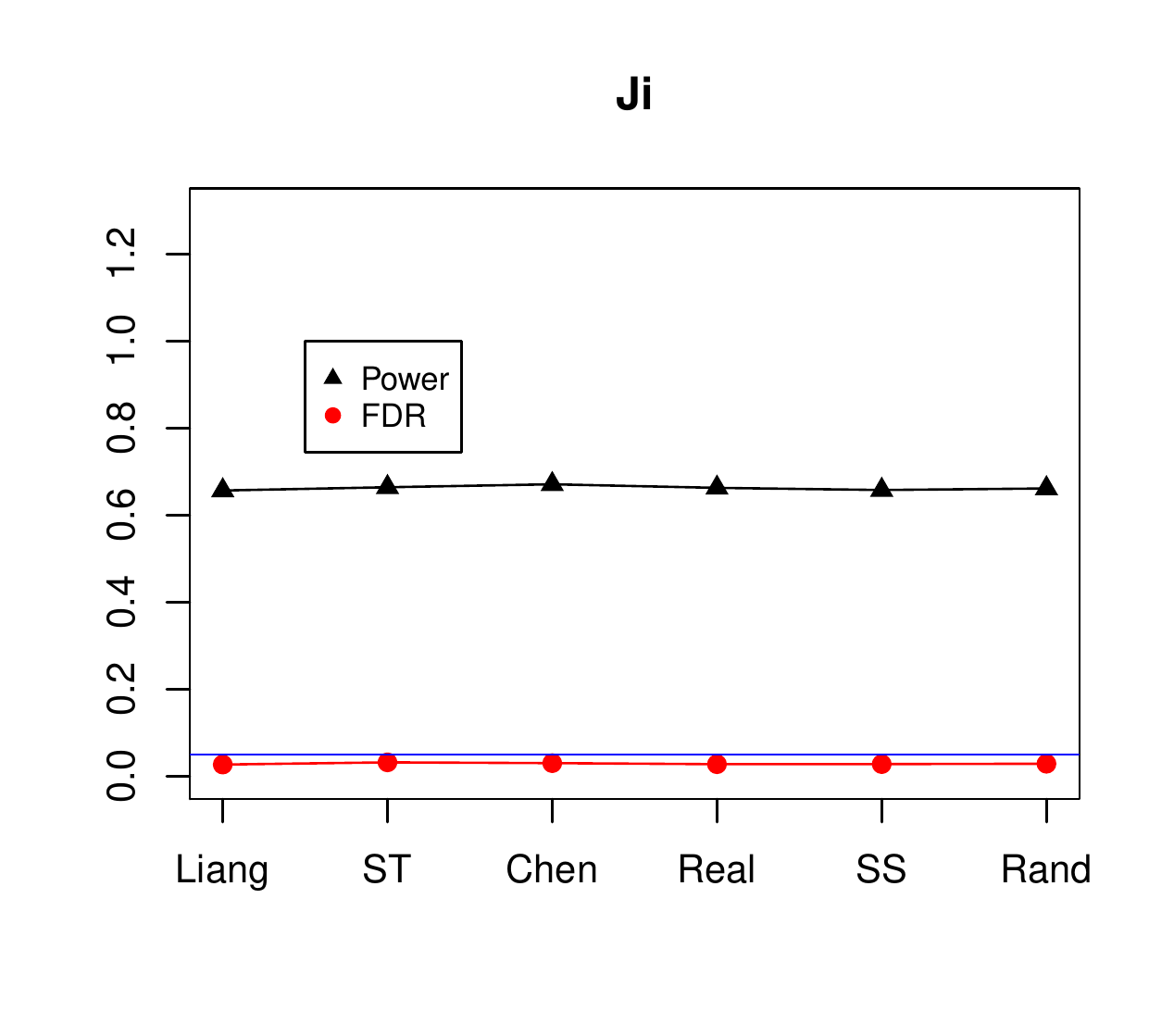}&
			\includegraphics[width=0.32\linewidth]{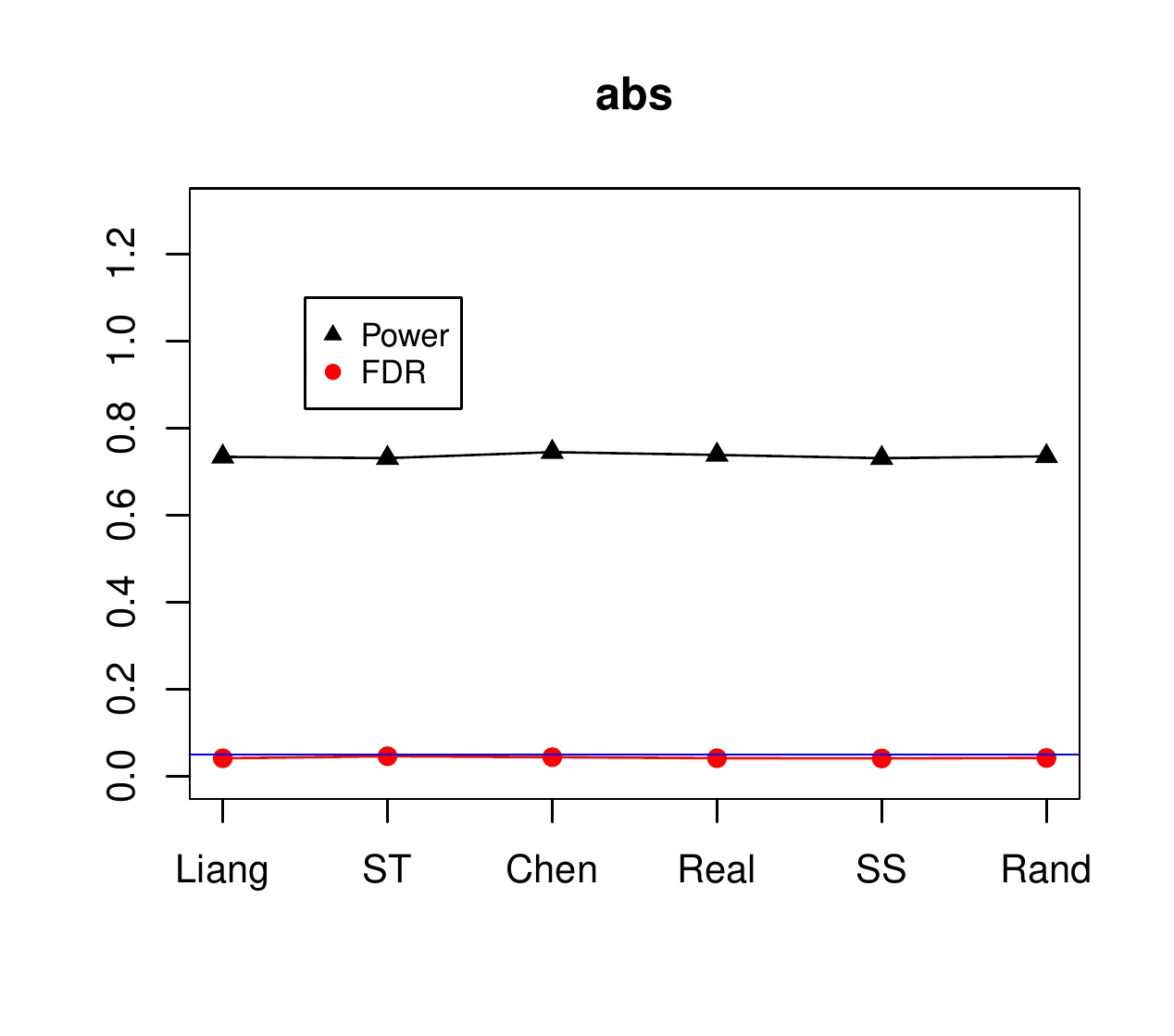} & \includegraphics[width=0.32\linewidth]{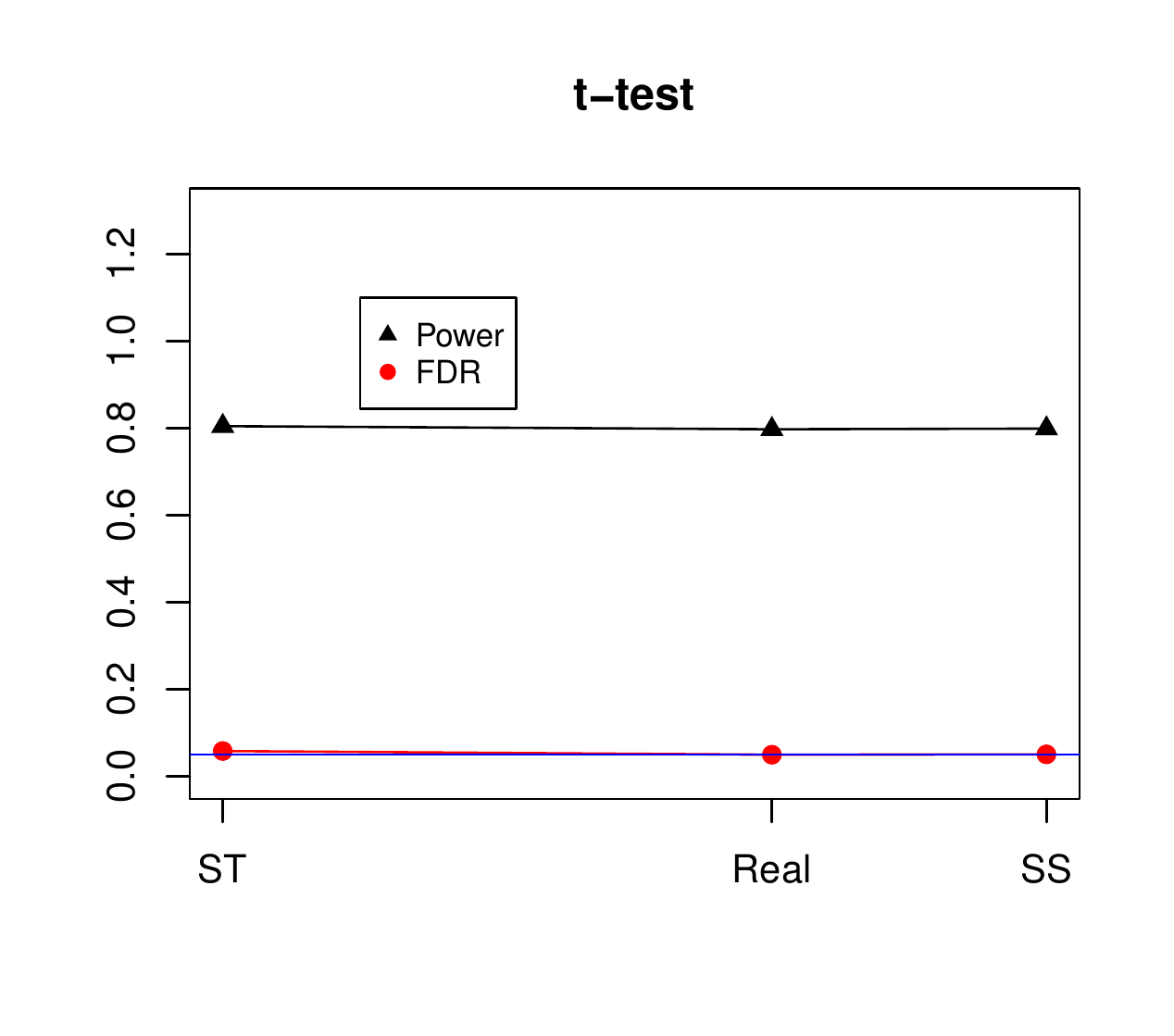}\\
			\end{array}$
			$\begin{array}{cc}
			\includegraphics[width=0.32\linewidth]{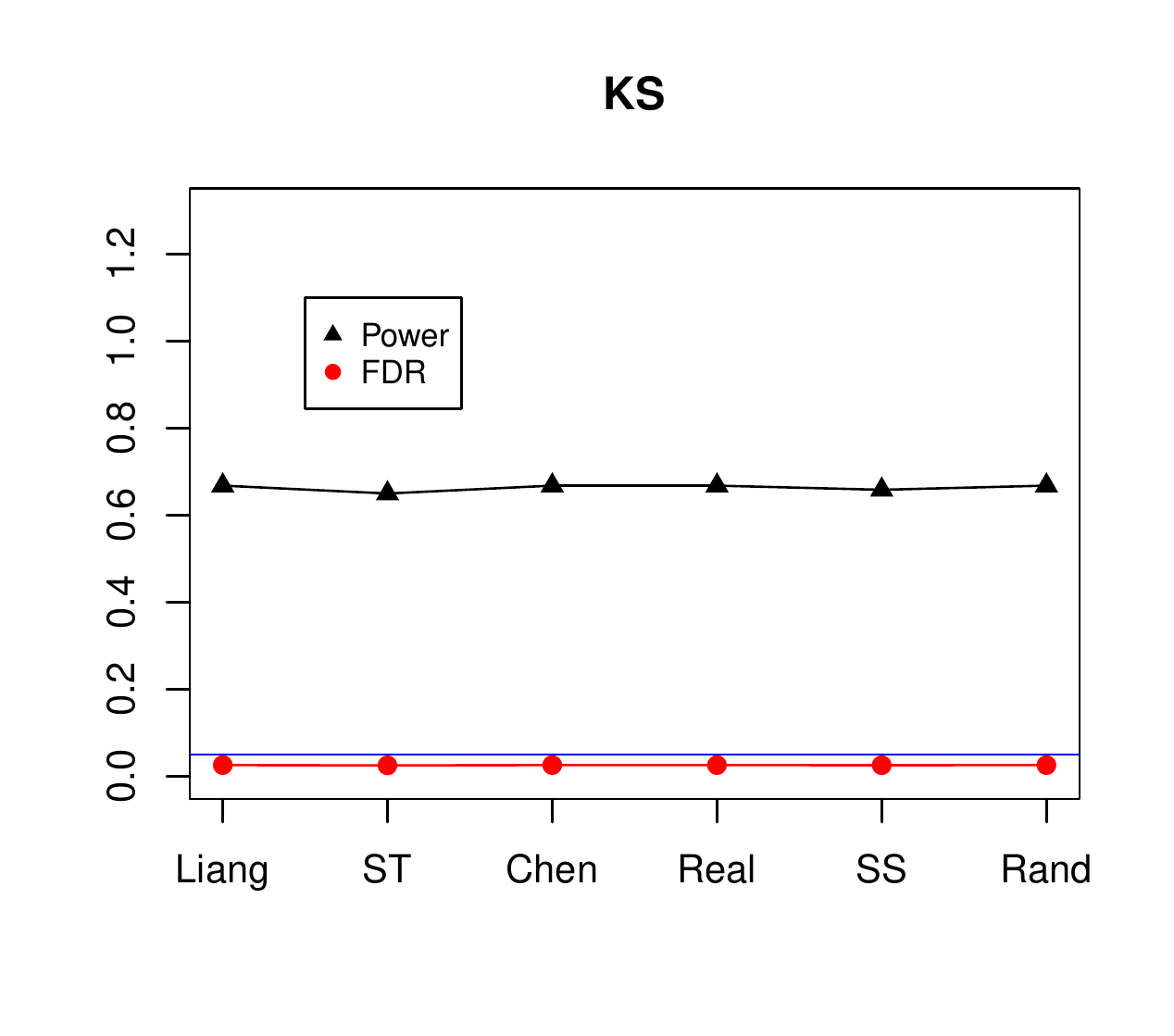}&
			\includegraphics[width=0.32\linewidth]{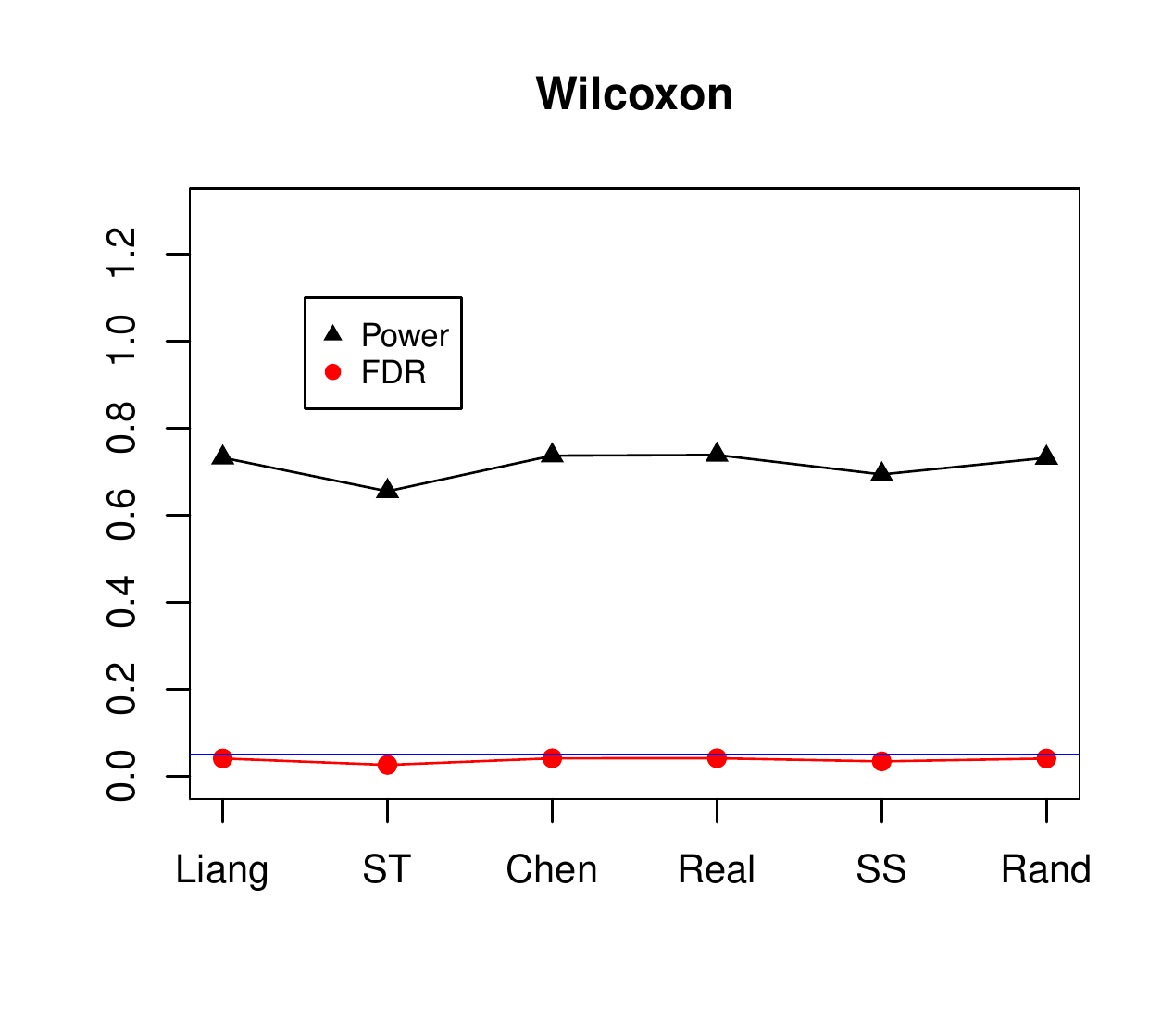}
			\end{array}$
		\end{center}
		\caption{Location differences with ${n}_1={n}_2=5$, $m=100$, $\delta=0.3$, $\mu=2$ and $A$ given by (\ref{ec:A13}). The Monte Carlo estimator of the FDR and power are reported for each test and $q$-value method. The blue line corresponds to $\alpha=0.05$.}
		\label{fig:Table8}
	\end{figure}

Among the several $q$-value procedures, the best results for hdu $P$-values are achieved by the Chen method. Indeed, the power of the Chen method is comparable to (and sometimes larger than) that corresponding to the benchmark method which uses the true $\pi_0$ (labelled as \textit{Real} in Figures and Tables). Liang and Rand methods perform correctly too. However, the $q$-value methods for continuous $P$-values, SS and ST, perform badly when applied to discrete uniform $P$-values; an exception is found in settings where the  discreteness of the $P$-values is weak. Generally speaking, it is seen that the discrete methods improve their continuous counterparts regardless the particular permutation test which is employed.

\begin{table}[htb]
	\begin{center}
		\scalebox{0.7}[0.7]{ 
			\begin{tabular}{cccccccccccccccc}
				\hline
				& 	& \multicolumn{14}{c}{Two-sample tests ($m=100$)} \\
				&  & \multicolumn{2}{c}{$J_i$} && \multicolumn{2}{c}{abs}
				&& \multicolumn{2}{c}{t-test}
				&& \multicolumn{2}{c}{KS}
				&& \multicolumn{2}{c}{Wilcoxon}\\
				\cline{3-4}\cline{6-7} \cline{9-10} \cline{12-13} \cline{15-16}\\
				$\widehat{\pi}_0$	& &  Bias & Sd &&	Bias & Sd && Bias & Sd && Bias & Sd && Bias & Sd\\	
				Linag&&	0.0367&	0.0459	&&	0.0167&	0.0418	&&	-&-		&&	0.0137&	0.0523	&&	0.0215	&0.0467	\\
				ST&&	0.0430&	0.1954	&&	0.0315&	0.1972	&&	0.0121&	0.1871	&&	0.4969&	0.0192	&&	0.3855&	0.1670	\\
				Chen&&	-0.0032&	0.0536	&&	-0.0193&	0.0532	&&	-&	-	&&	0.0229&	0.0453	&&	0.0026&	0.0481	\\
				SS&&	0.0266	&0.0714	&&	0.0104	&0.0707	&&	0.0025&	0.0715	&&	0.1604&	0.0672	&&	0.0880&	0.0692	\\
				Rand&&	0.0185&	0.0718	&&	0.0028&	0.0711	&&	-&-		&&	0.0131&	0.0536	&&	0.0067&	0.0653	\\
				\hline 		
				& 	& \multicolumn{14}{c}{Two-sample tests ($m=1000$)} \\
				&  & \multicolumn{2}{c}{$J_i$} && \multicolumn{2}{c}{abs}
				&& \multicolumn{2}{c}{t-test}
				&& \multicolumn{2}{c}{KS}
				&& \multicolumn{2}{c}{Wilcoxon}\\
				\cline{3-4}\cline{6-7} \cline{9-10} \cline{12-13} \cline{15-16}\\
				$\widehat{\pi}_0$	& &  Bias & Sd &&	Bias & Sd && Bias & Sd && Bias & Sd && Bias & Sd\\	
				Liang&&	0.0266&	0.0199	&&	0.0099&	0.0172	&&	-&-		&&	0.0134	&0.0176	&&	0.0095&	0.0193	\\
				ST&&	0.0371&	0.0631	&&	0.0280&	0.0680	&&	0.0009&	0.0655	&&	0.5000	&0.0000	&&	0.4683&	0.0548	\\
				Chen&&	0.0097&	0.0190	&&	-0.0049&	0.0179	&&	-&-		&&	0.0149	&0.0175	&&	0.0044&	0.0163	\\
				SS&&	0.0254 &	0.0228	&&	0.0121&	0.0221	&&	0.0042&	0.0222	&&	0.1601&	0.0226	&&	0.0873&	0.0225	\\
				Rand &&	0.0170&	0.0229	&&	0.0041	&0.0219	&&	-&-		&&	0.0126&	0.0179	&&	0.0058&	0.0211	\\
				\hline
				
		\end{tabular}}	
	\end{center}
	\caption{
		Location differences with ${n}_1={n}_2=5$, $m=100$ and $m=1000$, $\delta=0.5$, $\mu=2$ and $A$ given by (\ref{ec:A13}).  The Monte Carlo bias and standard deviation of each $\pi_0$ estimator are provided.	}
	\label{ca:pi0}
\end{table}

%The t-test is not very sensitive to the specific $q$-value approach employed to take the multiplicity issue into account. However, other tests are really sensitive to the $q$-value method. For example, only the $q$-value methods for discrete $P$-values perform correctly for the KS test, whereas the SS and ST methods may report zero power (see Tables \ref{ca:L_2_1000_4_0.5} and \ref{ca:L_2_1000_4_0.5_dep}). 

With respect to the estimation of $\pi_0$ it is seen that, for continuous $P$-values (i.e. for the parametric tests), both the ST and the SS procedures report estimates with a small positive bias which decreases as $m$ increases, the standard deviation being decreasing too. The bias of the ST is somehow smaller than that of SS (this is particularly clear for $m=1000$), while the SS approach entails a smaller variance (see e.g. Table \ref{ca:pi0}). For the discrete tests, the behaviour of the ST and SS $q$-value procedures is not so promising. Even when their standard deviation decrease for an increasing $m$, they exhibit a large positive bias which remains roughly constant when moving from $m=100$ to $m=1000$. This suggests the inconsistency of such $\hat \pi_0$'s. On the other hand, among the three estimators proposed for discrete $P$-values, the method with the smallest bias is Chen, Rand being competitive in most of the scenarios. It should be noted however that Chen method shows a systematic bias in the simulated settings, 
although of small magnitude (Table \ref{ca:pi0}).

	\begin{figure}[H]
	\begin{center}$
		\begin{array}{ccc}
		\includegraphics[width=0.32\linewidth]{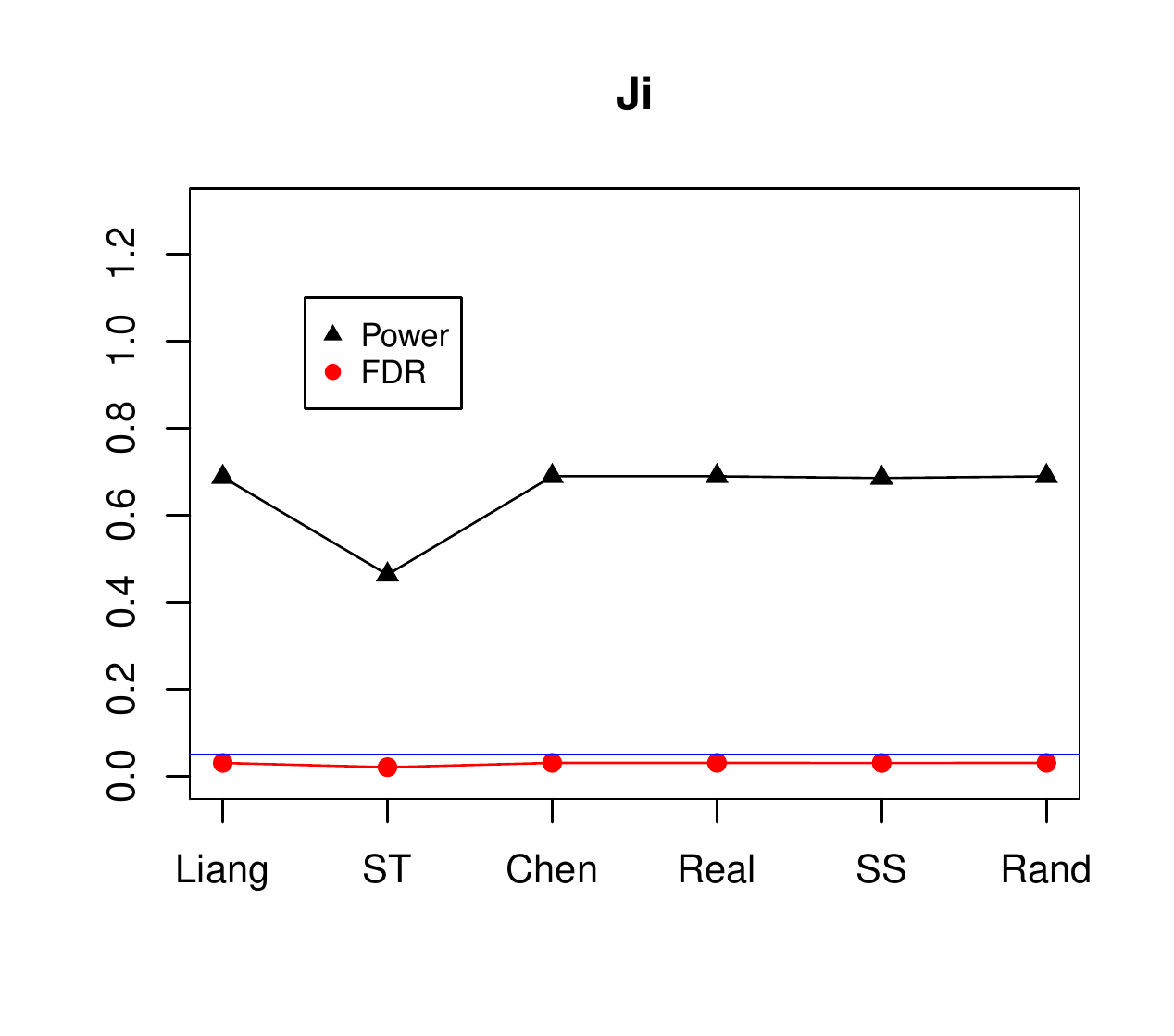}&
		\includegraphics[width=0.32\linewidth]{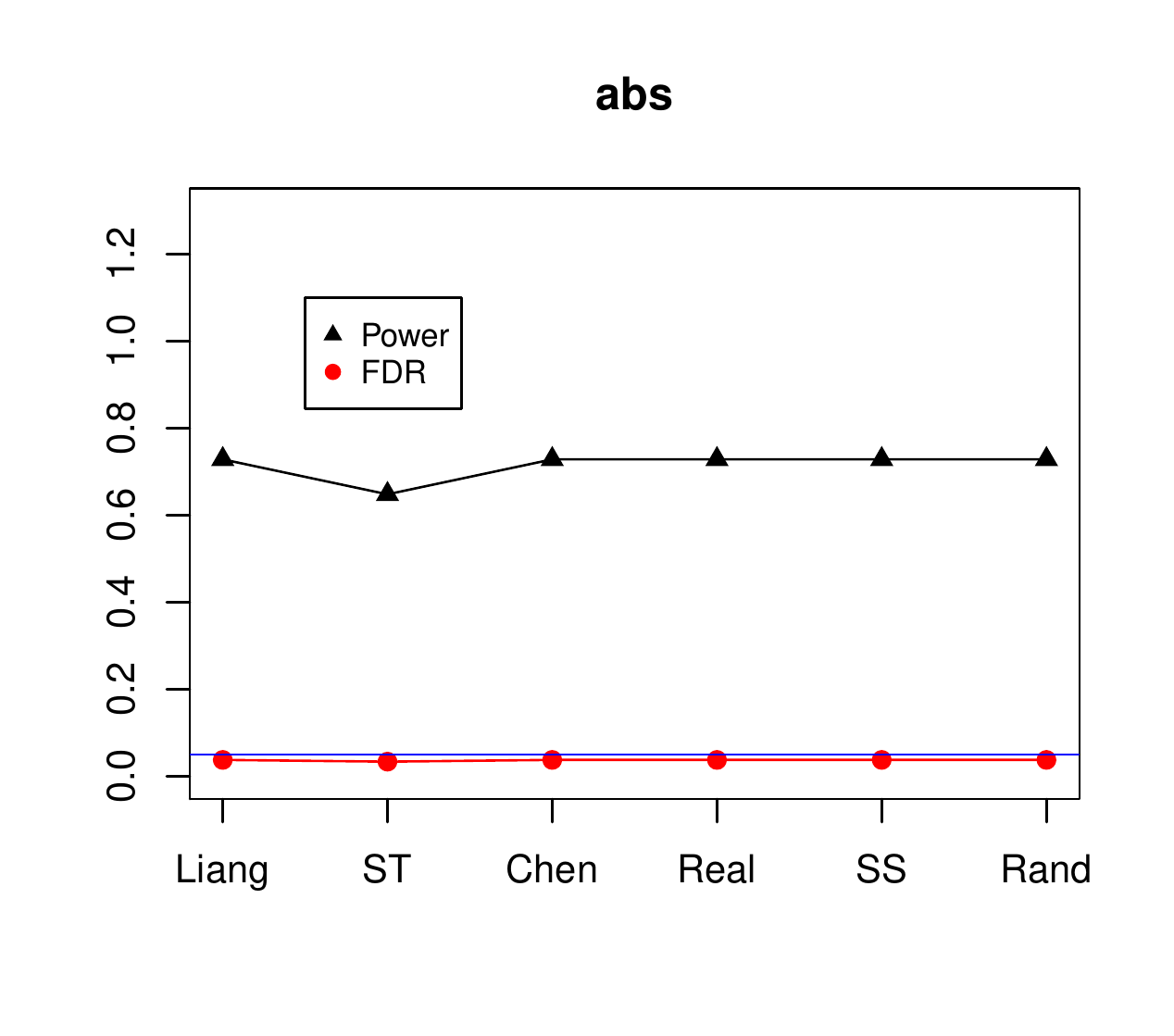}&\includegraphics[width=0.32\linewidth]{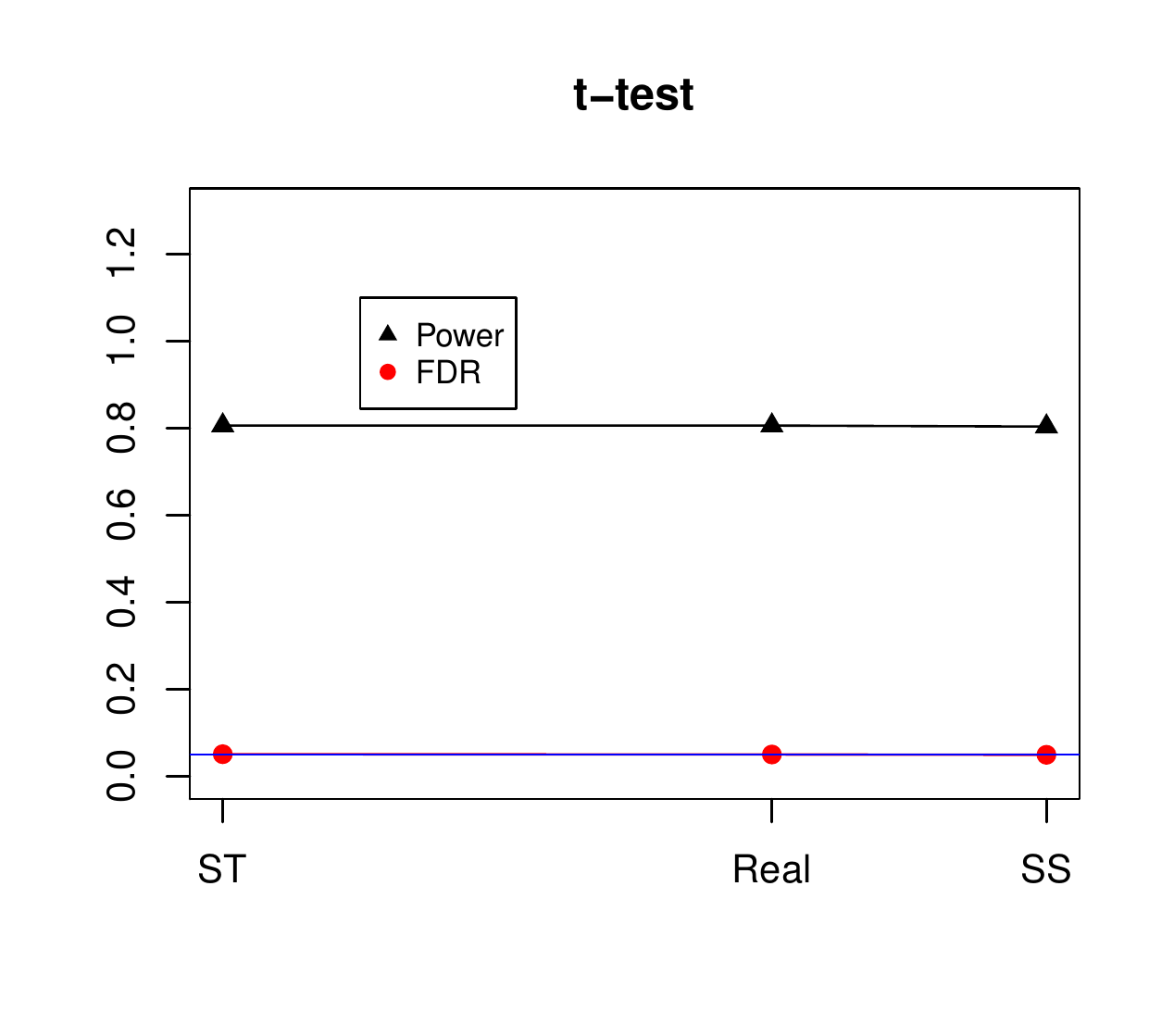}\\
		\end{array}$
		$
		\begin{array}{cc}
		\includegraphics[width=0.32\linewidth]{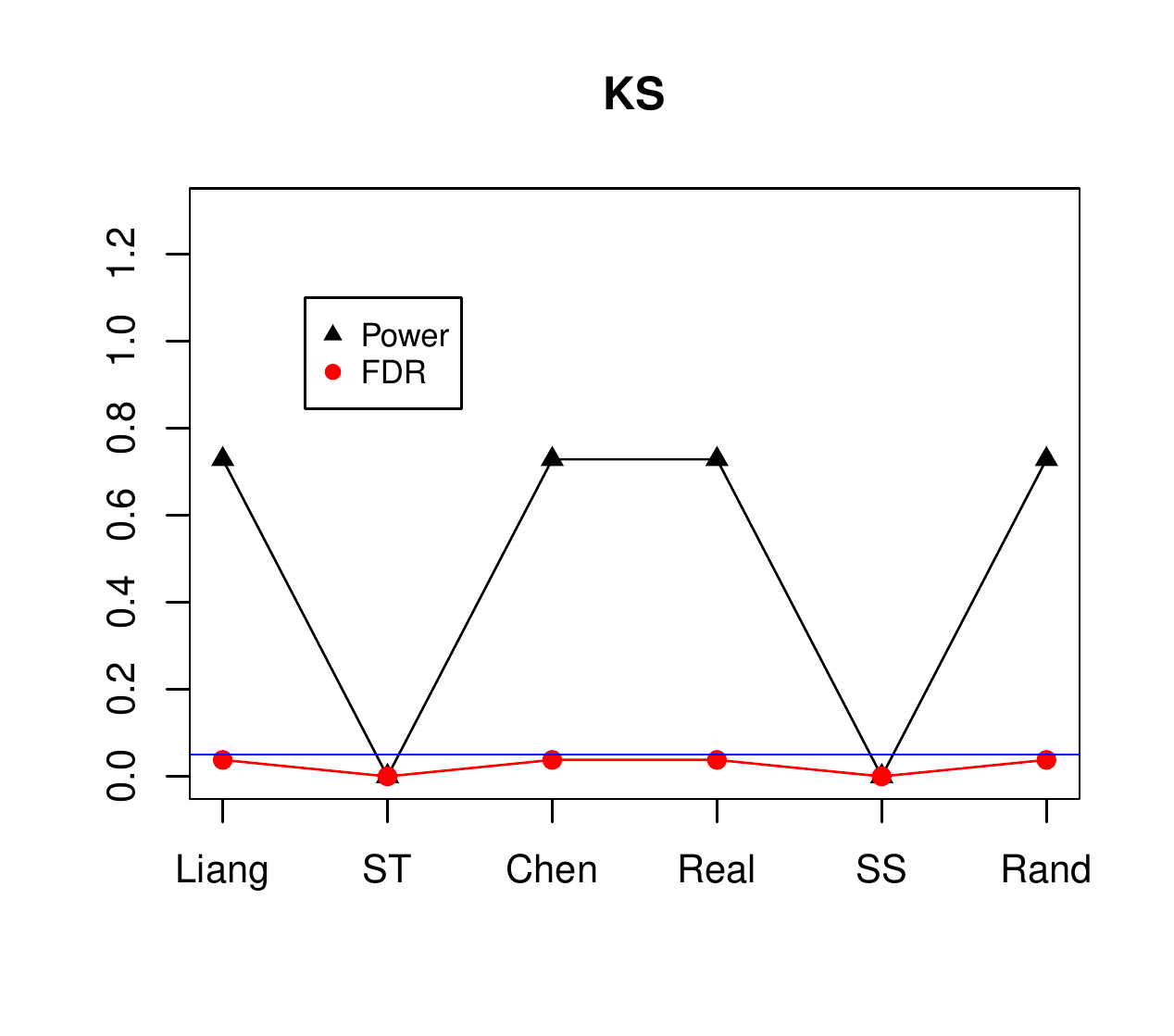}&\includegraphics[width=0.32\linewidth]{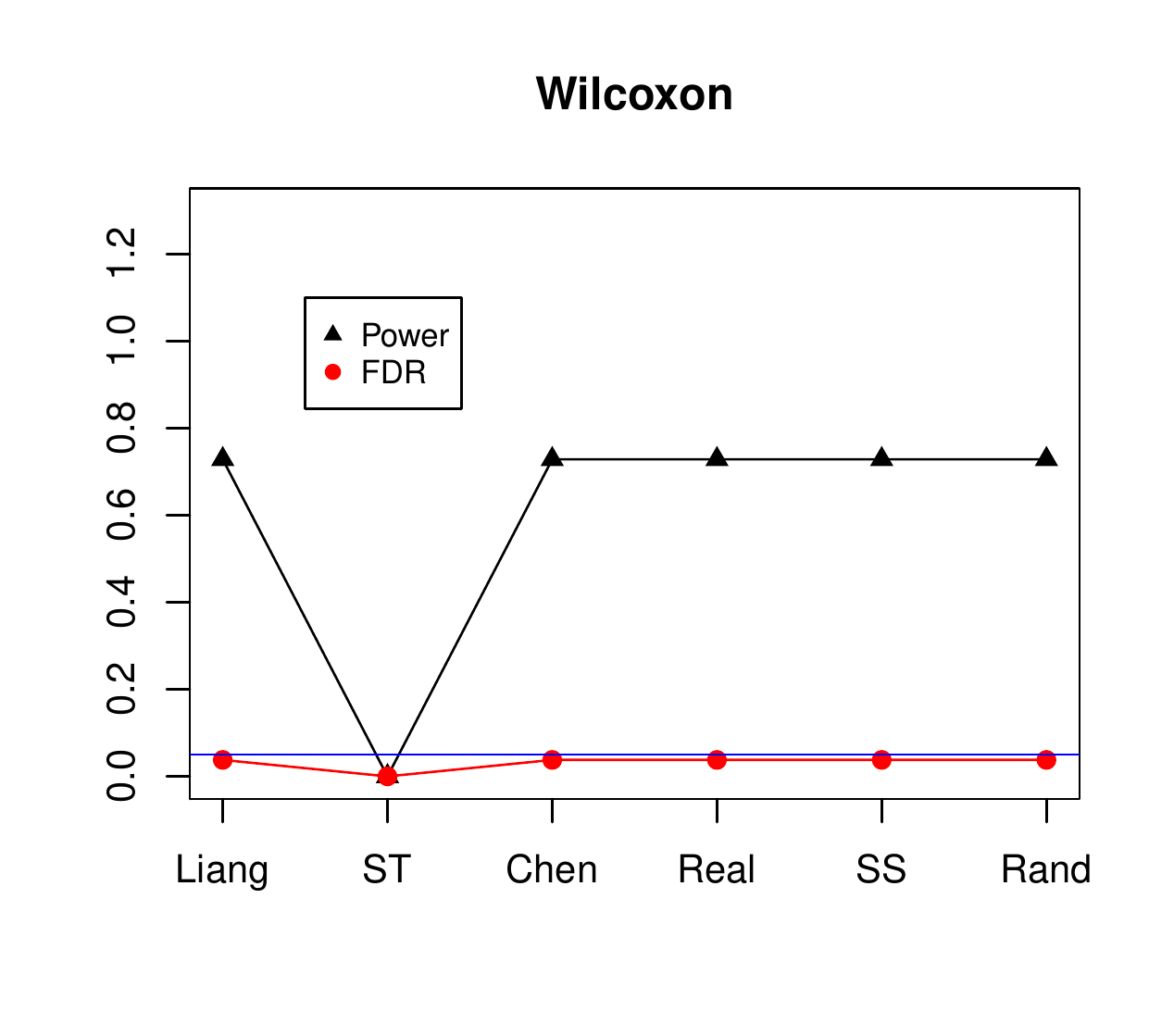}\\
		\end{array}$
		
	\end{center}
	\caption{Location differences with ${n}_1={n}_2=4$, $m=1000$, $\delta=0.5$, $\mu=2$ and $A$ given by (\ref{ec:A13}). The Monte Carlo estimator of the FDR and power are reported for each test and $q$-value method. The blue line corresponds to $\alpha=0.05$.}
	\label{fig:Table10}
\end{figure}

From our simulation results, interesting conclusions on the relative performance of the tests can be obtained. For differences in location, the optimal procedure is the t-test, as expected. The power of the abs and the Wilcoxon tests is uniformly larger than that of the local test based on the $J_i$ while, depending on the setting, the KS may provide larger, roughly equal, or smaller power relative to the $J_i$ test (see Figures \ref{fig:Table8} and \ref{fig:Table10}). 

\begin{figure}[H]
	\begin{center}$
		\begin{array}{ccc}
		\includegraphics[width=0.32\linewidth]{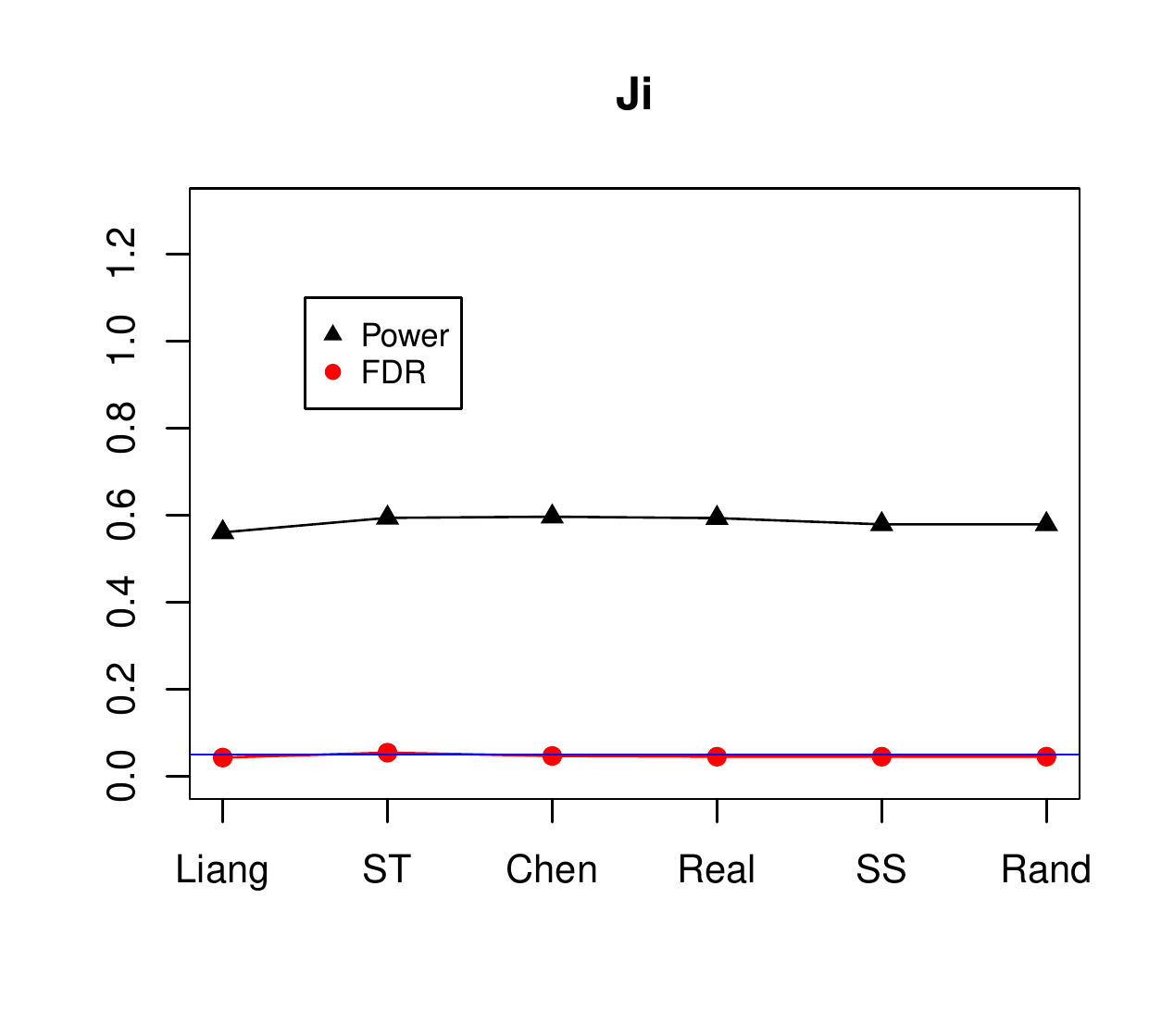}&
		\includegraphics[width=0.32\linewidth]{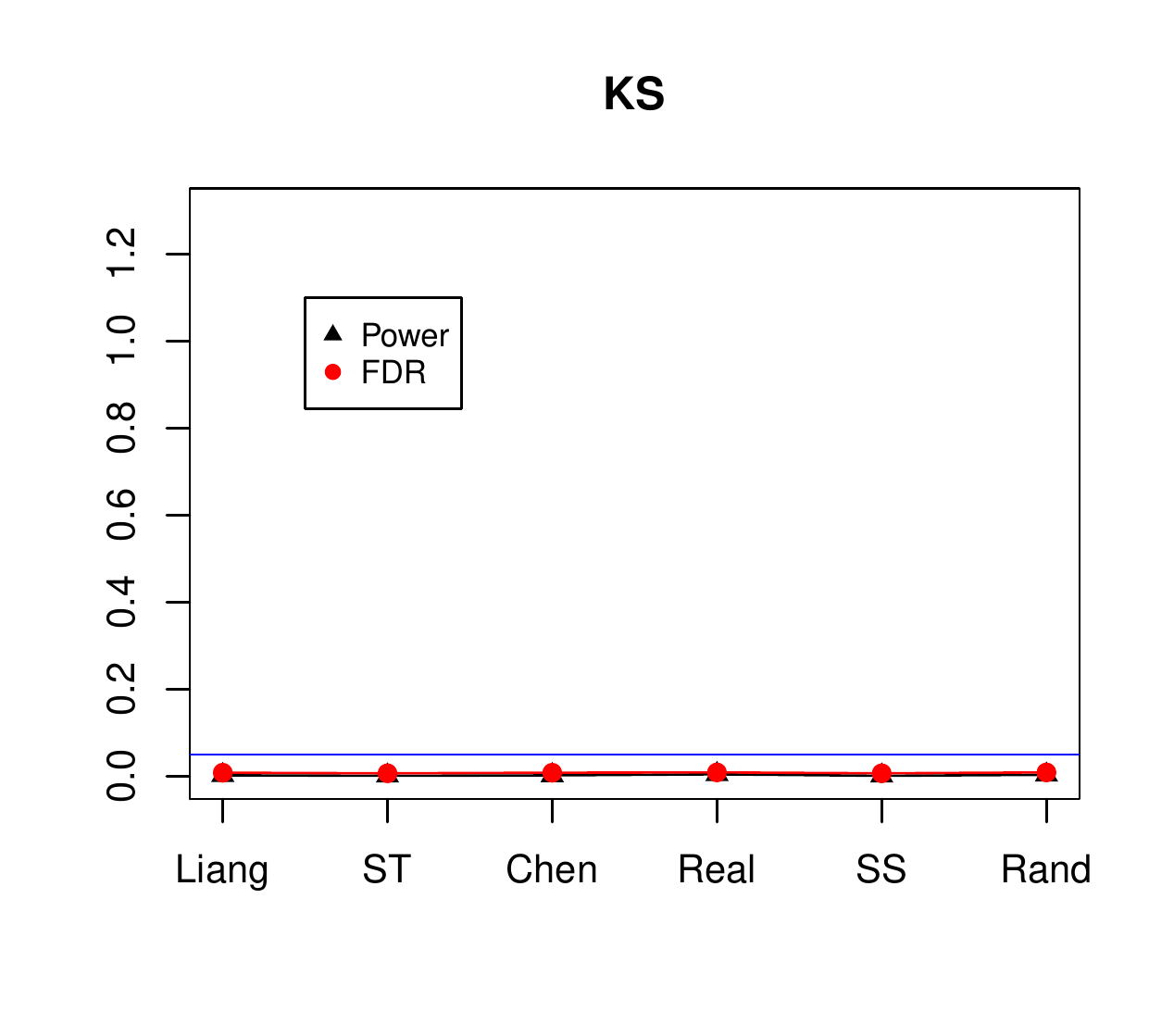} & \includegraphics[width=0.32\linewidth]{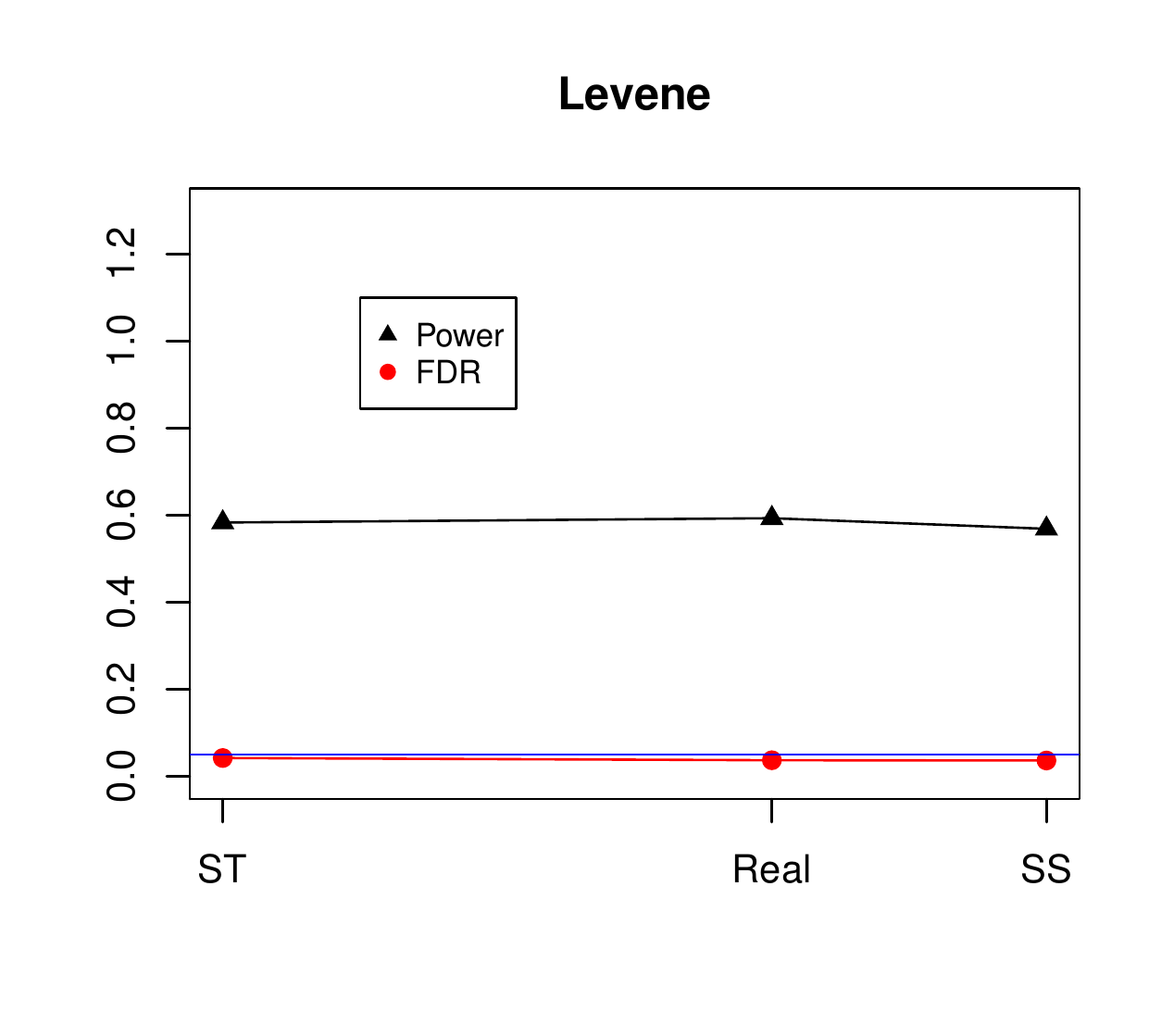}\\
		\end{array}$
		$\begin{array}{ccc}
		\includegraphics[width=0.32\linewidth]{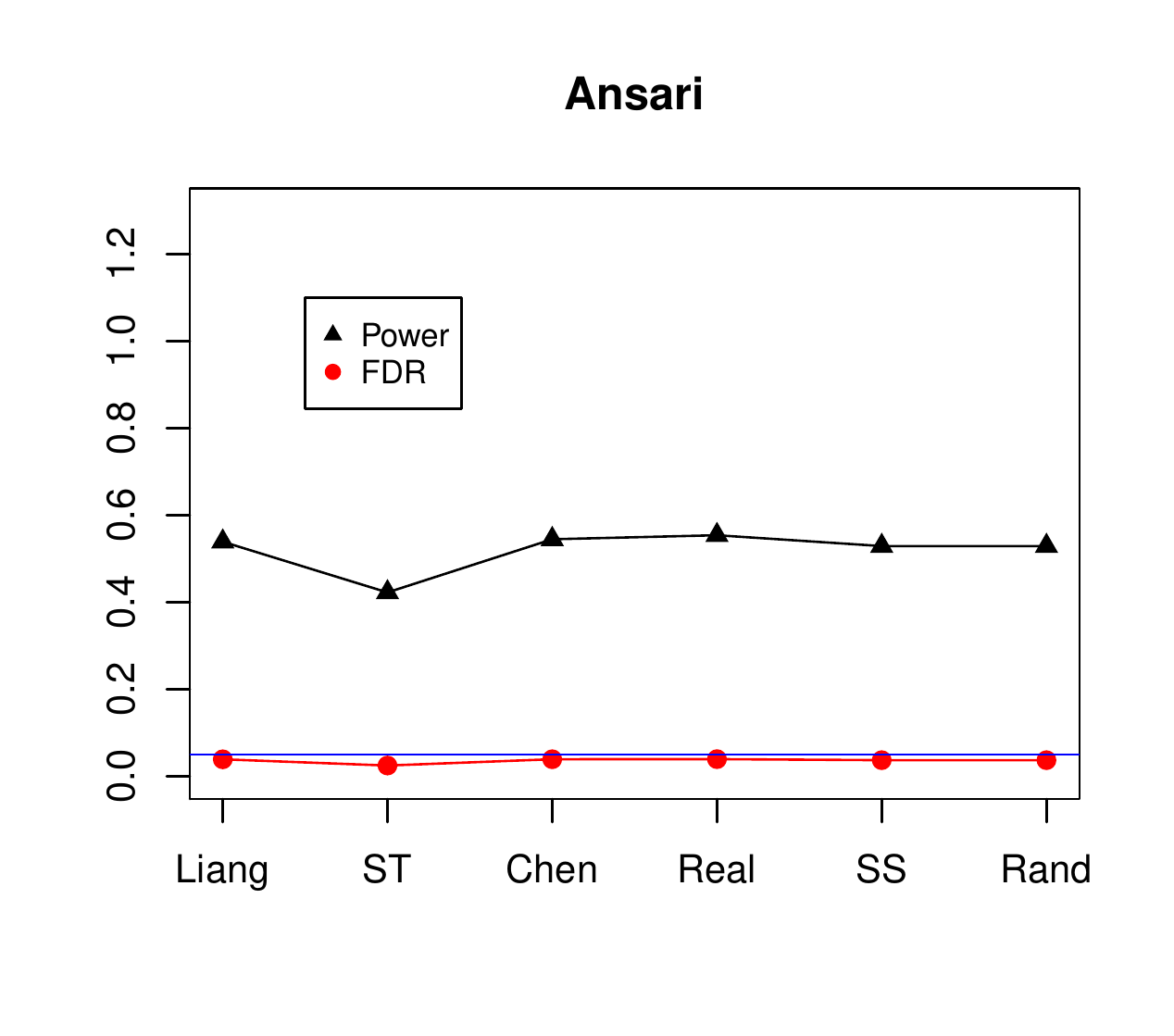}&
		\includegraphics[width=0.32\linewidth]{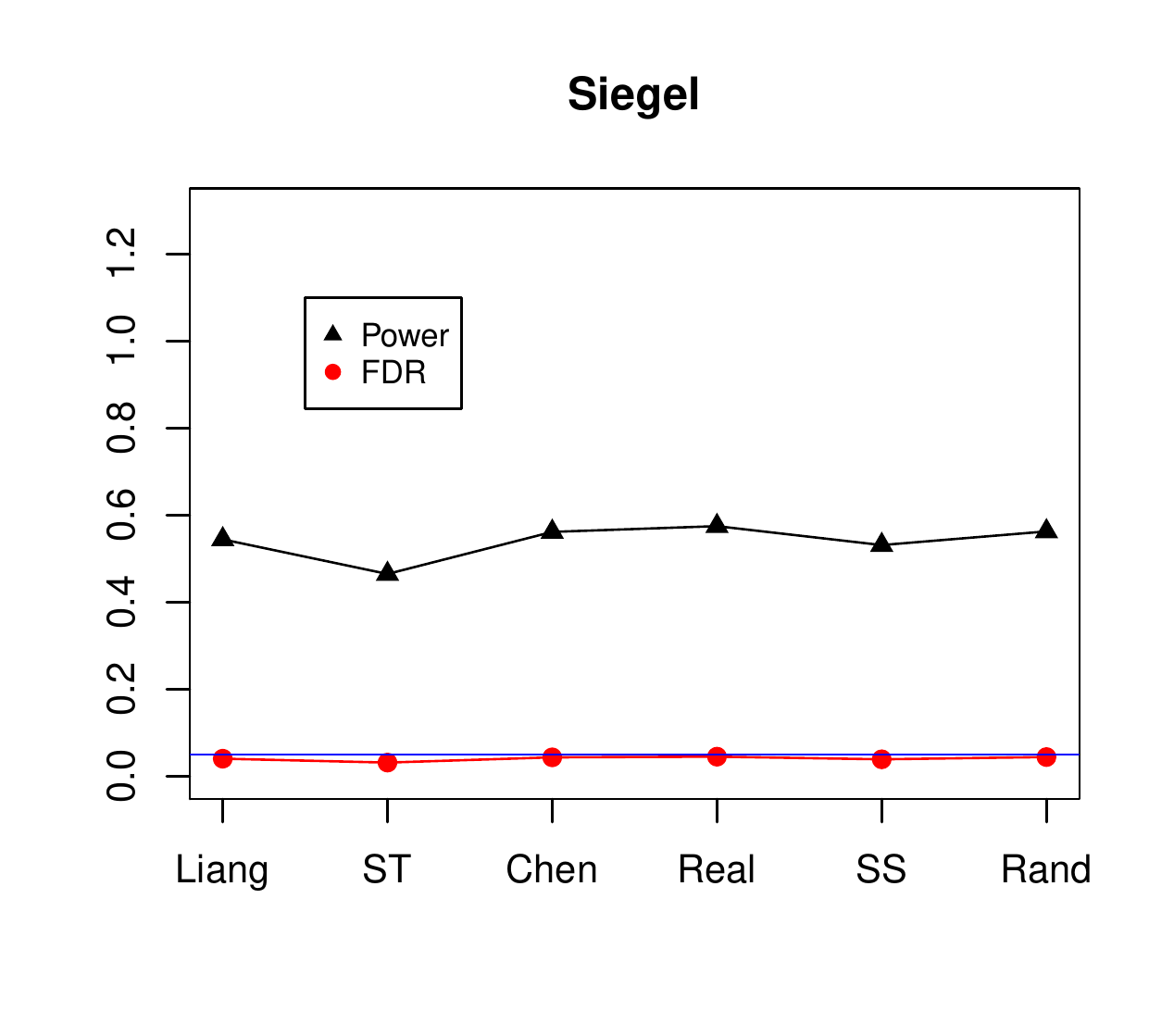}&
		\includegraphics[width=0.32\linewidth]{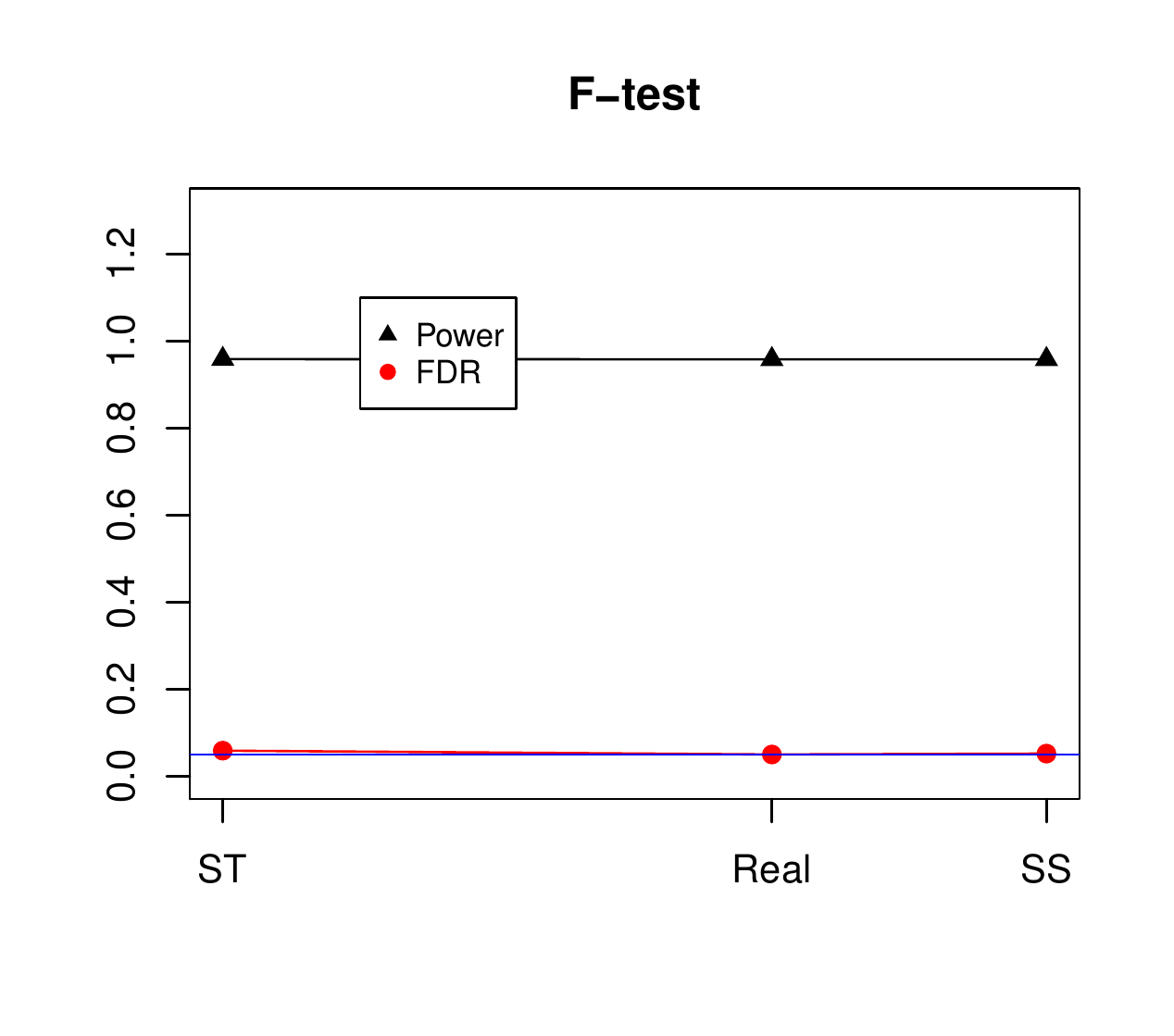}\\
		\end{array}$
	\end{center}
	\caption{Scale differences with ${n}_1={n}_2=8$, $m=100$, $\delta=0.5$ and $A$ given by (\ref{ec:A13}). The Monte Carlo estimator of the FDR and power are reported for each test and $q$-value method. The blue line corresponds to $\alpha=0.05$.}
	\label{fig:Table16}
\end{figure}

On the other hand, for scale differences, not surprisingly the parametric test ($F$-test) is the optimal procedure. In this setting, the $J_i$ permutation test is competitive with respect to Ansari-Bradley,  Siegel-Tukey and  Levene tests (see Figure \ref{fig:Table16}). Note that the results of Siegel-Tukey test are only reported for one of the settings since it behaves similarly to the Ansari-Bradley test; the latter avoids the drawbacks of Siegel-Tukey test as mentioned in Section \ref{se:tests}.

\begin{figure}[H]
	\begin{center}$
		\begin{array}{ccc}
		\includegraphics[width=0.32\linewidth]{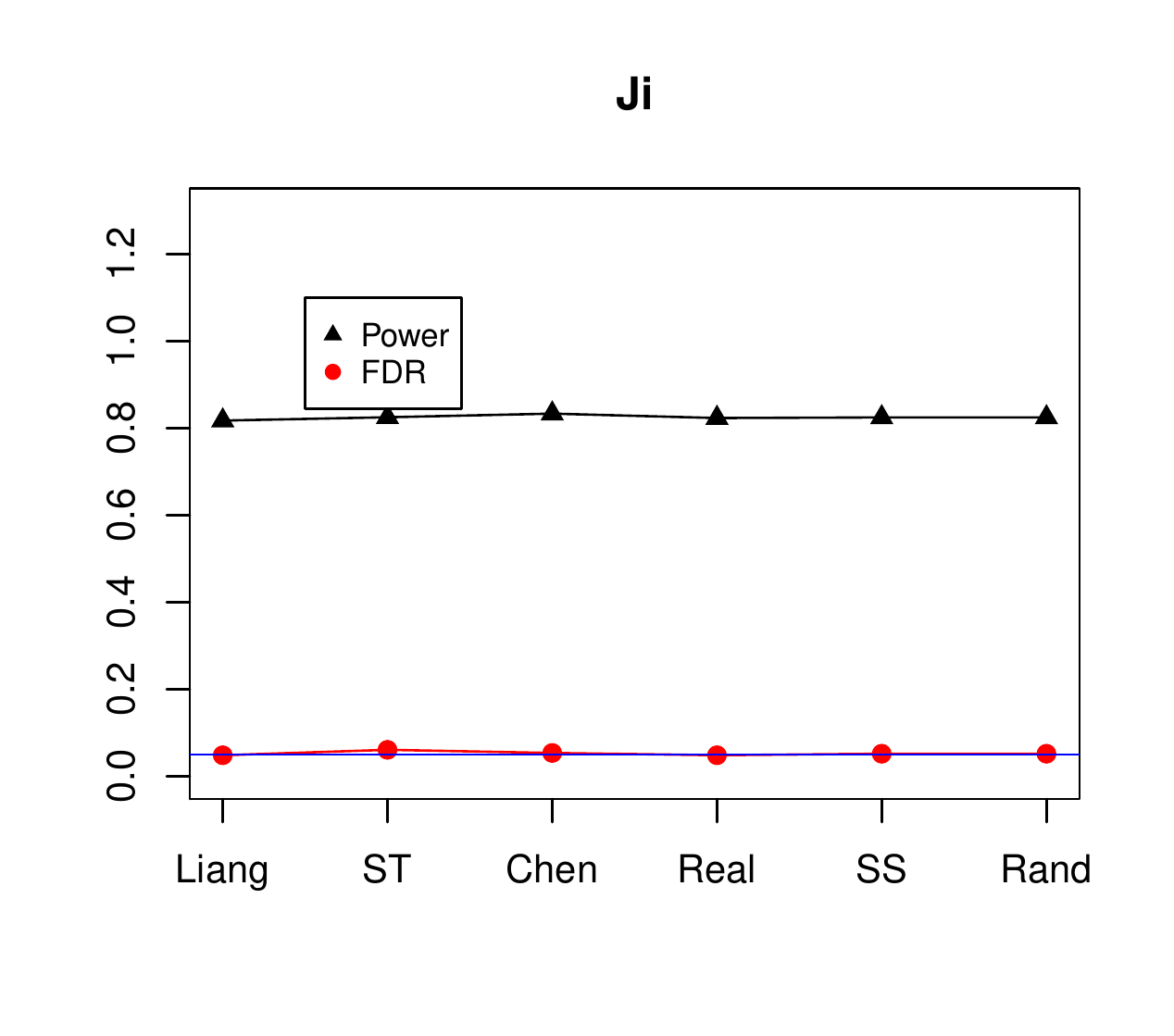}&\includegraphics[width=0.32\linewidth]{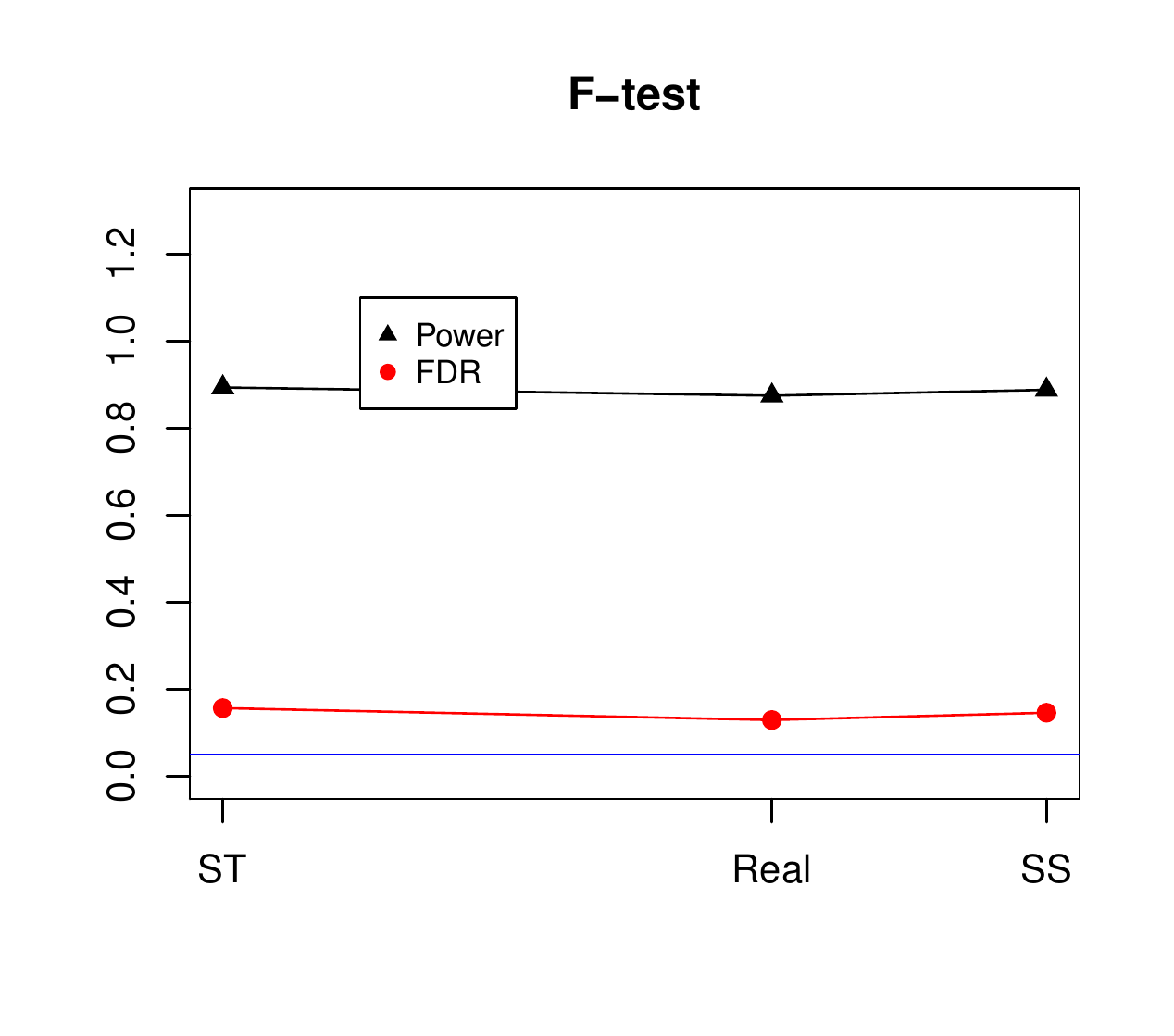}&\includegraphics[width=0.32\linewidth]{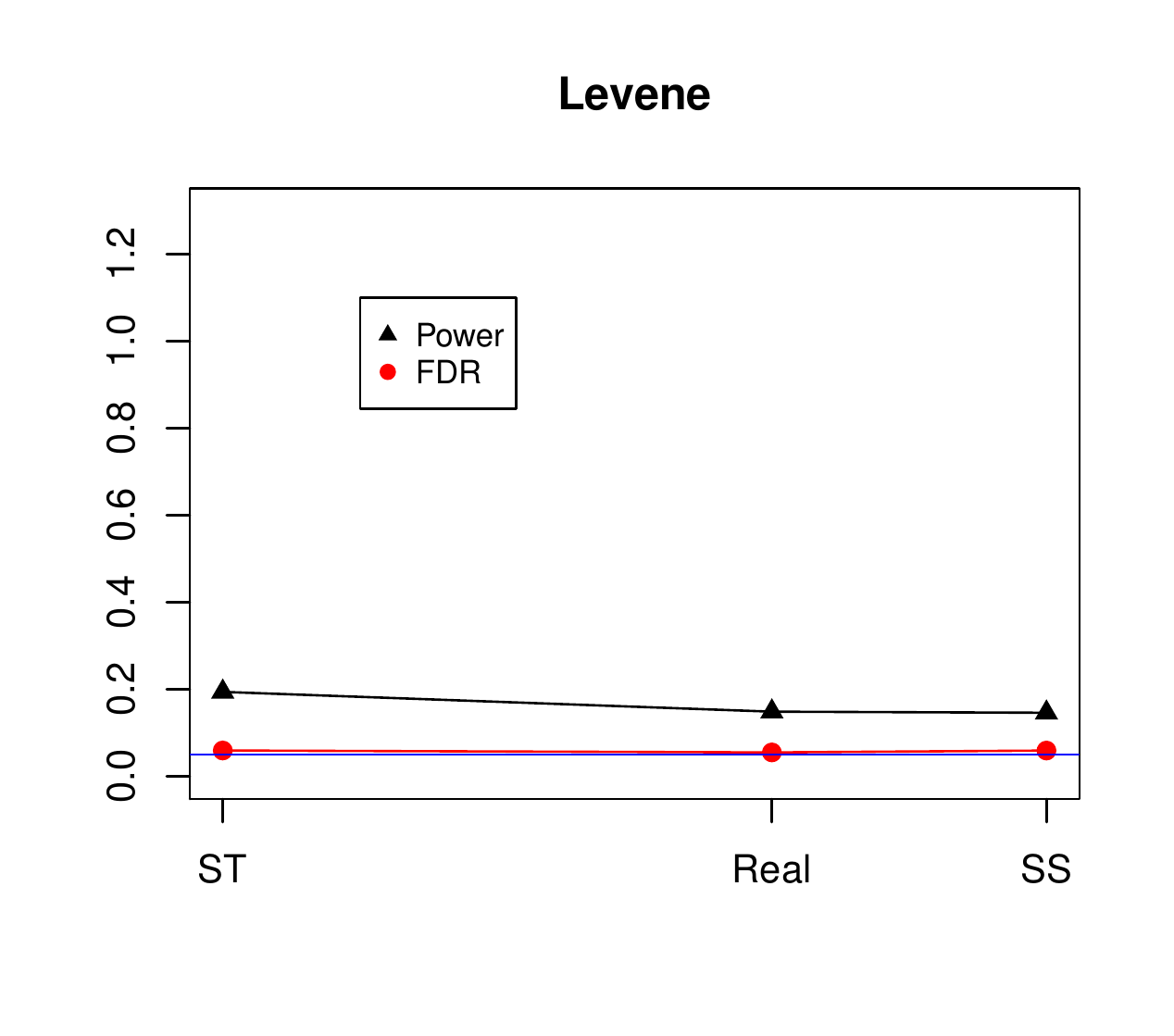}	\\
		\end{array}$
		$\begin{array}{cc}
		\includegraphics[width=0.32\linewidth]{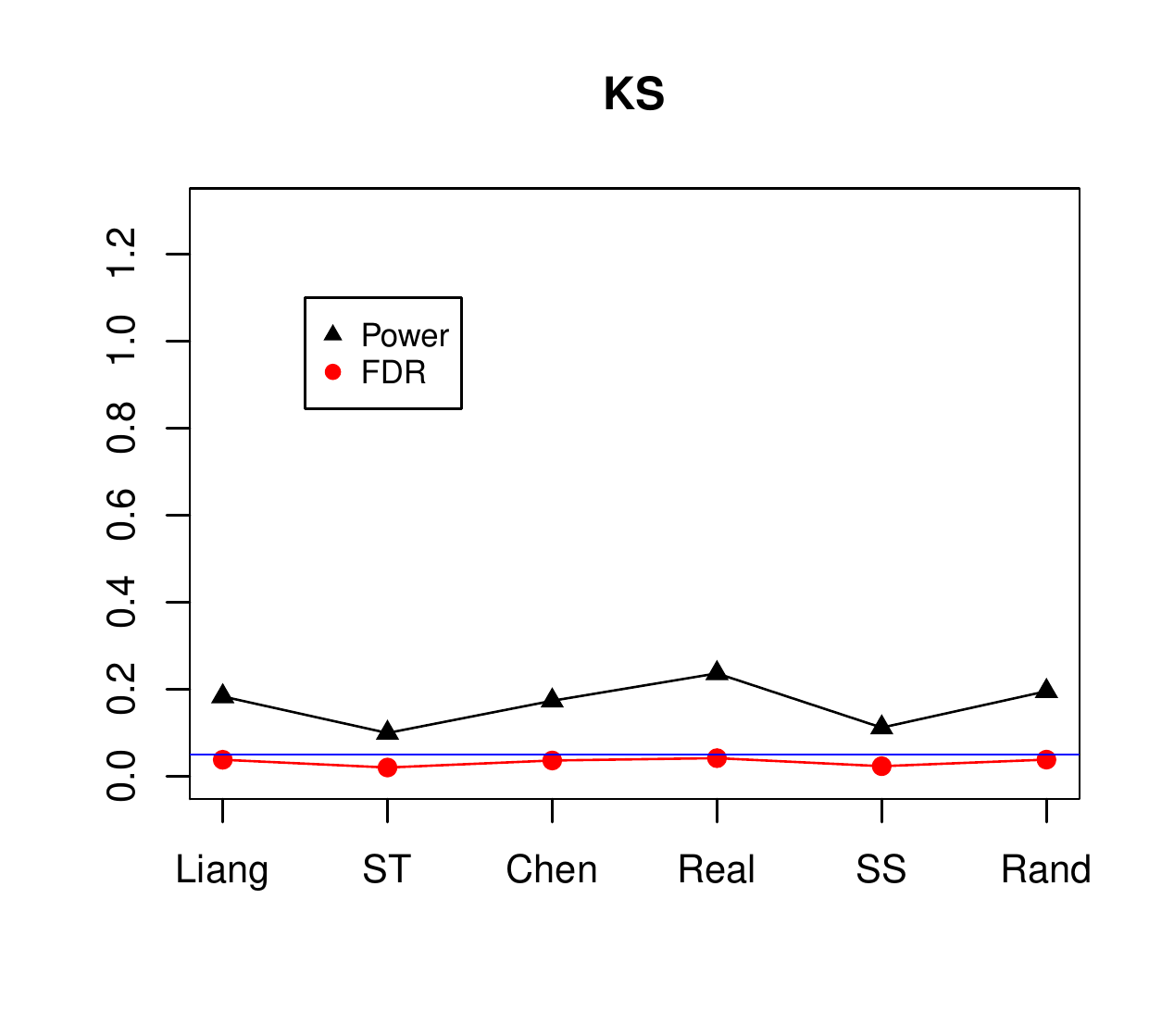}&	\includegraphics[width=0.32\linewidth]{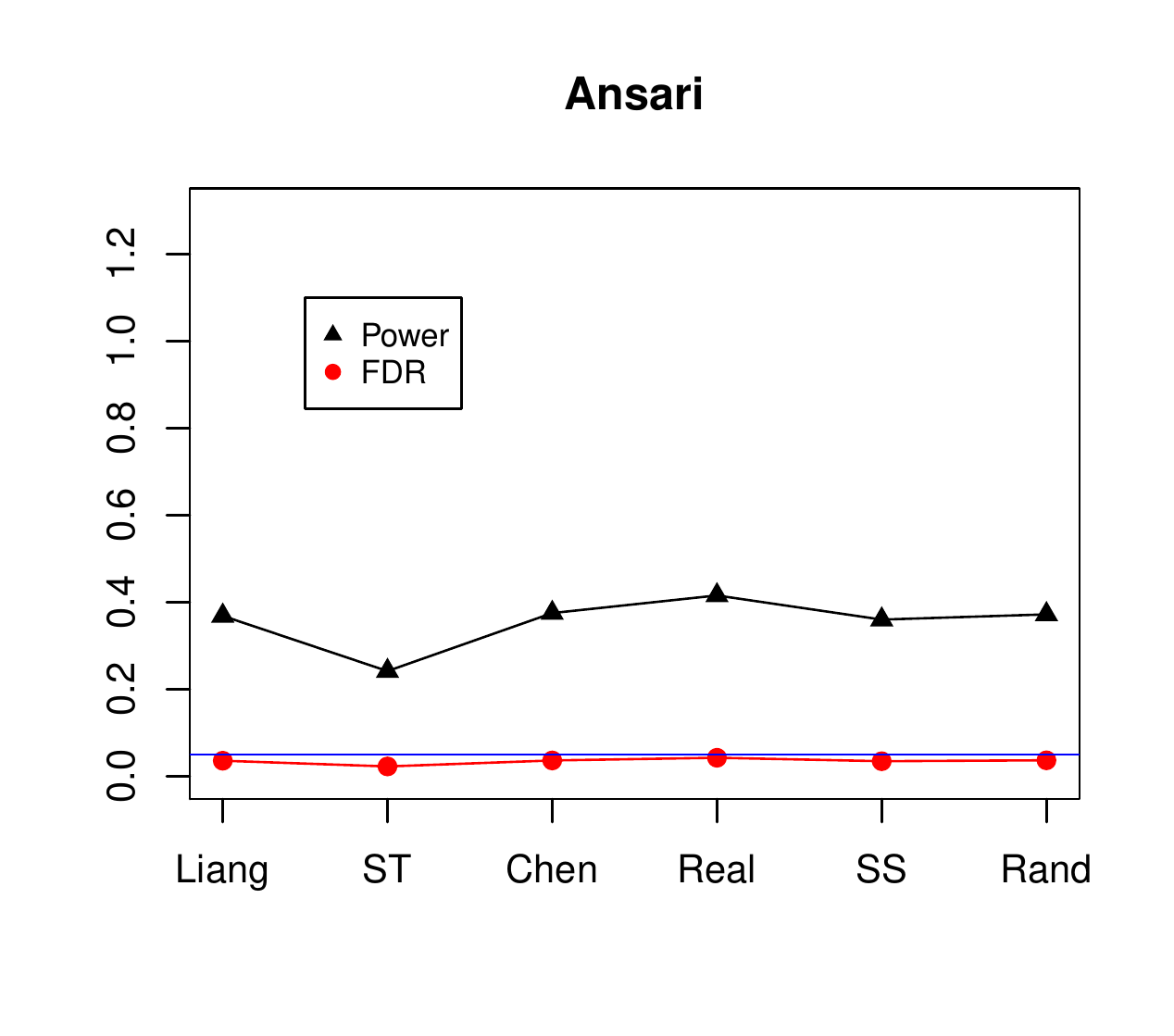}\\
		\end{array}$
	\end{center}
	\caption{Shape differences with ${n}_1={n}_2=8$, $m=100$, $\delta=0.5$ and $A$ given by (\ref{ec:A23}). The Monte Carlo estimator of the FDR and power are reported for each test and $q$-value method. The blue line corresponds to $\alpha=0.05$.}
	\label{fig:Table22}
\end{figure}

Finally, in the setting with differences in shape the most powerful test is the $F$-test; however, this test may exhibit an FDR above the nominal level and, hence, it is not recommended. The $J_i$ permutation test reports a power very close to that achieved by the $F$-test while respecting the FDR nominal level (see Figure \ref{fig:Table22}). 
%The power of the $J_i$ test statistics relative to the $F$-test is between $94\%$ and $95\%$ when the true value of $\pi_0$ is used. 
Hence, one may conclude that the test based on the $J_i$ permutation $P$-values is the optimal test for the scenarios with differences in shape. It is worth to mention that the KS test reports a very poor (almost zero) power in all settings except in the first one (location setting). Interestingly, it is seen that the omnibus test based on the $J_i$ statistics may be competitive or even better than other well-known two-sample tests. More precisely, the $J_i$ test is a good option to detect any type of differences in distribution instead of the KS test which may perform poorly when the sample sizes are small and the differences are other than location. 

The additional simulation results obtained for the one sample problem (Supplementary Material) were in agreement to those of the two-sample setting. The only exception was a relatively smaller bias of Liang estimator for $\pi_0$ compared to Chen approach.

	}
	
}

	\section{Real data analysis}\label{se:real3}
In this section we consider two real data examples. The first is a genetic data set which consists of a large number of gene expression levels measured on two groups of patients with breast cancer, classified according to BRCA mutation type. Then, the framework in this first real data set is the two-sample problem setting considered in Sections \ref{se:tests} and \ref{se:simul3}. The second real data example is a economic data set which have the daily log return of the five Spanish banks with highest capitalization for approximately one thousand days. In this case we have a one-sample setting since the aim is to test whether or not the expectation of the log returns is zero (more details in Section \ref{se:real32}). As we mentioned previously, simulations based on the one-sample setting, where the aim is to test a null hypothesis related with the mean of each of the $m$ variables, are available in the Suplementary Material.

\subsection{Genetic data}\label{se:real31}
We consider the microarray study of
hereditary breast cancer in \cite{He}. The data set
consists of $m=3170$ logged gene 
expression levels measured on
${n}_1=7$ patients with breast tumors
having BRCA1 mutations, on ${n}_2=8$
patients with breast tumors
having BRCA2 mutations and on
patients with sporadic breast
cancer, which we did not
use. Following \cite{S2} we eliminate all the genes whose measurement exceed 20; the final number of genes is $m=3170$. We are interested in testing the null
hypothesis that the distribution of each of the $m=3170$ genes is the same
for the two types of tumor, BRCA1 tumor 
and BRCA2 tumor.

\begin{table}[H]
	\begin{center}
		\scalebox{0.7}[0.7]{ 
			\begin{tabular}{ccccccccccc}
				\hline
				& 	& \multicolumn{9}{c}{Two-sample tests} \\
				& & $J_i$ & abs &t-test & $F$-test& KS& Wilcoxon & Ansari & Siegel & Levene\\
				$\widehat{\pi}_0$-method	& &  &&&&&&&&\\
				Liang&&	0.7513	&	0.6907	&	-	&	-	&	0.8648	&	0.7568	&	1	&	1	&	-	\\
				ST&&	0.6705	&	0.6888	&	0.6885	&	0.9297	&	0.7558	&	1	&	1	&	1	&	1	\\
				Chen&&	0.7508	&	0.6891	&	-	&	-	&	0.7635	&	0.7254	&	1	&	1	&-		\\
				SS&&	0.7514	&	0.6909	&	0.6871	&	0.9495	&	0.8259	&	0.7470	&	1	&	1	&	1	\\
				Rand&&	0.7511	&	0.6908	&	-	&	-	&	0.8259	&	0.7467	&	1	&	1	&	-	\\
		\end{tabular}}	
	\end{center}
	\caption{The $\pi_0$ estimates obtained by each method for the Hedenfalk data.}
	\label{ca:realpi0}
\end{table}

Previous analyses of this data set rejected the complete null hypothesis, so one or more genes out of the 3170 are differently distributed; see \cite{Marta2} and references therein. Table \ref{ca:realpi0} reports the $\pi_0$ estimates for the several methods investigated in this paper. {Note that the $P$-values derived from the application of the t-test and $F$-test are continuous and hence only the ST and SS estimators can be applied.} Table \ref{ca:realpi0} shows that the tests designed to detect scale differences report very conservative results, with $\widehat{\pi}_0=1$ or $\widehat \pi_0>0.9$, thus suggesting that the main differences between the distributions are not in scale. The number of rejections for such tests at FDR level $\alpha=0.05$ is zero for any of the $q$-value approaches. On the other hand, the values $\widehat{\pi}_0$ for the remaining tests indicate that the proportion of true null hypotheses is rather large. The number of rejections of each of the remaining methods are 9 for $J_i$, 96 for abs, 75 for t-test and 18 for KS (all the $q$-value methods report the same value), whereas Wilcoxon test resports 61 rejections for all the $q$-value methods except ST for which the result is zero rejections. 

Based on Table \ref{ca:realpi0} and on the aforementioned number of rejections for each test one may conclude that the differences between the distribution of the genes are basically due to location. For this reason, the more powerful tests are the ones designed to detect only location differences, whereas the tests that are able to detect any type of difference are less powerful. However, as we pointed out in our simulation study, these latter tests are powerful when the differences between the distributions are not only due to their location. {Then, we may also conclude that the final result depends mainly of which individual test is applied instead of the selected method for estimating $\pi_0$ (except if we apply the ST method to discrete uniform distributed $P$-values.)
}.

\begin{figure}[h!]
	\begin{center}
		\captionsetup{width=1\textwidth}
		\includegraphics[width=0.55\linewidth]{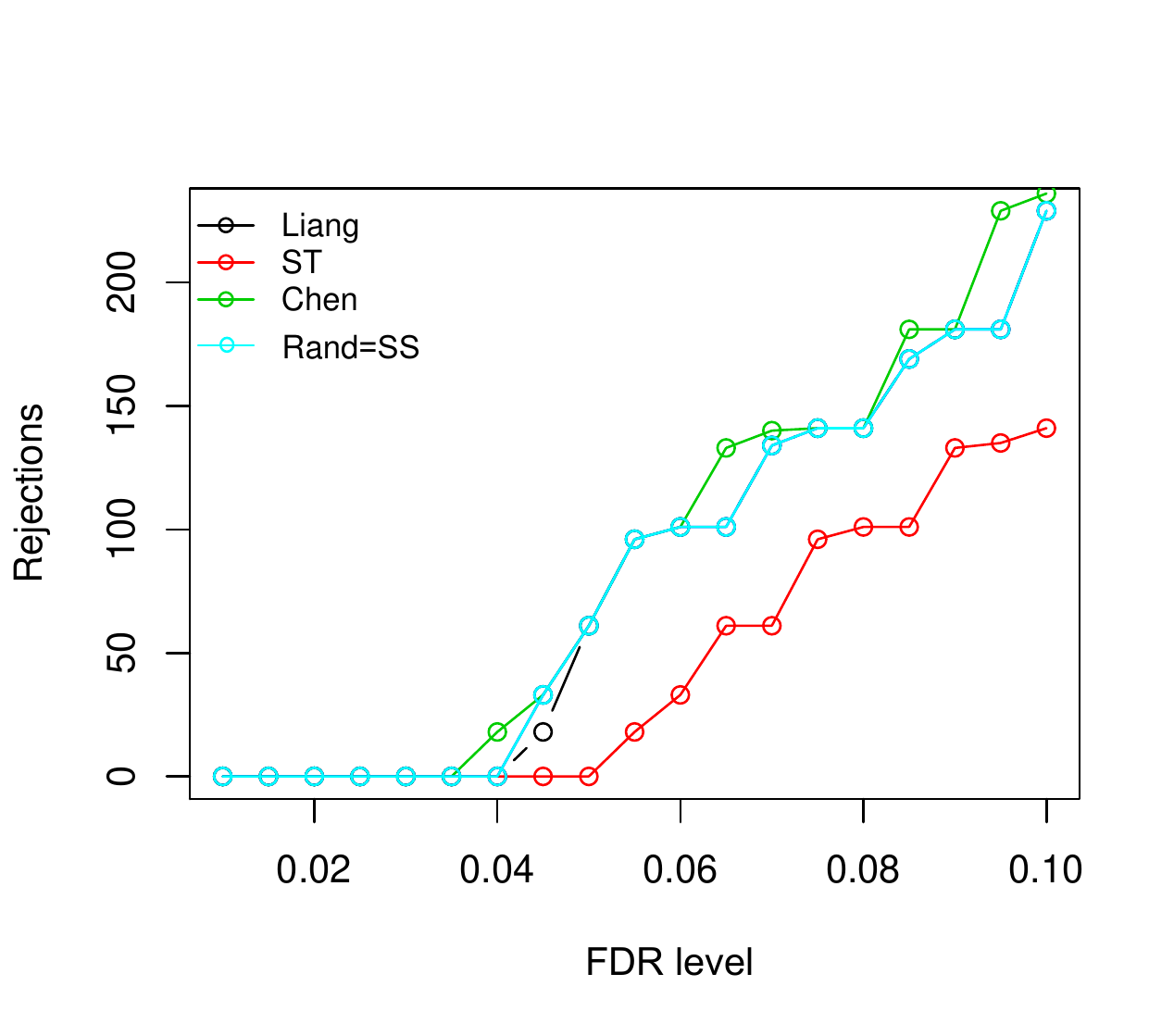}
		\caption{Number of rejections of Wilcoxon test depending on the nominal FDR level. The number of rejections of Liang, Chen and ST methods overlap the corresponding to  Rand and SS methods when are not shown.
				} 
		\label{fi:alphas}
	\end{center}
\end{figure}

Regarding the $q$-value method, in this application the number of rejections is the same for all tests regardless of the $q$-value method, except for Wilcoxon test. This is explained by the fact that, when ${n}_1=7$ and ${n}_2=8$, the total number of permutations $N$ is $6435$ and then the discreteness of the $P$-values of the tests is not very strong. However, Wilcoxon test has a ``more pronounced discreteness'' than the $J_i$ permutation test or the absolute value test, so it is not surprising that the ST method performs badly reporting zero rejections. Figure \ref{fi:alphas} depicts the number of rejections reported by Wilcoxon test for each of the $q$-value methods along a sequence of nominal levels ($\alpha= 0.010, 0.015, \dots, 0.095, 0.100$). From Figure \ref{fi:alphas} it is seen that the ST method is too conservative, whereas the SS method behaves surprisingly well in this case; this does not happen in the second real data application considered in Section \ref{se:real32}, were the application of SS method is misleading too. 

\subsection{Financial data}\label{se:real32}

In this Section we provide a real data illustration, corresponding to the one sample setting. We consider daily log returns of the five Spanish banks with highest market capitalization (Santander, BBVA, Bankinter, Caixabank, and Sabadell) from January 1, 2015, (first date registered) to June 4, 2018, and from June 4, 2018, to December 1, 2018. The first period corresponds to the term of a right-wing party in the Spanish government, while the second period relates the term of a left-wing party. The data are available at \url{https://finance.yahoo.com/q?s=ibm}. The variable log return of an asset at time $i$ is defined as $r_i=\log(P_i) -\log(P_{i-1})$ where $P_i$ is the price of an asset at time $i$. The first goal of our illustrative application is to explore if for any of these two terms (right-wing, left-wing) the efficiency of the financial market is violated, and to which extent.
The second goal is to identify the particular period of time where 
the financial market lived the worst situation in terms of effiency; this could allow for association studies with respect to economic or political events.

A classical assumption in finance is that the markets are efficient. This means that the price of assets contains all the information available \citep{fama}. However, this theoretical assumption is not always true in practice. For example, inefficiency can be a consequence of transactions costs or due to arrival information about the assets \citep[see][]{grossman,french} The expectation of the returns must be close to zero if the market is efficient. {For this reason the aforementioned goals are addressed by testing if the expectation of the log returns is zero or not for each time instant.} More specifically, we conclude that the market is efficient on day $i$ if $\mu_i \equiv E(r_i) =0$ where $r_i$ is the log return of the asset at time $i$ \citep[see][]{economic}.

We fix some notation. The data set with the information of the right-wing term is denoted by $X=\left[X_1, \dots, 
X_m\right]^T$ where $X_i=(X_{i1}, \dots, X_{i5})$ contains the log returns of the $5$ banks at time $i$ which are considered as observations (sample) of the same variable $r_i$, $i=1, \dots, m$, for $m=873$ (the length of the right-wing party period, after a data cleaning process). 
On the other hand the data set with the information of the left-wing term is denoted by $Y=\left[Y_1, \dots, 
Y_q\right]^T$ where $Y_i=(Y_{i1}, \dots, Y_{i5})$  contains the log returns of the $5$ banks at time $i$ which are considered as observations (sample) of the same variable $r_i$, $i=1, \dots, q$, $q=128$  (the length of the left-wing party term, after a data cleaning process). Note that we assume that the observations in $X_i$ are independent for $i=1, \dots, m$. This assumption has sense in this economic example since the log return of a bank at time $i$ depends, among others, on the behaviour of the banks at previous time instants but not on the situation at time $i$. 
%The effect of the situation of one of the banks on the others is not immediate, we need at least a time instant to see it. 
In other words the financial contagion, that is, the spread of market disturbances, does not
occur immediately.
% the failure of a bank or financial intermediary produces the undermining confidence in similar banks in the next time instants.

In order to test for $E(r_i)=0$ we consider two different test statistics: the parametric one sample t-test and the nonparametric one sample Wilcoxon test. The results attained by the several $q$-values at FDR level $\alpha=0.05$ for the $X$ and $Y$ samples are reported in Table  \ref{ca:ecopi}. We can see that the parametric test reports the largest number of rejections for both samples. However, 
the $t$-test assumes that the sample is normally distributed, and it seems that this assumption is violated in this setting. {
	Applying the Shapiro-Wilk normality test to the pooled sample of standardized daily log returns yields a $P$-value smaller than $2.2\times 10^{-16}$.
	} This is why a nonparametric test such as Wilcoxon is of interest. 
%{Note also that the vectors $X_1, \dots, X_m$ can be dependent, nevertheless, this is not a issue for the performance of the $q$-value method. We have verified in our extensive simulation study that the behaviour of the $q$-value method is correct even when the $m$ vectors are highly dependent.

%	\begin{figure}[h!]
%		\begin{center}
%			\captionsetup{width=1\textwidth}
%			\includegraphics[width=0.65\linewidth]{ecdf.pdf}
%		\end{center}
%		\caption{{Normality checking. Histogram of the pooled sample, the kernel density estimate based on such sample is the red line, whereas the blue curve is the standard normal density.}} 
%		\label{fi:normality}
%		
%	\end{figure}
	The number of rejections reported by the nonparametric test may be as low as zero when the $q$-values for continuous tests are naively applied; however, the discrete $q$-values give almost as many rejections as with the parametric t-test. In this illustrative application, Liang, Chen and Rand corrections report the same amount of rejections. These results are in agreement with what we have observed in our simulated scenarios (Supplementary Material). Summarizing, one may say that the application of the improved $q$-values may be critical whenever the $P$-values are discrete, which is the situation with nonparametric tests and small sample sizes; SS and ST methods for continuous tests cannot be recommended in such a setting. 
	
		\begin{table}[H]
		\begin{center}
			\scalebox{0.7}[0.7]{ 
				\begin{tabular}{cccccc|cccccc}
					\hline
					& 	& \multicolumn{2}{c}{Right-wing party ($X$)} & \multicolumn{2}{c}{Left-wing party ($Y$)} &	& 	& \multicolumn{2}{c}{Right-wing party ($X$)} & \multicolumn{2}{c}{Left-wing party ($Y$)}\\
					&  & t-test & Wilcoxon & t-test & Wilcoxon & & & t-test & Wilcoxon & t-test & Wilcoxon\\
					$\widehat{\pi}_0$-method	& & & & & & $\widehat{\pi}_0$-method	& & & & &\\
					Liang&&	-	&	0.2704		&	-	&	0.3611	&	Liang&&	-	&	578		&	-	&	62		\\
					ST&&	0.2084	&	0.5121		&	0.2303	&	1		&ST&&	612	&	0		&	78	&	0		\\
					Chen&&	-	&	0.2725	&	-	&	0.3750&	Chen&&	-	&	578	&	-	&	62	\\
					SS&&	0.2467	&	0.3042		&	0.4219	&	0.4062&		SS&&	582	&	499		&	56	&	0		\\
					Rand&&	-	&	0.2708	&	-	&	0.3802	&Rand&&	-	&	578	&	-	&	62	\\
			\end{tabular}}	
		\end{center}
		\caption{The estimates for $\pi_0$ (left) and the number of rejections (right) given by each method. Financial data.}
		\label{ca:ecopi}
	\end{table}
	
	We have compared the proportion of true null hypothesis for the right-wing party and left-wing party. The estimates of $\pi_0$ corresponding to the Wilcoxon test with improved $q$-values are 0.27 (right-wing party) and $0.36-0.38$ (left-wing). {Hence, the proportion of inefficient days in each period, $1-\widehat{\pi}_0$, is  0.73 (right-wing party) and $0.62-0.64$ (left-wing). This result could suggest an association between efficiency of the Spanish financial market and the particular party in the Government. Regarding the particular time period in which the market efficiency is violated, the inspection of the $q$-values reveals that the period between December 4, 2015, and August 28, 2016, reports the largest number of inefficient days. Interestingly, during this period two successive elections took place (due to failed negotiations), with a new government agreed precisely by August 28, 2016. Therefore, the political instability would have influenced the performance of the market along these nine months.}

	\section{Discussion}\label{se:conclu}

Standard $q$-values for continuous tests may be inaccurate when applied to discrete $P$-values. In this paper we have investigated $q$-value methods {for hdu tests}. The three methods (Liang, Chen and Rand) performed correctly in our simulated one sample and two-sample scenarios, with a slightly better behaviour of Chen method. It is worth to mention that, in the case of the $J_i$ test, the performance of SS and ST methods improved when the sample size increased, i.e, when the degree of discreteness was reduced. However, SS and ST still performed poorly for other nonparametric test (such as Ansari, Siegel and KS tests), for which the discreteness is relatively stronger.
Regarding the estimation of $\pi_0$, the conclusions are similar: Chen estimator is a good option for hdu $P$-values. {Therefore, our practical recommendation for discrete uniform and homogeneous $P$-values is to apply Chen $\pi_0$ estimator and its corresponding $q$-value.} The recommendation holds both independent and dependent tests since, in our simulations, the relative behaviour of the different estimators of FDR and $\pi_0$ and $q$-value methods were unaffected by the correlation. 

As a by-product, our simulation study has revealed that, in the setting of MCP, the test based on the $J_i$ statistics is competitive, and may perform even better than other well-known two-sample tests. For example, our simulation results suggest that the KS test should not be used when the sample sizes are small and the differences are other than location {\citep[see also][]{KSpoor}}. {In general, the accuracy of the results will depend not only on a suitable choice of the $q$-value method but also on the selection of an appropriate test, so particular attention should be paid to this regard.}

Other existing methods for discrete $P$-values as those in {\cite{sebas}} and \cite{Heller2012} reduce to their continuous counterparts when the null distribution of the $P$-values is discrete uniform. Therefore, they are not an option in our hdu setting. This also applies to other discrete corrections which are available in the literature, since most of them follow ideas similar to those in the aforementioned two papers. {This does not apply however to the randomization approach, which has served to introduce non trivial corrections for hdu $P$-values.} 

\vspace*{0.5cm}

{\bf Acknowledgements:}
This work has received financial support of the Call 2015 Grants for PhD contracts for training
of doctors of the Ministry of Economy and Competitiveness, cofinanced by the European Social
Fund (Ref. BES-2015-074958). We acknowledge support from MTM2014-55966-P project, Ministry
of Economy and Competitiveness, and MTM2017-89422-P project, Ministry of Economy, Industry
and Competitiveness, State Research Agency, and Regional Development Fund, UE. We also
acknowledge the financial support provided by the SiDOR research group through the grant Competitive
Reference Group, 2016-2019 (ED431C 2016/040), funded by the ``Conseller\'ia de Cultura,
Educaci\'on e Ordenaci\'on Universitaria. Xunta de Galicia''. To finish, the first author would like to thank the University of Vigo, and its Escola Internacional de Doutoramento (EIDO) by the financial support
provided through mobility doctorate grants.  

{The authors also thank Jos\'e Carlos Soage, research support technician in SiDOR group, for helping them in the analysis of the financial data.}

\newpage
\bibliographystyle{chicago}
\bibliography{bibliocap2}

\end{document}

\maketitle

\section{Simulation Study}

In this section, firstly Section \ref{se:one} we report verify the performance of the different methods in the one-sample setting, whereas in \ref{se:two} we report all results corresponding to the simulation study in Section 4 of the manuscript.

\subsection{One sample setting}\label{se:one}
{
In this section we investigate through simulations the behavior of the different methods, introduced in Section 2 of the main paper, in the setting of one-sample
problem with low sample size, for many variables. This setting is the one addressed in the financial application in Section 5.2 of the main paper.

This simulation study follows the design of the one in Section 4 of the main paper. Hence, first of all we introduce a reminder.

We consider a vector autoregressive model of order 1 VAR$(1)$ defined as 
\begin{equation*}\label{ec:autoregresivom2}
W_{t}=A W_{t-1}+ \varepsilon_t,
\end{equation*}
where $W_t=(W_{t1}, \cdots, W_{tn_1})^T$, $A=(a_{ij})$ is an $n_1
\times n_1$ design matrix such that the proccess $(W_t)_{t \in \mathbb{N}}$
is stationary and
$\varepsilon_t \in \mathbb{R}^{n_1}$ are i.i.d 
random vectors, termed innovations.

We generate a time series of length $m$ from the vector autoregressive 
model with innovations, $\varepsilon_t \sim N_{n_1}(0, I_{n_1})$ and initial
point $W_0 \sim
N_{n_1}(0, \Sigma)$ where $\Sigma$ is the stationary covariance matrix,
i.e,  $\Sigma= A^T \Sigma A+ I_{n_1}$. Denote the vectors generated in this way by $W_{i}=(W_{i1}, \cdots, W_{i{n_1}})^T$, $i=1, \dots, m$. 
{Then the data set $X$ is based on $W=[W_1, \dots, W_m]$. Each of the vectors $X_i$, $i=1, \dots, m$ must consist on independent and identical distributed observations. Hence, $X$ is based on a standarization of $W$.

Depending of the choice of the design matrix $A$ of the above VAR model we obtain a particular degree of dependence. 
In this study, we consider two possibilities for $A$,
each of which
is an $n_1 \times n_1$ lower triangular matrix with elements $a_{ij}$
satisfying $a_{ij}=0$ for $i-j>1$:
\begin{itemize}
	\item {Independence} is simulated by setting
	\begin{equation}\label{ec:A13}
	a_{ii}=0, \;i=1, \dots,n_1,\; \text{and} \;a_{i,i-1}=0, \;i=2, \dots,n_1. 
	\end{equation} 
	\item {Medium dependence}
	of $X_{ij}$ and $X_{kj}$ for $i\not=k$ and {strong dependence} of $X_{ij}$ and $X_{lk}$ for $i \not= l$ and $j \not= k$ is  is simulated by setting
	\begin{equation}\label{ec:A23}
	a_{ii}=0.5, \;i=1, \dots,n_1,\; \text{and} \;a_{i,i-1}=0.4, \;i=2, \dots,n_1. 
	\end{equation}
\end{itemize}

We define $X_i^{(0)}=\Sigma^{-1/2} W_i$, where $W_i$ are the vectors generated from the VAR$(1)$ model
with stationary covariance matrix $\Sigma$. Then we consider the
collection of densities $\{f_1,f_2\}$, where $f_1$
is $N(0,1)$ and $f_2$ is $N(\mu,1)$, and a sequence $I = \{I_j : j \in \{1, \dots,m\}\}$ which is a
sequence of i.i.d random variables such that $P(I_1=1)=1-\rho$ and $P(I_1=2)=\rho$. Finally,  
${X}_i=F_{I_i}^{-1}(\Phi(X_i^{(0)}))$, where $F_i$, $i=1,2$, are
the cumulative distribution functions corresponding to the densities
$f_i, i=1,2$. Hence, our aim is  to test the null hypotheses $H_{0i}:\mu_i=0$ for $i=1, \dots, m$ where $\mu_i=E(X_i)$. We consider the two different tests used in the financial application: the parametric one sample t-test and the nonparametric one sample Wilcoxon test.

In each setting we estimate the Monte Carlo FDR and the power of each $q$-value method, as we explained in Section 4 of the main paper. 
We also compute the average Monte Carlo bias and standard deviation of each $\pi_0$ estimator. We draw 1000 Monte Carlo trials and we take $0.05$ as the nominal level $\alpha$ for the FDR.

In Table \ref{table:parameters1} we specify all the different parameters and their respective ranges to help the understanding of the simulation settings.

{
	\begin{table}[htb]
		\begin{center}
			\scalebox{0.9}[0.9]{ 
				\begin{tabular}{|c|c|c|c|l|}
					\hline
					Sample size $n_1$ & Dimension $m$ & Location parameter $\mu$ & $\delta=1-\pi_0$ & Design matrix $A$\\
					6 & 100& 2& 0 & (\ref{ec:A13}) Independence\\
					 & 1000 & 3 & 0.3 & (\ref{ec:A23}) Dependence\\
					 	& & &0.4& \\
					& & &0.5& \\
					\hline
			\end{tabular}}	
		\end{center}
		\caption{The different parameters considered in the simulation settings  and their respective ranges.}
		\label{table:parameters1}
	\end{table}

Under the global null hypothesis, i.e.,  $\delta=0$, all the tests control the FDR at the nominal level. The FDR values for the Wilcoxon test are approximately zero. The fact that the FDR is approximately zero indicates that the test is overly conservative. The same happens for $\delta=0.3$ the Wilcoxon test is too much conservative reporting almost zero power, for this reason the results are not shown since no discussion about them is possible.

Tables \ref{ca:mu2100} and \ref{ca:mu3100} contain the
results for the settings ${n}_1=6$, $\mu=2$ and $\mu=3$, respectively, where $m=100$, $\delta=0.4,0.5$ when $A$ is given by (\ref{ec:A13}) (independent setting). 
Tables \ref{ca:mu21000} and \ref{ca:mu31000} contain the such results 
for $m=1000$. Finally, Table \ref{ca:mu2100dep}  contain the
results for the settings ${n}_1=6$, $\mu=2$ where $m=100$, $\delta=0.4,0.5$ when $A$ is given by (\ref{ec:A23}) (dependent setting). 

The relevant conclusions drawn from this simulation study match the ones obtained in the two-sample problem setting (Section 4 of the main paper).

Regarding the behaviour of the one-sample tests considered, the optimal procedure is the t-test, as expected. However, the power of  Wilcoxon test is roughly equal to the power of the  t-test for ${n}_1=6$, $\mu=3$, $m=100$, $\delta=0.5$ and ${n}_1=6$, $\mu=2,3$, $m=1000$, $\delta=0.4$ for $A$ given by (\ref{ec:A13}) (independent setting). Note that except in the settings ${n}_1=6$, $\mu=2$, $m=100$, $\delta=0.4$ for $A$ given by (\ref{ec:A13}) and (\ref{ec:A23}) the difference between the power of both tests is not large.

In relation to the comparison of the performance of the different $q$-value method we conclude that the three discrete methods, Liang, Chen and Rand, perform correctly, the differences in behaviour between the three $q$-value approaches are 
soft. The ST method reports a poor power compared to  three discrete methods in all the simulating settings. Whereas the performance of the SS method  is poor for $\delta=0.4$ and improves when the proportion of false null hypotheses increases.

Finally, if we compare the relative performance of the $\pi_0$ estimators the main conclusions match the ones mentioned in Section 4 of the main paper. In such section we explained that three estimators proposed for discrete $P$-values report similar results in most secenarios, the method with the smallest bias is Chen. However, in the current setting we can see that Liang method reports the smallest bias in almost all the cases.
}

	\begin{table}[H]
	\begin{center}
		\scalebox{0.7}[0.7]{ 
			\begin{tabular}{ccccccccccccc}
				\hline
				& 	& \multicolumn{11}{c}{One-sample tests} \\
				&  & \multicolumn{5}{c}{$\delta=0.4$} & & \multicolumn{5}{c}{$\delta=0.5$}\\
				\cline{3-7} \cline{9-13}\\
				& & \multicolumn{2}{c}{t-test}
				&& \multicolumn{2}{c}{Wilcoxon}
				&& \multicolumn{2}{c}{t-test} && \multicolumn{2}{c}{Wilcoxon}\\
				$\widehat{\pi}_0$	& &  Bias & Sd &&	Bias & Sd && Bias & Sd && Bias & Sd \\	
				Liang	& &    - & -	&&  0.0020   &  0.0423   				&& -&- 				&& 0.0020  & 0.0375 \\
		    	ST && 0.0033 & 0.2121	&&      0.3448& 0.1162		            && -0.0020& 0.2020  && 0.3797 & 0.1720\\
				Chen&& - & -	&&      -0.0102& 0.0528	         				&& -& - 			&& -0.0095    & 0.0489\\
				SS&& 0.0006 & 0.0815	&&      0.0767& 0.0806	                && 0.0013& 0.0709   &&  0.0619   & 0.0735\\
			   Rand && - & -	&&      0.0021& 0.0779		                    && -& - 			&&  0.0015   & 0.0688\\
								\hline 		
					& 	& \multicolumn{11}{c}{One-sample tests} \\
	&  & \multicolumn{5}{c}{$\delta=0.4$} & & \multicolumn{5}{c}{$\delta=0.5$}\\
	\cline{3-7} \cline{9-13}\\
	& & \multicolumn{2}{c}{t-test}
	&& \multicolumn{2}{c}{Wilcoxon}
	&& \multicolumn{2}{c}{t-test} && \multicolumn{2}{c}{Wilcoxon}\\
	MCP	& &  FDR & POWER &&	FDR & POWER&& FDR & POWER && FDR & POWER \\	
	Liang	& &    - & -	&&  0.0248   &  0.3136   				&& -&- 				&& 0.0359 &0.8736 \\
	ST && 0.0608 & 0.9346	&&      0.0015& 0.0224		            && 0.0664& 0.9663  && 0.0087& 0.1731\\
	Chen&& - & -	&&      0.0286& 0.3948         				&& -& - 			&& 0.0380& 0.8798\\
	Real && 0.0500 & 0.9386	&&      0.0250& 0.3310		                    && 0.0502 &0.9704			&&  0.0343& 0.8692\\
	SS&& 0.0518 & 0.9382	&&      0.0112& 0.1360	                && 0.0522 &0.9697   &&  0.0351 &0.8421\\
	Rand && - & -	&&      0.0273& 0.3829		                    && -& - 			&&  0.0386 &0.8770\\
				\hline
		\end{tabular}}	
	\end{center}
	\caption{One-sample setting. The Monte Carlo estimator of the FDR and power are reported for each test and $q$-value method obtained for ${n}_1=6$, $\mu=2$, $m=100$, $\delta=0.4,0.5$ when $A$ is given by (\ref{ec:A13}) (independent setting). The Monte Carlo bias and standard deviation of each $\pi_0$ estimator are also provided.}
	\label{ca:mu2100}
\end{table}

	\begin{table}[H]
	\begin{center}
		\scalebox{0.7}[0.7]{ 
			\begin{tabular}{ccccccccccccc}
				\hline
				& 	& \multicolumn{11}{c}{One-sample tests} \\
				&  & \multicolumn{5}{c}{$\delta=0.4$} & & \multicolumn{5}{c}{$\delta=0.5$}\\
				\cline{3-7} \cline{9-13}\\
				& & \multicolumn{2}{c}{t-test}
				&& \multicolumn{2}{c}{Wilcoxon}
				&& \multicolumn{2}{c}{t-test} && \multicolumn{2}{c}{Wilcoxon}\\
				$\widehat{\pi}_0$	& &  Bias & Sd &&	Bias & Sd && Bias & Sd && Bias & Sd \\	
				Liang	& &    - & -	&&  0.0002   &  0.0370   				&& -&- 				&& -0.0006 &0.0325 \\
				ST && 0.0034 & 0.2120	&&      0.3449& 0.1162		            && -0.0019& 0.2020  && 0.3798 &0.1719\\
				Chen&& - & -	&&      -0.0104& 0.0527         				&& -& - 			&&  -0.0097& 0.0489\\
				SS&& 0.0006 & 0.0815	&&      0.0767& 0.0805	                && 0.0013& 0.0709   &&  0.0618 &0.0735\\
				Rand && - & -	&&      0.0021& 0.0778		                    && -& - 			&&  0.0014 &0.0688\\
					& 	& \multicolumn{11}{c}{One-sample tests} \\
&  & \multicolumn{5}{c}{$\delta=0.4$} & & \multicolumn{5}{c}{$\delta=0.5$}\\
\cline{3-7} \cline{9-13}\\
& & \multicolumn{2}{c}{t-test}
&& \multicolumn{2}{c}{Wilcoxon}
&& \multicolumn{2}{c}{t-test} && \multicolumn{2}{c}{Wilcoxon}\\
MCP	& &  FDR & POWER &&	FDR & POWER&& FDR & POWER && FDR & POWER \\	
Liang	& &    - & -	&&  0.0438   &  0.9106  				&& -&- 				&&0.0323& 0.9921 \\
ST && 0.0607 & 0.9993	&&      0.0028& 0.0497		            && 0.0663& 0.9998  && 0.0106 &0.2847\\
Chen&& - & -	&&      0.0426& 0.8945         				&& -& - 			&&  0.0354 &0.9929\\
Real && 0.0501 & 0.9994	&&      0.0442& 0.9914		                    && 0.0503& 0.9999			&&  0.0302& 0.9918\\
SS&& 0.0518 & 0.9995	&&      0.0240& 0.4192	                && 0.0522& 0.9999   &&  0.0320 &0.9922\\
Rand && - & -	&&      0.0380& 0.7754		                    && -& - 			&&  0.0358 &0.9932\\
\hline	
		\end{tabular}}	
	\end{center}
	\caption{One-sample setting. The Monte Carlo estimator of the FDR and power are reported for each test and $q$-value method obtained for ${n}_1=6$, $\mu=3$, $m=100$, $\delta=0.4,0.5$ when $A$ is given by (\ref{ec:A13}) (independent setting). The Monte Carlo bias and standard deviation of each $\pi_0$ estimator are also provided.}
	\label{ca:mu3100}
\end{table}

	\begin{table}[H]
	\begin{center}
		\scalebox{0.7}[0.7]{ 
			\begin{tabular}{ccccccccccccc}
				\hline
				& 	& \multicolumn{11}{c}{One-sample tests} \\
				&  & \multicolumn{5}{c}{$\delta=0.4$} & & \multicolumn{5}{c}{$\delta=0.5$}\\
				\cline{3-7} \cline{9-13}\\
				& & \multicolumn{2}{c}{t-test}
				&& \multicolumn{2}{c}{Wilcoxon}
				&& \multicolumn{2}{c}{t-test} && \multicolumn{2}{c}{Wilcoxon}\\
				$\widehat{\pi}_0$	& &  Bias & Sd &&	Bias & Sd && Bias & Sd && Bias & Sd \\	
				Liang	& &    - & -	&&    -0.0004& 0.0119   				&& -&- 				&&  0.0007 &0.0142 \\
				ST && 0.0006 & 0.0667	&&     0.3990 &0.0095		            && -0.0039& 0.0626  &&  0.4617 &0.0545\\
				Chen&& - & -	&&       -0.0079 &0.0175	         				&& -& - 			&&  -0.0054& 0.0147\\
				SS&& 0.0008 & 0.0253	&&      0.0735& 0.0257	                && 0.0012& 0.0229   &&  0.0635 &0.0222\\
				Rand && - & -	&&        -0.0017& 0.0238		                    && -& - 			&&  0.0010& 0.0206\\
				\hline 		
				& 	& \multicolumn{11}{c}{One-sample tests} \\
				&  & \multicolumn{5}{c}{$\delta=0.4$} & & \multicolumn{5}{c}{$\delta=0.5$}\\
				\cline{3-7} \cline{9-13}\\
				& & \multicolumn{2}{c}{t-test}
				&& \multicolumn{2}{c}{Wilcoxon}
				&& \multicolumn{2}{c}{t-test} && \multicolumn{2}{c}{Wilcoxon}\\
				MCP	& &  FDR & POWER &&	FDR & POWER&& FDR & POWER && FDR & POWER \\	
				Liang	& &    - & -	&&  0.0448& 0.9999   				&& -&- 				&& 0.0345& 0.8697 \\
				ST && 0.0511 & 1	&&      0.0000 &0.0000		            && 0.0513& 0.9712  &&  0.0000 &0.0000\\
				Chen&& - & -	&&       0.0448 &0.9999         				&& -& - 			&&0.0345& 0.8697\\
				Real && 0.0504 & 1	&&      0.0448& 0.9999		                    && 0.0500 &0.9710			&&   0.0345& 0.8697
				\\
				SS&& 0.0507 & 1	&&      0.0214& 0.4420	                && 0.0502 &0.9708   &&  0.0345& 0.8697\\
				Rand && - & -	&&      0.0446 &0.9959		                    && -& - 			&&   0.0345& 0.8697\\
				\hline
		\end{tabular}}	
	\end{center}
	\caption{One-sample setting. The Monte Carlo estimator of the FDR and power are reported for each test and $q$-value method obtained for ${n}_1=6$, $\mu=2$, $m=1000$, $\delta=0.4,0.5$ when $A$ is given by (\ref{ec:A13}) (independent setting). The Monte Carlo bias and standard deviation of each $\pi_0$ estimator are also provided. }
	\label{ca:mu21000}
\end{table}

	\begin{table}[H]
	\begin{center}
		\scalebox{0.7}[0.7]{ 
			\begin{tabular}{ccccccccccccc}
				\hline
				& 	& \multicolumn{11}{c}{One-sample tests} \\
				&  & \multicolumn{5}{c}{$\delta=0.4$} & & \multicolumn{5}{c}{$\delta=0.5$}\\
				\cline{3-7} \cline{9-13}\\
				& & \multicolumn{2}{c}{t-test}
				&& \multicolumn{2}{c}{Wilcoxon}
				&& \multicolumn{2}{c}{t-test} && \multicolumn{2}{c}{Wilcoxon}\\
				$\widehat{\pi}_0$	& &  Bias & Sd &&	Bias & Sd && Bias & Sd && Bias & Sd \\	
				Liang	& &    - & -	&&    -0.0004& 0.0119   				&& -&- 				&&  0.0000 &0.0097 \\
				ST && 0.0006 & 0.0667	&&     0.3990 &0.0095		            && -0.0038& 0.0626  &&  0.4619 &0.0544\\
				Chen&& - & -	&&       -0.0079 &0.0175	         				&& -& - 			&&  -0.0057& 0.0146\\
				SS&& 0.0008 & 0.0253	&&      0.0735& 0.0257	                && 0.0012& 0.0229   &&  0.0634 &0.0222\\
				Rand && - & -	&&        -0.0017& 0.0238		                    && -& - 			&&  0.0009 &0.0206\\
				\hline 		
				& 	& \multicolumn{11}{c}{One-sample tests} \\
				&  & \multicolumn{5}{c}{$\delta=0.4$} & & \multicolumn{5}{c}{$\delta=0.5$}\\
				\cline{3-7} \cline{9-13}\\
				& & \multicolumn{2}{c}{t-test}
				&& \multicolumn{2}{c}{Wilcoxon}
				&& \multicolumn{2}{c}{t-test} && \multicolumn{2}{c}{Wilcoxon}\\
				MCP	& &  FDR & POWER &&	FDR & POWER&& FDR & POWER && FDR & POWER \\	
				Liang	& &    - & -	&&  0.0448&1  				&& -&- 				&& 0.0304& 0.9916 \\
				ST && 0.0511 & 1	&&      0.0000 &0.0000		            && 0.0513& 0.9999  &&  0.0009 &0.0277\\
				Chen&& - & -	&&       0.0448 &1         				&& -& - 			&&0.0304 &0.9916\\
				Real && 0.0504 & 1	&&      0.0448& 1		                    && 0.0498 &0.9999			&&   0.0304& 0.9916\\
				SS&& 0.0507 & 1	&&      0.0214& 0.4420	                && 0.0502 &0.9999   && 0.0304 &0.9916\\
				Rand && - & -	&&      0.0446 &0.9959		                    && -& - 			&&   0.0304& 0.9916\\
				\hline
		\end{tabular}}	
	\end{center}
	\caption{One-sample setting. The Monte Carlo estimator of the FDR and power are reported for each test and $q$-value method obtained for ${n}_1=6$, $\mu=3$, $m=1000$, $\delta=0.4,0.5$ when $A$ is given by (\ref{ec:A13}) (independent setting). The Monte Carlo bias and standard deviation of each $\pi_0$ estimator are also provided.}
	\label{ca:mu31000}
\end{table}

	\begin{table}[H]
	\begin{center}
		\scalebox{0.7}[0.7]{ 
			\begin{tabular}{ccccccccccccc}
				\hline
				& 	& \multicolumn{11}{c}{One-sample tests} \\
				&  & \multicolumn{5}{c}{$\delta=0.4$} & & \multicolumn{5}{c}{$\delta=0.5$}\\
				\cline{3-7} \cline{9-13}\\
				& & \multicolumn{2}{c}{t-test}
				&& \multicolumn{2}{c}{Wilcoxon}
				&& \multicolumn{2}{c}{t-test} && \multicolumn{2}{c}{Wilcoxon}\\
				$\widehat{\pi}_0$	& &  Bias & Sd &&	Bias & Sd && Bias & Sd && Bias & Sd \\	
				Liang	& &    - & -	&&    -0.0030& 0.0581  				&& -&- 				&& -0.0018  &0.0496 \\
				ST && 0.0046 & 0.2331	&&     0.3323 &0.1427		            && 0.0056& 0.2141  &&  0.3670 &0.1878\\
				Chen&& - & -	&&       -0.0089 &0.0724	         				&& -& - 			&&  -0.0074& 0.0620\\
				SS&& 0.0052 & 0.1068	&&      0.0778& 0.1042	                && 0.0030& 0.0906   &&  0.0661 &0.0886\\
				Rand && - & -	&&        0.0019& 0.0987		                    && -& - 			&&  0.0030 &0.0850\\
				\hline 		
				& 	& \multicolumn{11}{c}{One-sample tests} \\
				&  & \multicolumn{5}{c}{$\delta=0.4$} & & \multicolumn{5}{c}{$\delta=0.5$}\\
				\cline{3-7} \cline{9-13}\\
				& & \multicolumn{2}{c}{t-test}
				&& \multicolumn{2}{c}{Wilcoxon}
				&& \multicolumn{2}{c}{t-test} && \multicolumn{2}{c}{Wilcoxon}\\
				MCP	& &  FDR & POWER &&	FDR & POWER&& FDR & POWER && FDR & POWER \\	
				Liang	& &    - & -	&&  0.0307& 0.3574				&& -&- 				&& 0.0382& 0.8817 \\
				ST && 0.0649 & 0.9364	&&      0.0042 &0.0432		            && 0.0678& 0.9683  &&  0.0122 &0.1885\\
				Chen&& - & -	&&       0.0341 &0.4226        				&& -& - 			&&0.0398 &0.8850\\
				Real && 0.0478 & 0.9400	&&      0.0263& 0.3634		                    && 0.0492 &0.9718			&&   0.0337& 0.8712\\
				SS&& 0.0521 & 0.9401	&&      0.0184& 0.1960	                && 0.0537 &0.9712   && 0.0358 &0.8225\\
				Rand && - & -	&&      0.0330 &0.4164		                    && -& - 			&&   0.0413& 0.8812\\
				\hline
		\end{tabular}}	
	\end{center}
	\caption{One-sample setting. The Monte Carlo estimator of the FDR and power are reported for each test and $q$-value method obtained for ${n}_1=6$, $\mu=2$, $m=100$, $\delta=0.4,0.5$ when $A$ is given by (\ref{ec:A23}) (dependent setting). The Monte Carlo bias and standard deviation of each $\pi_0$ estimator are also provided. }
	\label{ca:mu2100dep}
\end{table}

\newpage
\subsection{Two sample setting}\label{se:two}

Below the set of tables corresponding to the different setting simulated in Section 4 of the manuscript are shown.

\begin{table}[htb]
	\begin{center}
		\scalebox{0.7}[0.7]{ 
			\begin{tabular}{cccccccccccccccc}
				\hline
				& 	& \multicolumn{14}{c}{Two-sample tests} \\
				&  & \multicolumn{2}{c}{$J_i$} && \multicolumn{2}{c}{abs}
				&& \multicolumn{2}{c}{t-test}
				&& \multicolumn{2}{c}{KS}
				&& \multicolumn{2}{c}{Wilcoxon}\\
				\cline{3-4}\cline{6-7} \cline{9-10} \cline{12-13} \cline{15-16}\\
				$\widehat{\pi}_0$	& &  Bias & Sd &&	Bias & Sd && Bias & Sd && Bias & Sd && Bias & Sd\\	
				Liang	& &    0.0584 & 0.0532	&&      0.0280& 0.0498&& -&- &&  0.0353 & 0.0449&&     0.0238 & 0.0622\\
				ST && 0.0883 & 0.1200	&&      0.0733& 0.1990		&& 0.0124& 0.1908 &&    0.4923 &0.0419		&&     0.3476& 0.1778\\
				Chen&&     0.0177 & 0.0567&&	      -0.0034& 0.0534	&&-   & - &&        0.0488 & 0.0449	&&       0.0089& 0.0533\\
				SS&& 0.0412& 0.0724	&& 0.0256 & 0.0711		&&  0.0123 & 0.0711&&	 0.3256 & 0.0685		&&  0.0281 &0.0710\\
				Rand &&        0.0262& 0.0707	&&        0.0115 &0.0700	&& -&- &&		        0.0300 & 0.0521	&&       0.0133& 0.0690\\	
				
				\hline 		
				& 	& \multicolumn{14}{c}{Two-sample tests} \\
				&  & \multicolumn{2}{c}{$J_i$} && \multicolumn{2}{c}{abs}
				&& \multicolumn{2}{c}{t-test}
				&& \multicolumn{2}{c}{KS}
				&& \multicolumn{2}{c}{Wilcoxon}\\
				\cline{3-4}\cline{6-7} \cline{9-10} \cline{12-13} \cline{15-16}\\
				MCP	& &  FDR & POWER &&	FDR & POWER && FDR & POWER && FDR & POWER && FDR & POWER\\	
				Liang&&	0.0257&	0.5419	&&	0.0377&	0.6966	&&	-&-		&&	0.0360&	0.6850	&&	0.0380&	0.6893	\\
				ST&&	0.0234&	0.4192	&&	0.0345&	0.5245	&&	0.0589&	0.8127	&&	0.0002&	0.0046	&&	0.0080&	0.1391	\\
				Chen&&	0.0286&	0.6192	&&	0.0395&	0.7125	&&	-&-		&&	0.0355&	0.6739	&&	0.0386&	0.7039	\\
				Real&&	0.0306&	0.6793	&&	0.0377&	0.7265	&&	0.0512&	0.8090	&&	0.0377&	0.7265	&&	0.0377&	0.7265	\\
				SS&&	0.0263&	0.5653	&&	0.0378&	0.6757	&&	0.0514&	0.8063	&&	0.0037&	0.0487	&&	0.0373&	0.6751	\\
				Rand &&	0.0273&	0.5880	&&	0.0389&	0.6959	&&	-&-		&&	0.0359 &	0.6832	&&	0.0384&	0.6920	\\		
				\hline
		\end{tabular}}	
	\end{center}
	\caption{Two-sample setting. Location differences with ${n}_1={n}_2=4$, $m=100$, $\delta=0.5$, $\mu=2$ and $A$ given by (\ref{ec:A13}). The Monte Carlo estimator of the FDR and power are reported for each test and $q$-value method.  The Monte Carlo bias and standard deviation of each $\pi_0$ estimator are also provided. }
	\label{ca:L_2_100_4_0.5}
\end{table}

\vspace*{3cm}

\begin{table}[htb]
	\begin{center}
		\scalebox{0.7}[0.7]{ 
			\begin{tabular}{cccccccccccccccc}
				\hline
				& 	& \multicolumn{14}{c}{Two-sample tests} \\
				&  & \multicolumn{2}{c}{$J_i$} && \multicolumn{2}{c}{abs}
				&& \multicolumn{2}{c}{t-test}
				&& \multicolumn{2}{c}{KS}
				&& \multicolumn{2}{c}{Wilcoxon}\\
				\cline{3-4}\cline{6-7} \cline{9-10} \cline{12-13} \cline{15-16}\\
				$\widehat{\pi}_0$	& &  Bias & Sd &&	Bias & Sd && Bias & Sd && Bias & Sd && Bias & Sd\\	
				Liang &&	0.0299&	0.0462	&&	0.0105&	0.0454	&&	-&-		&&	0.0071&	0.0636	&&	0.0119&	0.0508	\\
				ST&&	0.0246&	0.2106	&&	0.0198&	0.2036	&&	-0.0102	&0.2033	&&	0.3000 &	0.0000	&&	0.2800&	0.0687	\\
				Chen&&	-0.0122&0.0648	&&	-0.0268&	0.0624	&&		-&-	&&	0.0085&	0.0495	&&	-0.0054	&0.0560	\\
				SS&&	0.0284&	0.0847	&&	0.0147&	0.0842	&&	0.0037&	0.0837	&&	0.2042&	0.0735	&&	0.1127&	0.0802	\\
				Rand &&	0.0166&	0.0848	&&	0.0038&	0.0841	&&	-&-		&&	0.0065&	0.0652	&&	0.0018&	0.0781	\\
				
				\hline 		
				& 	& \multicolumn{14}{c}{Two-sample tests} \\
				&  & \multicolumn{2}{c}{$J_i$} && \multicolumn{2}{c}{abs}
				&& \multicolumn{2}{c}{t-test}
				&& \multicolumn{2}{c}{KS}
				&& \multicolumn{2}{c}{Wilcoxon}\\
				\cline{3-4}\cline{6-7} \cline{9-10} \cline{12-13} \cline{15-16}\\
				MCP	& &  FDR & POWER &&	FDR & POWER && FDR & POWER && FDR & POWER && FDR & POWER\\	
				Liang&&	0.0273&	0.6572	&&	0.0416&	0.7345	&&	-&-		&&	0.0260	&0.6683	&&	0.0412&	0.7326	\\
				ST&&	0.0323	&0.6646	&&	0.0464&	0.7316	&&	0.0583&	0.8050	&&	0.0254	&0.6502	&&	0.0265&	0.6555	\\
				Chen&&	0.0305&	0.6714	&&	0.0443&	0.7453	&&	-&-		&&	0.0260&	0.6683	&&	0.0416&	0.7373	\\
				Real&&	0.0281&	0.6632	&&	0.0418	&0.7389	&&	0.0499&	0.7979	&&	0.0260&	0.6683	&&	0.0417&	0.7388	\\
				SS&&	0.0283&	0.6583	&&	0.0413&	0.7314	&&	0.0507	&0.7994	&&	0.0258&	0.6588	&&	0.0345&	0.6940	\\
				Rand&&	0.0291&	0.6618	&&	0.0424&	0.7356	&&	-&-		&&	0.0260	&0.6683	&&	0.0411&	0.7323	\\		
				\hline
		\end{tabular}}	
	\end{center}
	\caption{Two-sample setting. 
		Location differences with ${n}_1={n}_2=5$, $m=100$, $\delta=0.3$, $\mu=2$ and $A$ given by (\ref{ec:A13}). The Monte Carlo estimator of the FDR and power are reported for each test and $q$-value method.  The Monte Carlo bias and standard deviation of each $\pi_0$ estimator are also provided. }
	\label{ca:L_2_100_5_0.3}
\end{table}

\begin{table}[htb]
	\begin{center}
		\scalebox{0.7}[0.7]{ 
			\begin{tabular}{cccccccccccccccc}
				\hline
				& 	& \multicolumn{14}{c}{Two-sample tests} \\
				&  & \multicolumn{2}{c}{$J_i$} && \multicolumn{2}{c}{abs}
				&& \multicolumn{2}{c}{t-test}
				&& \multicolumn{2}{c}{KS}
				&& \multicolumn{2}{c}{Wilcoxon}\\
				\cline{3-4}\cline{6-7} \cline{9-10} \cline{12-13} \cline{15-16}\\
				$\widehat{\pi}_0$	& &  Bias & Sd &&	Bias & Sd && Bias & Sd && Bias & Sd && Bias & Sd\\	
				Linag&&	0.0367&	0.0459	&&	0.0167&	0.0418	&&	-&-		&&	0.0137&	0.0523	&&	0.0215	&0.0467	\\
				ST&&	0.0430&	0.1954	&&	0.0315&	0.1972	&&	0.0121&	0.1871	&&	0.4969&	0.0192	&&	0.3855&	0.1670	\\
				Chen&&	-0.0032&	0.0536	&&	-0.0193&	0.0532	&&	-&	-	&&	0.0229&	0.0453	&&	0.0026&	0.0481	\\
				SS&&	0.0266	&0.0714	&&	0.0104	&0.0707	&&	0.0025&	0.0715	&&	0.1604&	0.0672	&&	0.0880&	0.0692	\\
				Rand&&	0.0185&	0.0718	&&	0.0028&	0.0711	&&	-&-		&&	0.0131&	0.0536	&&	0.0067&	0.0653	\\
				\hline 		
				& 	& \multicolumn{14}{c}{Two-sample tests} \\
				&  & \multicolumn{2}{c}{$J_i$} && \multicolumn{2}{c}{abs}
				&& \multicolumn{2}{c}{t-test}
				&& \multicolumn{2}{c}{KS}
				&& \multicolumn{2}{c}{Wilcoxon}\\
				\cline{3-4}\cline{6-7} \cline{9-10} \cline{12-13} \cline{15-16}\\
				MCP	& &  FDR & POWER &&	FDR & POWER && FDR & POWER && FDR & POWER && FDR & POWER\\	
				Liang&&	0.0331&	0.8115	&&	0.0439	&0.8725	&&	-&-		&&	0.0107&	0.6674	&&	0.0357&	0.8402	\\
				ST&&	0.0395&	0.8128	&&	0.0500&	0.8709	&&	0.0560	&0.8910	&&	0.0106&	0.6672	&&	0.0221&	0.7741	\\
				Chen&&	0.0367&	0.8247	&&	0.0481&	0.8807	&&	-&-		&&	0.0107	&0.6674	&&	0.0372&	0.8425	\\
				Real&&	0.0349&	0.8237	&&	0.0448	&0.8759	&&	0.0478&	0.8903	&&	0.0106&	0.6672	&&	0.0340&	0.8374	\\
				SS&&	0.0344	&0.8156	&&	0.0450	&0.8738	&&	0.0496&	0.8897	&&	0.0106	&0.6672	&&	0.0330&	0.8281	\\
				Rand &&	0.0351&	0.8192	&&	0.0458&	0.8757	&&	-&-		&&	0.0107&	0.6674	&&	0.0375&	0.8430	\\
				\hline
		\end{tabular}}	
	\end{center}
	\caption{Two-sample setting. 
		Location differences with ${n}_1={n}_2=5$, $m=100$, $\delta=0.5$, $\mu=2$ and $A$ given by (\ref{ec:A13}). The Monte Carlo estimator of the FDR and power are reported for each test and $q$-value method.  The Monte Carlo bias and standard deviation of each $\pi_0$ estimator are also provided.	}
	\label{ca:L_2_100_5_0.5}
\end{table}

\begin{table}[htb]
	\begin{center}
		\scalebox{0.7}[0.7]{ 
			\begin{tabular}{cccccccccccccccc}
				\hline
				& 	& \multicolumn{14}{c}{Two-sample tests} \\
				&  & \multicolumn{2}{c}{$J_i$} && \multicolumn{2}{c}{abs}
				&& \multicolumn{2}{c}{t-test}
				&& \multicolumn{2}{c}{KS}
				&& \multicolumn{2}{c}{Wilcoxon}\\
				\cline{3-4}\cline{6-7} \cline{9-10} \cline{12-13} \cline{15-16}\\
				$\widehat{\pi}_0$	& &  Bias & Sd &&	Bias & Sd && Bias & Sd && Bias & Sd && Bias & Sd\\	
				Liang&&	0.0427&	0.0254	&&	0.0154&	0.0202	&&	-&-		&&	0.0376&	0.0139	&&	0.0140&	0.0235	\\
				ST&&	0.0915&	0.0710	&&	0.0731&	0.0687	&&	0.0022	&0.0640	&&	0.5000&	0.0000	&&	0.4135&	0.0716	\\
				Chen&&	0.0254&	0.0199	&&	0.0039&	0.0187	&&	-&-		&&	0.0390&	0.0139	&&	0.0109&	0.0199	\\
				SS&&	0.0437	&0.0234	&&	0.0240&	0.0233	&&	0.0084&	0.0236	&&	0.3295	&0.0215	&&	0.0277&	0.0238	\\
				Rand&&	0.0281&	0.0232	&&	0.0089&	0.0228	&&	-&-		&&	0.0314&	0.0160	&&	0.0128&	0.0231	\\
				\hline 		
				& 	& \multicolumn{14}{c}{Two-sample tests} \\
				&  & \multicolumn{2}{c}{$J_i$} && \multicolumn{2}{c}{abs}
				&& \multicolumn{2}{c}{t-test}
				&& \multicolumn{2}{c}{KS}
				&& \multicolumn{2}{c}{Wilcoxon}\\
				\cline{3-4}\cline{6-7} \cline{9-10} \cline{12-13} \cline{15-16}\\
				MCP	& &  FDR & POWER &&	FDR & POWER && FDR & POWER && FDR & POWER && FDR & POWER\\	
				Liang&&	0.0310&	0.6879	&&	0.0379&	0.7287	&&	-&-		&&	0.0379&	0.7287	&&	0.0379&	0.7287	\\
				ST&&	0.0212	&0.4636	&&	0.0339&	0.6484	&&	0.0511&	0.8066	&&	0.0000&	0.0000	&&	0.0000&	0.0000	\\
				Chen&&	0.0311&	0.6898	&&	0.0379&	0.7287	&&	-&-		&&	0.0379&	0.7287	&&	0.0379&	0.7287	\\
				Real &&	0.0311&	0.6898	&&	0.0379&	0.7287	&&	0.0504&	0.8064	&&	0.0379&	0.7287	&&	0.0379&	0.7287	\\
				SS&&	0.0309&	0.6859	&&	0.0379&	0.7287	&&	0.0497&	0.8041	&&	0.0000&	0.0000	&&	0.0379&	0.7287	\\
				Rand&&	0.0311&	0.6898	&&	0.0379&	0.7287	&&		-&-	&&	0.0379&	0.7287	&&	0.0379&	0.7287	\\
				\hline
		\end{tabular}}	
	\end{center}
	\caption{Two-sample setting. 
		Location differences with ${n}_1={n}_2=4$, $m=1000$, $\delta=0.5$, $\mu=2$ and $A$ given by (\ref{ec:A13}). The Monte Carlo estimator of the FDR and power are reported for each test and $q$-value method.  The Monte Carlo bias and standard deviation of each $\pi_0$ estimator are also provided.}
	\label{ca:L_2_1000_4_0.5}
\end{table}

\vspace*{5cm}

\begin{table}[htb]
	\begin{center}
		\scalebox{0.7}[0.7]{ 
			\begin{tabular}{cccccccccccccccc}
				\hline
				& 	& \multicolumn{14}{c}{Two-sample tests} \\
				&  & \multicolumn{2}{c}{$J_i$} && \multicolumn{2}{c}{abs}
				&& \multicolumn{2}{c}{t-test}
				&& \multicolumn{2}{c}{KS}
				&& \multicolumn{2}{c}{Wilcoxon}\\
				\cline{3-4}\cline{6-7} \cline{9-10} \cline{12-13} \cline{15-16}\\
				$\widehat{\pi}_0$	& &  Bias & Sd &&	Bias & Sd && Bias & Sd && Bias & Sd && Bias & Sd\\	
				Liang &&	0.0260&	0.0184	&&	0.0080&	0.0166	&&	-&-		&&	0.0079&	0.0193	&&	0.0089&	0.0201	\\
				ST &&	0.0541	&0.0772	&&	0.0373&	0.0759	&&	0.0008&	0.0731	&&	0.3000&	0.0000	&&	0.3000&	0.0000	\\
				Chen&&	0.0067	&0.0207	&&	-0.0087&	0.0189	&&	-&-		&&	0.0089&	0.0184	&&	-0.0002	&0.0176	\\
				SS &&	0.0284&	0.0260	&&	0.0150&	0.0248	&&	0.0035&	0.0249	&&	0.2102&	0.0249	&&	0.1168&	0.0251	\\
				Rand &&	0.0168&	0.0262	&&	0.0036&	0.0250	&&	-&-		&&	0.0077&	0.0197	&&	0.0044&	0.0236	\\
				\hline 		
				& 	& \multicolumn{14}{c}{Two-sample tests} \\
				&  & \multicolumn{2}{c}{$J_i$} && \multicolumn{2}{c}{abs}
				&& \multicolumn{2}{c}{t-test}
				&& \multicolumn{2}{c}{KS}
				&& \multicolumn{2}{c}{Wilcoxon}\\
				\cline{3-4}\cline{6-7} \cline{9-10} \cline{12-13} \cline{15-16}\\
				MCP	& &  FDR & POWER &&	FDR & POWER && FDR & POWER && FDR & POWER && FDR & POWER\\	
				Liang &&	0.0230 &	0.6329	&&	0.0458	&0.7529	&&	-&-		&&	0.0271&	0.6652	&&	0.0454&	0.7504	\\
				ST&&	0.0237&	0.6388	&&	0.0389&	0.7202	&&	0.0508&	0.7957	&&	0.0271&	0.6652	&&	0.0271&	0.6652	\\
				Chen&&	0.0258&	0.6472	&&	0.0465&	0.7570	&&	-&-		&&	0.0271&	0.6652	&&	0.0462&	0.7550	\\
				Real&&	0.0265&	0.6517	&&	0.0465&	0.7570	&&	0.0500&	0.7957	&&	0.0271&	0.6652	&&	0.0465&	0.7570	\\
				SS&&	0.0231&	0.6333	&&	0.0441&	0.7449	&&	0.0499&	0.7950	&&	0.0271&	0.6652	&&	0.0281&	0.6693	\\
				Rand&&	0.0247&	0.6413	&&	0.0453&	0.7512	&&	-&-		&&	0.0271&	0.6652	&&	0.0452&	0.7511	\\
				\hline
		\end{tabular}}	
	\end{center}
	\caption{Two-sample setting. 
		Location differences with ${n}_1={n}_2=5$, $m=1000$, $\delta=0.3$, $\mu=2$ and $A$ given by (\ref{ec:A13}). The Monte Carlo estimator of the FDR and power are reported for each test and $q$-value method.  The Monte Carlo bias and standard deviation of each $\pi_0$ estimator are also provided.}
	\label{ca:L_2_1000_5_0.3}
\end{table}

\begin{table}[htb]
	\begin{center}
		\scalebox{0.7}[0.7]{ 
			\begin{tabular}{cccccccccccccccc}
				\hline
				& 	& \multicolumn{14}{c}{Two-sample tests} \\
				&  & \multicolumn{2}{c}{$J_i$} && \multicolumn{2}{c}{abs}
				&& \multicolumn{2}{c}{t-test}
				&& \multicolumn{2}{c}{KS}
				&& \multicolumn{2}{c}{Wilcoxon}\\
				\cline{3-4}\cline{6-7} \cline{9-10} \cline{12-13} \cline{15-16}\\
				$\widehat{\pi}_0$	& &  Bias & Sd &&	Bias & Sd && Bias & Sd && Bias & Sd && Bias & Sd\\	
				Liang&&	0.0266&	0.0199	&&	0.0099&	0.0172	&&	-&-		&&	0.0134	&0.0176	&&	0.0095&	0.0193	\\
				ST&&	0.0371&	0.0631	&&	0.0280&	0.0680	&&	0.0009&	0.0655	&&	0.5000	&0.0000	&&	0.4683&	0.0548	\\
				Chen&&	0.0097&	0.0190	&&	-0.0049&	0.0179	&&	-&-		&&	0.0149	&0.0175	&&	0.0044&	0.0163	\\
				SS&&	0.0254 &	0.0228	&&	0.0121&	0.0221	&&	0.0042&	0.0222	&&	0.1601&	0.0226	&&	0.0873&	0.0225	\\
				Rand &&	0.0170&	0.0229	&&	0.0041	&0.0219	&&	-&-		&&	0.0126&	0.0179	&&	0.0058&	0.0211	\\
				\hline 		
				& 	& \multicolumn{14}{c}{Two-sample tests} \\
				&  & \multicolumn{2}{c}{$J_i$} && \multicolumn{2}{c}{abs}
				&& \multicolumn{2}{c}{t-test}
				&& \multicolumn{2}{c}{KS}
				&& \multicolumn{2}{c}{Wilcoxon}\\
				\cline{3-4}\cline{6-7} \cline{9-10} \cline{12-13} \cline{15-16}\\
				MCP	& &  FDR & POWER &&	FDR & POWER && FDR & POWER && FDR & POWER && FDR & POWER\\	
				Liang&&	0.0359&	0.8140	&&	0.0451	&0.8722	&&	-&-		&&	0.0115&	0.6665	&&	0.0363&	0.8366	\\
				ST&&	0.0351&	0.8108	&&	0.0448&	0.8693	&&	0.0509&	0.8898	&&	0.0115&	0.6665	&&	0.0201&	0.7599	\\
				Chen&&	0.0375	&0.8199	&&	0.0468	&0.8765	&&	-&-		&&	0.0115&	0.6665	&&	0.0363&	0.8366	\\
				Real&&	0.0379&	0.8222	&&	0.0447&	0.8715	&&	0.0501&	0.8892	&&	0.0115&	0.6665	&&	0.0363&	0.8366	\\
				SS&&	0.0360&	0.8143	&&	0.0452&	0.8724	&&	0.0499&	0.8883	&&	0.0115&	0.6665	&&	0.0363&	0.8365	\\
				Rand &&	0.0368&	0.8174	&&	0.0459&	0.8742	&&	-&-		&&	0.0115&	0.6665	&&	0.0363&	0.8366	\\
				\hline
		\end{tabular}}	
	\end{center}
	\caption{ Two-sample setting. Location differences with ${n}_1={n}_2=5$, $m=1000$, $\delta=0.5$, $\mu=2$ and $A$ given by (\ref{ec:A13}). The Monte Carlo estimator of the FDR and power are reported for each test and $q$-value method.  The Monte Carlo bias and standard deviation of each $\pi_0$ estimator are also provided.}
	\label{ca:L_2_1000_5_0.5}
\end{table}

\vspace*{5cm}

\begin{table}[htb]
	\begin{center}
		\scalebox{0.7}[0.7]{ 
			\begin{tabular}{cccccccccccccccc}
				\hline
				& 	& \multicolumn{14}{c}{Two-sample tests} \\
				&  & \multicolumn{2}{c}{$J_i$} && \multicolumn{2}{c}{abs}
				&& \multicolumn{2}{c}{t-test}
				&& \multicolumn{2}{c}{KS}
				&& \multicolumn{2}{c}{Wilcoxon}\\
				\cline{3-4}\cline{6-7} \cline{9-10} \cline{12-13} \cline{15-16}\\
				$\widehat{\pi}_0$	& &  Bias & Sd &&	Bias & Sd && Bias & Sd && Bias & Sd && Bias & Sd\\	
				Liang&&	0.0562&	0.0609	&&	0.0269&	0.0607	&&	-&-		&&	0.0372&	0.0503	&&	0.0222	&0.0703	\\
				ST&&	0.0913&	0.2045	&&	0.0789&	0.2078	&&	0.0153&	0.1976	&&	0.4906	&0.0495	&&	0.3400&	0.1818	\\
				Chen&&	0.0173&	0.0653	&&	-0.0039&	0.0678	&&	-&-		&&	0.0508	&0.0503	&&	0.0084&	0.0623	\\
				SS&&	0.0431&	0.0828	&&	0.0248&	0.0873	&&	0.0098&	0.0871	&&	0.3288&	0.0773	&&	0.0298&	0.0821	\\
				Rand &&	0.0272&	0.0814	&&	0.0109&	0.0867	&&	-&-		&&	0.0313	&0.0585	&&	0.0148&	0.0799	\\
				\hline 		
				& 	& \multicolumn{14}{c}{Two-sample tests} \\
				&  & \multicolumn{2}{c}{$J_i$} && \multicolumn{2}{c}{abs}
				&& \multicolumn{2}{c}{t-test}
				&& \multicolumn{2}{c}{KS}
				&& \multicolumn{2}{c}{Wilcoxon}\\
				\cline{3-4}\cline{6-7} \cline{9-10} \cline{12-13} \cline{15-16}\\
				MCP	& &  FDR & POWER &&	FDR & POWER && FDR & POWER && FDR & POWER && FDR & POWER\\	
				Liang&&	0.0261&	0.5430	&&	0.0388&	0.6962	&&	-&-		&&	0.0363&	0.6849	&&	0.0391&	0.6967	\\
				ST&&	0.0243&	0.4166	&&	0.0377&	0.5091	&&	0.0616&	0.8137	&&	0.0008&	0.0077	&&	0.0112&	0.1564	\\
				Chen&&	0.0290&	0.6140	&&	0.0406&	0.7142	&&	-&-		&&	0.0357&	0.6683	&&	0.0395&	0.7103	\\
				Real&&	0.0296&	0.6800	&&	0.0374&	0.7309	&&	0.0505&	0.8082	&&	0.0374&	0.7309	&&	0.0374&	0.7309	\\
				SS&&	0.0271&	0.5510	&&	0.0392&	0.6733	&&	0.0534&	0.8070	&&	0.0056&	0.0599	&&	0.0383&	0.6699	\\
				Rand&&	0.0281&	0.5740	&&	0.0398&	0.6849	&&	-&-		&&	0.0363&	0.6876	&&	0.0394&	0.6876	\\	
				\hline
		\end{tabular}}	
	\end{center}
	\caption{Two-sample setting. Location differences with ${n}_1={n}_2=4$, $m=100$, $\delta=0.5$, $\mu=2$ and $A$ given by (\ref{ec:A23}). The Monte Carlo estimator of the FDR and power are reported for each test and $q$-value method.  The Monte Carlo bias and standard deviation of each $\pi_0$ estimator are also provided.}
	\label{ca:L_2_100_4_0.5_dep}
\end{table}

\begin{table}[htb]
	\begin{center}
		\scalebox{0.7}[0.7]{ 
			\begin{tabular}{cccccccccccccccc}
				\hline
				& 	& \multicolumn{14}{c}{Two-sample tests} \\
				&  & \multicolumn{2}{c}{$J_i$} && \multicolumn{2}{c}{abs}
				&& \multicolumn{2}{c}{t-test}
				&& \multicolumn{2}{c}{KS}
				&& \multicolumn{2}{c}{Wilcoxon}\\
				\cline{3-4}\cline{6-7} \cline{9-10} \cline{12-13} \cline{15-16}\\
				$\widehat{\pi}_0$	& &  Bias & Sd &&	Bias & Sd && Bias & Sd && Bias & Sd && Bias & Sd\\	
				Liang &&	0.0436&	0.0269	&&	0.0155&	0.0233	&&	-&-		&&	0.0383&	0.0156	&&	0.0143&	0.0254\\
				ST&&	0.0915&	0.0747	&&	0.0742&	0.0728	&&	0.0033	&0.0707	&&	0.5000&	0.0000	&&	0.4105&	0.0779\\
				Chen&&	0.0260&	0.0221	&&	0.0039&	0.0221	&&	-&-		&&	0.0397&	0.0156	&&	0.0116	&0.0221\\
				SS&&	0.0439&	0.0253	&&	0.0245&	0.0269	&&	0.0093&	0.0270	&&	0.3306&	0.0241	&&	0.0286&	0.0263\\
				Rand &&	0.0282&	0.0250	&&	0.0099&	0.0267	&&	-&-	&&	0.0319&	0.0187	&&	0.0138&	0.0256\\		
				\hline 		
				& 	& \multicolumn{14}{c}{Two-sample tests} \\
				&  & \multicolumn{2}{c}{$J_i$} && \multicolumn{2}{c}{abs}
				&& \multicolumn{2}{c}{t-test}
				&& \multicolumn{2}{c}{KS}
				&& \multicolumn{2}{c}{Wilcoxon}\\
				\cline{3-4}\cline{6-7} \cline{9-10} \cline{12-13} \cline{15-16}\\
				MCP	& &  FDR & POWER &&	FDR & POWER && FDR & POWER && FDR & POWER && FDR & POWER\\	
				Liang &&	0.0309&	0.6864	&&	0.0380&	0.7284	&&	-&-		&&	0.0380&0.7284	&&	0.0380&	0.7284	\\
				ST&&	0.0211&	0.4573	&&	0.0335&	0.6338	&&	0.0513&	0.8063	&&	0.0000&	0.0000	&&	0.0380&	0.0015	\\
				Chen&&	0.0310&	0.6897	&&	0.0380&	0.7284	&&	-&-		&&	0.0380&	0.7284	&&	0.0380	&0.7284	\\
				Real &&	0.0310&	0.6897	&&	0.0380&	0.7284	&&	0.0502&	0.8065	&&	0.0380	&0.7284	&&	0.0380&	0.7284	\\
				SS&&	0.0308&	0.6838	&&	0.0380&	0.7284	&&	0.0496&	0.8037	&&	0.0000&	0.0000	&&	0.0380&	0.7284	\\
				Rand &&	0.0309&	0.6877	&&	0.0380&	0.7284	&&	-&-		&&	0.0380&	0.7284	&&	0.0380	&0.7284	\\		
				\hline	
		\end{tabular}}	
	\end{center}
	\caption{Two-sample setting. Location differences with ${n}_1={n}_2=4$, $m=1000$, $\delta=0.5$, $\mu=2$ and $A$ given by (\ref{ec:A23}). The Monte Carlo estimator of the FDR and power are reported for each test and $q$-value method.  The Monte Carlo bias and standard deviation of each $\pi_0$ estimator are also provided.}
	\label{ca:L_2_1000_4_0.5_dep}
\end{table}

\vspace*{5cm}

\begin{table}[htb]
	\begin{center}
		\scalebox{0.7}[0.7]{ 
			\begin{tabular}{cccccccccccccccc}
				\hline
				& 	& \multicolumn{14}{c}{Two-sample tests} \\
				&  & \multicolumn{2}{c}{$J_i$} && \multicolumn{2}{c}{$F$-test}
				&& \multicolumn{2}{c}{KS}
				&& \multicolumn{2}{c}{Ansari}
				&& \multicolumn{2}{c}{Levene}\\
				\cline{3-4}\cline{6-7} \cline{9-10} \cline{12-13} \cline{15-16}\\
				$\widehat{\pi}_0$	& &  Bias & Sd &&	Bias & Sd && Bias & Sd && Bias & Sd && Bias & Sd\\	
				Liang&&	0.0268&	0.0510	&&	-&-		&&	0.1130&	0.0800	&&	0.040&8	0.0666	&&	-&-		\\
				ST&&	-0.0220	&0.2038	&&	-0.0172&	0.2048	&&	0.3000&	0.0000	&&	0.2793&	0.0626	&&	0.0284	&0.2005	\\
				Chen&&	-0.0235&	0.0597	&&	-&-		&&	0.1120	&0.0594	&&	0.0128	&0.0625	&&	-&-		\\
				SS&&	0.0084&	0.0817	&&	0.0031&	0.0831	&&	0.2983&	0.0089	&&	0.0652&	0.0860	&&	0.0386	&0.0834	\\
				Rand &&	0.0083&	0.0818	&&	-&-		&&	0.0651&	0.0674	&&	0.0208&	0.0815	&&	-&-		\\
				\hline 		
				&  & \multicolumn{2}{c}{$J_i$} && \multicolumn{2}{c}{$F$-test}
				&& \multicolumn{2}{c}{KS}
				&& \multicolumn{2}{c}{Ansari}
				&& \multicolumn{2}{c}{Levene}\\
				\cline{3-4}\cline{6-7} \cline{9-10} \cline{12-13} \cline{15-16}\\
				MCP	& &  FDR & POWER &&	FDR & POWER && FDR & POWER && FDR & POWER && FDR & POWER\\	
				Liang&&	0.0387&	0.1924	&&	-&-		&&	0.0088&	0.0019	&&	0.0391&	0.3272	&&	-&-		\\
				ST&&	0.0472&	0.2579	&&	0.0568&	0.9192	&&	0.0088&	0.0019	&&	0.0256	&0.2606	&&	0.0383&	0.2549	\\
				Chen&&	0.0426&	0.2225	&&	-&-		&&	0.0088 &	0.0019	&&	0.0400&	0.3384	&&		-&-	\\
				Real&&	0.0403&	0.2019	&&	0.0490&	0.9153	&&	0.0088	&0.0019	&&	0.0386	&0.3359	&&	0.0363	&0.2324	\\
				SS&&	0.0402&	0.2069	&&	0.0507&	0.9163	&&	0.0088	&0.0019	&&	0.0360	&0.3192	&&	0.0356	&0.2235	\\
				Rand&&	0.0402&	0.2069	&&	-&-		&&	0.0088&	0.0019	&&	0.0395&	0.3387	&&	-&-		\\
				\hline
		\end{tabular}}	
	\end{center}
	\caption{Two-sample setting. Scale differences with ${n}_1={n}_2=8$, $m=100$, $\delta=0.3$ and $A$ given by (\ref{ec:A13}). The Monte Carlo estimator of the FDR and power are reported for each test and $q$-value method.  The Monte Carlo bias and standard deviation of each $\pi_0$ estimator are also provided.}
	\label{ca:s_8_100_0.3_ind}
\end{table}

\begin{table}[htb]
	\begin{center}
		\scalebox{0.66}[0.68]{ 
			\begin{tabular}{ccccccccccccccccccc}
				\hline
				& 	& \multicolumn{14}{c}{Two-sample tests} \\
				&  & \multicolumn{2}{c}{$J_i$} && \multicolumn{2}{c}{$F$-test}
				&& \multicolumn{2}{c}{KS}
				&& \multicolumn{2}{c}{Ansari}
				&& \multicolumn{2}{c}{Siegel} && \multicolumn{2}{c}{Levene}\\
				\cline{3-4}\cline{6-7} \cline{9-10} \cline{12-13} \cline{15-16} \cline{18-19}\\
				$\widehat{\pi}_0$	& &  Bias & Sd &&	Bias & Sd && Bias & Sd && Bias & Sd && Bias & Sd && Bias & Sd\\	
				Liang &&	0.0345&	0.0548	&&	-&-		&&	0.1727	&0.0739	&&	0.0518&	0.0702	&&	0.0698&	0.0607	&&		-&-	\\
				ST&&	0.0081&	0.1956	&&	0.0039&	0.1832	&&	0.4995	&0.0117	&&	0.3729&	0.1563	&&	0.3008&	0.1987	&&	0.0039&	0.1832	\\
				Chen&&	-0.0052&	0.0553	&&	-&-		&&	0.1817&	0.0616	&&	0.0341	&0.0589	&&	0.0349	&0.0585	&&	-&-		\\
				SS&&	0.0165	&0.0704	&&	0.0020&	0.0728	&&	0.4396&	0.0645	&&	0.0697	&0.0780	&&	0.1023&	0.0785	&&	0.0020&	0.0728	\\
				Rand&&	0.0164&	0.0703	&&-&-			&&	0.1158	&0.0673	&&	0.0697&	0.0780	&&	0.0363&	0.0758	&&	-&-		\\
				
				\hline 		
				& 	& \multicolumn{14}{c}{Two-sample tests} \\
				&  & \multicolumn{2}{c}{$J_i$} && \multicolumn{2}{c}{$F$-test}
				&& \multicolumn{2}{c}{KS}
				&& \multicolumn{2}{c}{Ansari}
				&& \multicolumn{2}{c}{Siegel} && \multicolumn{2}{c}{Levene}\\
				\cline{3-4}\cline{6-7} \cline{9-10} \cline{12-13} \cline{15-16} \cline{18-19}\\
				MCP	& &  FDR & POWER &&	FDR & POWER && FDR & POWER && FDR & POWER && FDR & POWER && FDR & POWER\\	
				Liang&&	0.0433&	0.5609	&&	-&-		&&	0.0085&	0.0025	&&	0.0393&	0.5394	&&	0.0407&	0.5448	&&	-&-		\\
				ST&&	0.0546&	0.5943	&&	0.0592&	0.9591	&&	0.0072&	0.0014	&&	0.0250&	0.4229	&&	0.0318&	0.4653	&&	0.0423&	0.5836	\\
				Chen&&	0.0470&	0.5969	&&	-&-		&&	0.0085&	0.0024	&&	0.0395	&0.5451	&&	0.0438&	0.5620	&&	-&-		\\
				Real&&	0.0453	&0.5937	&&	0.0504&	0.9586	&&	0.0092&	0.0048	&&	0.0397&	0.5543	&&	0.0454&	0.5752	&&	0.0372&	0.5935	\\
				SS&&	0.0454&	0.5794	&&	0.0524	&0.9586	&&	0.0072	&0.0014	&&	0.0373&	0.5294	&&	0.0393	&0.5317	&&	0.0366	&0.5691	\\
				Rand&&	0.0454&	0.5794	&&	-&-		&&	0.0092	&0.0036	&&	0.0373	&0.5294	&&	0.0445&	0.5631	&&	-&-		\\
				\hline
		\end{tabular}}	
	\end{center}
	\caption{Two-sample setting. Scale differences with ${n}_1={n}_2=8$, $m=100$, $\delta=0.5$ and $A$ given by (\ref{ec:A13}). The Monte Carlo estimator of the FDR and power are reported for each test and $q$-value method.  The Monte Carlo bias and standard deviation of each $\pi_0$ estimator are also provided.}
	\label{ca:s_8_100_0.5_ind}
\end{table}

\vspace*{6cm}

\begin{table}[htb]
	\begin{center}
		\scalebox{0.7}[0.7]{ 
			\begin{tabular}{cccccccccccccccc}
				\hline
				& 	& \multicolumn{14}{c}{Two-sample tests} \\
				&  & \multicolumn{2}{c}{$J_i$} && \multicolumn{2}{c}{$F$-test}
				&& \multicolumn{2}{c}{KS}
				&& \multicolumn{2}{c}{Ansari}
				&& \multicolumn{2}{c}{Levene}\\
				\cline{3-4}\cline{6-7} \cline{9-10} \cline{12-13} \cline{15-16}\\
				$\widehat{\pi}_0$	& &  Bias & Sd &&	Bias & Sd && Bias & Sd && Bias & Sd && Bias & Sd\\
				Liang			&&	0.0305&	0.0675	&&	-&-		&&	0.1716	&0.0955	&&	0.0511	&0.0733	&&	-&-		\\
				ST			&&	0.0131&	0.2085	&&	-0.0010	&0.1801	&&	0.4979	&0.0248	&&	0.3714&	0.1643	&&	0.0293	&0.1925	\\
				Chen			&&	-0.0099&	0.0749	&&	-&-		&&	0.1798	&0.0823	&&	0.0313	&0.0592	&&		-&-	\\
				SS			&&	0.0106&0.0928	&&	-0.0018	&0.0743	&&	0.4279&	0.0875	&&	0.0630&	0.0760	&&	0.0268	&0.0699	\\
				Rand			&&	0.0105&	0.0928	&&	-&-		&&	0.1132&	0.0891	&&	0.0300	&0.0723	&&	-&-		\\	
				\hline 		
				&  & \multicolumn{2}{c}{$J_i$} && \multicolumn{2}{c}{$F$-test}
				&& \multicolumn{2}{c}{KS}
				&& \multicolumn{2}{c}{Ansari}
				&& \multicolumn{2}{c}{Levene}\\
				\cline{3-4}\cline{6-7} \cline{9-10} \cline{12-13} \cline{15-16}\\
				MCP	& &  FDR & POWER &&	FDR & POWER && FDR & POWER && FDR & POWER && FDR & POWER\\	
				Liang&&	0.0475&	0.5489	&&		-&-	&&	0.0199 &	0.0053	&&	0.0385	&0.5332	&&			-&-	\\
				ST&&	0.0608&	0.5750	&&	0.0577&	0.9602	&&	0.0052&	0.0018	&&	0.0260	&0.4161	&&	0.0446	&0.5772	\\
				Chen &&	0.0520&	0.5844	&&			-&-	&&	0.0193&	0.0052	&&	0.0399&	0.5416	&&		-&-		\\
				Real&&	0.0469&	0.5848	&&	0.0479&	0.9587	&&	0.0239&	0.0064	&&	0.0398	&0.5495	&&	0.0368&	0.5894	\\
				SS&&	0.0507&	0.5634	&&	0.0508&	0.9595	&&	0.0067	&0.0022	&&	0.0380&	0.5295	&&	0.0371&	0.5669	\\
				Rand&&	0.0507&	0.5634	&&		-&-		&&	0.0214&	0.0062	&&	0.0405&	0.5449	&&		-&-		\\
				\hline
		\end{tabular}}	
	\end{center}
	\caption{Two-sample setting. Scale differences with ${n}_1={n}_2=8$, $m=100$, $\delta=0.5$ and $A$ given by (\ref{ec:A23}). The Monte Carlo estimator of the FDR and power are reported for each test and $q$-value method.  The Monte Carlo bias and standard deviation of each $\pi_0$ estimator are also provided.}
	\label{ca:s_8_100_0.5_dep}
\end{table}

\begin{table}[htb]
	\begin{center}
		\scalebox{0.7}[0.7]{ 
			\begin{tabular}{cccccccccccccccc}
				\hline
				& 	& \multicolumn{14}{c}{Two-sample tests} \\
				&  & \multicolumn{2}{c}{$J_i$} && \multicolumn{2}{c}{$F$-test}
				&& \multicolumn{2}{c}{KS}
				&& \multicolumn{2}{c}{Ansari}
				&& \multicolumn{2}{c}{Levene}\\
				\cline{3-4}\cline{6-7} \cline{9-10} \cline{12-13} \cline{15-16}\\
				$\widehat{\pi}_0$	& &  Bias & Sd &&	Bias & Sd && Bias & Sd && Bias & Sd && Bias & Sd\\
				Liang &&	0.0208&	0.0234	&&	-&-		&&	0.1677&	0.0197	&&	0.0391&	0.0231	&&	-&-		\\
				ST&&	-0.0005&	0.0672	&&	0.0048&	0.0617	&&	0.5000&	0.0000	&&	0.4497	&0.0567	&&	0.0283&	0.0657	\\
				Chen&&	0.0059&	0.0207	&&	-&-		&&	0.1691&	0.0197	&&	0.0380	&0.0210	&&	-&-		\\
				SS&&	0.0127 &	0.0242	&&	0.0032	&0.0217	&&	0.4571&	0.0267	&&	0.0701&	0.0234	&&	0.0296	&0.0222	\\
				Rand&&	0.0126 &	0.0242	&&	-&-		&&	0.1155	&0.0209	&&	0.0367&	0.0227	&&	-&-		\\
				\hline 		
				&  & \multicolumn{2}{c}{$J_i$} && \multicolumn{2}{c}{$F$-test}
				&& \multicolumn{2}{c}{KS}
				&& \multicolumn{2}{c}{Ansari}
				&& \multicolumn{2}{c}{Levene}\\
				\cline{3-4}\cline{6-7} \cline{9-10} \cline{12-13} \cline{15-16}\\
				MCP	& &  FDR & POWER &&	FDR & POWER && FDR & POWER && FDR & POWER && FDR & POWER\\	
				Liang&&	0.0427&	0.5673	&&	-&-		&&	0.0070	&0.0003	&&	0.0404&	0.5516	&&	-&-		\\
				ST&&	0.0454&	0.5870	&&	0.0507&	0.9580	&&	0.0010&	0.0000	&&	0.0219&	0.3894	&&	0.0333&	0.5587	\\
				Chen&&	0.0439&	0.5813	&&	-&-		&&	0.0070	&0.0003	&&	0.0405&	0.5518	&&	-&-		\\
				Real &&	0.0444&	0.5872	&&	0.0501	&0.9585	&&	0.0210&	0.0009	&&	0.0405	&0.5521	&&	0.0348&	0.5848	\\
				SS&&	0.0435&	0.5747	&&	0.0499	&0.9581	&&	0.0050&	0.0002	&&	0.0398&	0.5461	&&	0.0328&	0.5550	\\
				Rand&&	0.0435&	0.5748	&&	-&-		&&	0.0200	&0.0008	&&	0.0405	&0.5521	&&	-&-		\\		
				\hline
		\end{tabular}}	
	\end{center}
	\caption{Two-sample setting. Scale differences with ${n}_1={n}_2=8$, $m=1000$, $\delta=0.5$ and $A$ given by (\ref{ec:A13}). The Monte Carlo estimator of the FDR and power are reported for each test and $q$-value method.  The Monte Carlo bias and standard deviation of each $\pi_0$ estimator are also provided.}
	\label{ca:s_8_1000_0.5_ind}
\end{table}

\vspace*{7cm}

\begin{table}[htb]
	\begin{center}
		\scalebox{0.7}[0.7]{ 
			\begin{tabular}{cccccccccccccccc}
				\hline
				& 	& \multicolumn{14}{c}{Two-sample tests} \\
				&  & \multicolumn{2}{c}{$J_i$} && \multicolumn{2}{c}{abs}
				&& \multicolumn{2}{c}{Welch's test}
				&& \multicolumn{2}{c}{$F$-test}
				&& \multicolumn{2}{c}{KS}\\
				\cline{3-4}\cline{6-7} \cline{9-10} \cline{12-13} \cline{15-16}\\
				$\widehat{\pi}_0$	& &  Bias & Sd &&	Bias & Sd && Bias & Sd && Bias & Sd && Bias & Sd\\	
				Liang&&	0.0188 &	0.0439	&&	0.3939	&0.0531	&&	-&-		&&	-&-		&&	0.0670	&0.0528	\\
				ST&&	0.0085&	0.1954	&&	0.4096&	0.1405	&&	0.3277&	0.1785	&&	0.0163	&0.1862	&&	0.4993&	0.0133	\\
				Chen&&	-0.0151&	0.0523	&&	0.3616&	0.0712	&&	-&-		&&	-&-		&&	0.0810&	0.0528	\\
				SS&&	0.0098&	0.0691	&&	0.4030&	0.0809	&&	0.3923&	0.0841	&&	0.0189	&0.0745	&&	0.3132&	0.0751	\\
				Rand &&	0.0097	&0.0690	&&	0.4029&	0.0808	&&	-&-		&&	-&-		&&	0.0449&	0.0605	\\
				\hline
				& 	& \multicolumn{14}{c}{Two-sample tests} \\
				&  & \multicolumn{2}{c}{Wilcoxon} && \multicolumn{2}{c}{Ansari}
				&& \multicolumn{2}{c}{Siegel}
				&& \multicolumn{2}{c}{Levene}
				&& \multicolumn{2}{c}{\;}\\
				\cline{3-4}\cline{6-7} \cline{9-10} \cline{12-13} \\
				$\widehat{\pi}_0$	& &  Bias & Sd &&	Bias & Sd && Bias & Sd && Bias & Sd &&  & \\
				Liang&&	0.2870&	0.0680	&&	0.1002&	0.0769	&&	0.1242&	0.0715	&&	-&-		&&	&	\\
				ST&&	0.0000&	0.0000	&&	0.4461&	0.1079	&&	0.3958&	0.1545	&&	0.0436&	0.1890	&&	&	\\
				Chen&&	0.0000&	0.0000	&&	0.0815&	0.0621	&&	0.0816&	0.0619	&&	-&-		&&	&	\\
				SS&&	0.0000&	0.0000	&&	0.1223&	0.0831	&&	0.1565&	0.0833	&&	0.0564&	0.0757	&&	&	\\
				Rand&&	0.0000&	0.0000	&&	0.1223&	0.0831	&&	0.0881&	0.0820	&&	-&-		&&	&	\\
				\hline
		\end{tabular}}	
	\end{center}
	\caption{Two-sample setting. Location, Scale and Shape differences with ${n}_1={n}_2=8$, $m=100$, $\delta=0.5$ and $A$ given by (\ref{ec:A13}).  The Monte Carlo bias and standard deviation of each $\pi_0$ estimator are provided.}
	\label{ca:LSS_8_100_0.5_ind_pi0}
\end{table}

\begin{table}[htb]
	\begin{center}
		\scalebox{0.7}[0.7]{ 
			\begin{tabular}{cccccccccccccccc}
				\hline
				& 	& \multicolumn{14}{c}{Two-sample tests} \\
				&  & \multicolumn{2}{c}{$J_i$} && \multicolumn{2}{c}{abs}
				&& \multicolumn{2}{c}{Welch's test}
				&& \multicolumn{2}{c}{$F$-test}
				&& \multicolumn{2}{c}{KS}\\
				\cline{3-4}\cline{6-7} \cline{9-10} \cline{12-13} \cline{15-16}\\
				MCP	& &  FDR & POWER &&	FDR & POWER && FDR & POWER && FDR & POWER &&FDR &POWER\\
				Liang &&	0.0445&	0.8195	&&	0.0290	&0.0702	&&	-&-		&&	-&-		&&	0.0280	&0.1779	\\
				ST &&	0.0552&	0.8284	&&	0.0274	&0.0695	&&	0.0298	&0.0701	&&	0.0573&	0.8674	&&	0.0152&	0.1038	\\
				Chen &&	0.0487&	0.8338	&&	0.0298&	0.0723	&&	-&-		&&	-&-		&&	0.0263&	0.1678	\\
				Real &&	0.0448&	0.8271	&&	0.0495&	0.1031	&&	0.0431&	0.0909	&&	0.0506&	0.8672	&&	0.0336&	0.2241	\\
				SS&&	0.0464	&0.8249	&&	0.0285&	0.0693	&&	0.0257&	0.0655	&&	0.0512&	0.8643	&&	0.0163&	0.1086	\\
				Rand &&	0.0464&	0.8250	&&	0.0285&	0.0693	&&	-&-		&& -&-	&&	0.0292&	0.1886	\\
				
				\hline 		
				& 	& \multicolumn{14}{c}{Two-sample tests} \\
				&  & \multicolumn{2}{c}{Wilcoxon} && \multicolumn{2}{c}{Ansari}
				&& \multicolumn{2}{c}{Siegel}
				&& \multicolumn{2}{c}{Levene}
				&& \multicolumn{2}{c}{\;}\\
				\cline{3-4}\cline{6-7} \cline{9-10} \cline{12-13} \\
				MCP	& &  FDR & POWER &&	FDR & POWER && FDR & POWER && FDR & POWER && &\\	
				Liang &&	0.0245&	0.0534	&&	0.0361&	0.3676	&&	0.0378&	0.3784	&&		-&-	&&	&	\\
				ST&&	0.0000&	0.0000	&&	0.0208&	0.2382	&&	0.0272&	0.2789	&&	0.0387&	0.1713	&&	&	\\
				Chen&&	0.0000&	0.0000	&&	0.0371&	0.3758	&&	0.0402&	0.3953	&&	-&-		&&	&	\\
				Real&&	0.0000&	0.0000	&&	0.0427&	0.4201	&&	0.0441&	0.4266	&&0.0349&	0.1494	&&	&	\\
				SS&&	0.0000&	0.0000	&&	0.0350&	0.3593	&&	0.0360&	0.3659	&& 0.0309 & 0.1350		&&	&	\\
				Rand &&	0.0000&	0.0000	&&	0.0350&	0.3593	&&	0.0397&	0.3949	&&	-&-		&&	&	\\
				\hline
		\end{tabular}}	
	\end{center}
	\caption{Two-sample setting. Location, Scale and Shape differences with ${n}_1={n}_2=8$, $m=100$, $\delta=0.5$ and $A$ given by (\ref{ec:A13}). The Monte Carlo estimator of the FDR and power are reported for each test and $q$-value method.}
	\label{ca:LSS_8_100_0.5_ind_power}
\end{table}

\vspace*{8cm}

\begin{table}[htb]
	\begin{center}
		\scalebox{0.7}[0.7]{ 
			\begin{tabular}{cccccccccccccccc}
				\hline
				& 	& \multicolumn{14}{c}{Two-sample tests} \\
				&  & \multicolumn{2}{c}{$J_i$} && \multicolumn{2}{c}{$F$-test}
				&& \multicolumn{2}{c}{KS}
				&& \multicolumn{2}{c}{Ansari}
				&& \multicolumn{2}{c}{Levene}\\
				\cline{3-4}\cline{6-7} \cline{9-10} \cline{12-13} \cline{15-16}\\
				$\widehat{\pi}_0$	& &  Bias & Sd &&	Bias & Sd && Bias & Sd && Bias & Sd && Bias & Sd\\	
				Liang&&	0.0131&	0.0459	&&	-&-		&&	0.0386&	0.0586	&&	0.0720&	0.0738	&&	-&-		\\
				ST&&	0.0004&	0.2050	&&	-0.1152&	0.1983	&&	0.3000&	0.0000	&&	0.2885&	0.0451	&&	-0.0207&	0.2026	\\
				Chen&&	-0.0244	&0.0623	&&	-&-		&&	0.0505&	0.0556	&&	0.0395&	0.0633	&&	-&-		\\
				SS&&	0.0092&	0.0846	&&	-0.1117&	0.0821	&&	0.2887&	0.0296	&&	0.0948	&0.0851	&&	0.0190	&0.0847	\\
				Rand &&	0.0091	&0.0847	&&	-&-		&&	0.0254&	0.0668	&&	0.0508&	0.0815	&&	-&-		\\	
				\hline 		
				&  & \multicolumn{2}{c}{$J_i$} && \multicolumn{2}{c}{$F$-test}
				&& \multicolumn{2}{c}{KS}
				&& \multicolumn{2}{c}{Ansari}
				&& \multicolumn{2}{c}{Levene}\\
				\cline{3-4}\cline{6-7} \cline{9-10} \cline{12-13} \cline{15-16}\\
				MCP	& &  FDR & POWER &&	FDR & POWER && FDR & POWER && FDR & POWER && FDR & POWER\\	
				Liang&&	0.0464&	0.6004	&&	-&-		&&	0.0315&	0.0859	&&	0.0364&	0.1970	&&	-&-		\\
				ST&&	0.0528&	0.6058	&&	0.2105&	0.8346	&&	0.0210&	0.0672	&&	0.0248&	0.1561	&&	0.0721&	0.0759	\\
				Chen&&	0.0487&	0.6180	&&	-&-		&&	0.0310&	0.0852	&&	0.0370&	0.2025	&&		-&-	\\
				Real&&	0.0466&	0.6067	&&	0.1833&	0.8090	&&	0.0315&	0.0881	&&	0.0382&	0.2022	&&	0.0685&	0.0634	\\
				SS&&	0.0468&	0.6007	&&	0.2031&	0.8279	&&	0.0210	&0.0672	&&	0.0348	&0.1888	&&	0.0717&	0.0637	\\
				Rand&&	0.0468&	0.6009	&&	-&-		&&	0.0310	&0.0862	&&	0.0368&	0.1990	&&	-&-		\\	
				\hline
		\end{tabular}}	
	\end{center}
	\caption{Two-sample setting. Location, Scale and Shape differences with ${n}_1={n}_2=8$, $m=100$, $\delta=0.3$ and $A$ given by (\ref{ec:A13}). The Monte Carlo estimator of the FDR and power are reported for each test and $q$-value method.  The Monte Carlo bias and standard deviation of each $\pi_0$ estimator are also provided.}
	\label{ca:lss_8_100_0.3_ind}
\end{table}

\vspace*{0.5cm}

\begin{table}[htb]
	\begin{center}
		\scalebox{0.7}[0.7]{ 
			\begin{tabular}{cccccccccccccccc}
				\hline
				& 	& \multicolumn{14}{c}{Two-sample tests} \\
				&  & \multicolumn{2}{c}{$J_i$} && \multicolumn{2}{c}{$F$-test}
				&& \multicolumn{2}{c}{KS}
				&& \multicolumn{2}{c}{Ansari}
				&& \multicolumn{2}{c}{Levene}\\
				\cline{3-4}\cline{6-7} \cline{9-10} \cline{12-13} \cline{15-16}\\
				$\widehat{\pi}_0$	& &  Bias & Sd &&	Bias & Sd && Bias & Sd && Bias & Sd && Bias & Sd\\	
				Liang&&	0.0158&	0.0523	&&	-&-		&&	0.0661&	0.0624	&&	0.0976	&0.0805	&&	-&-		\\
				ST&&	0.0037&	0.1997	&&	-0.0784&	0.1717	&&	0.4983&	0.0157	&&	0.4396&	0.1140	&&	-0.0019&	0.1943	\\
				Chen&&	-0.0219&	0.0662	&&	-&-		&&	0.0800&	0.0623	&&	0.0794	&0.0658	&&	-&-		\\
				SS&&	0.0015&	0.0849	&&	-0.0774	&0.0767	&&	0.3114&	0.0875	&&	0.1184&	0.0850	&&	0.0211&	0.0776	\\
				Rand&&	0.0014&	0.0849	&&	-&-		&&	0.0432	&0.0707	&&	0.0842	&0.0810	&&	-&-		\\
				\hline 		
				&  & \multicolumn{2}{c}{$J_i$} && \multicolumn{2}{c}{$F$-test}
				&& \multicolumn{2}{c}{KS}
				&& \multicolumn{2}{c}{Ansari}
				&& \multicolumn{2}{c}{Levene}\\
				\cline{3-4}\cline{6-7} \cline{9-10} \cline{12-13} \cline{15-16}\\
				MCP	& &  FDR & POWER &&	FDR & POWER && FDR & POWER && FDR & POWER && FDR & POWER\\	
				Liang		&&	0.0486	&0.8179	&&	-&-		&&	0.0383&	0.1838	&&	0.0361&	0.3690	&&	-&-		\\
				ST		&&	0.0611	&0.8255	&&	0.1570&	0.8940	&&	0.0204&	0.0998	&&	0.0229	&0.2421	&&	0.0596&	0.1943	\\
				Chen		&&	0.0539	&0.8338	&&	-&-		&&	0.0366	&0.1739	&&	0.0367&	0.3754	&&	-&-		\\
				Real		&&	0.0487	&0.8237	&&	0.1296&	0.8751	&&	0.0420&	0.2368	&&	0.0429&	0.4159	&&	0.0551&	0.1488	\\
				SS		&&	0.0521	&0.8251	&&	0.1464&	0.8885	&&	0.0235	&0.1121	&&	0.0350&	0.3604	&&	0.0594&	0.1463	\\
				Rand		&&	0.0521	&0.8251	&&	-&-		&&	0.0386&	0.1958	&&	0.0370	&0.3724	&&	-&-		\\
				\hline
		\end{tabular}}	
	\end{center}
	\caption{Two-sample setting. Location, Scale and Shape differences with ${n}_1={n}_2=8$, $m=100$, $\delta=0.5$ and $A$ given by (\ref{ec:A23}). The Monte Carlo estimator of the FDR and power are reported for each test and $q$-value method.  The Monte Carlo bias and standard deviation of each $\pi_0$ estimator are also provided.}
	\label{ca:lss_8_100_0.5_dep}
\end{table}

\begin{table}[H]
	\begin{center}
		\scalebox{0.7}[0.7]{ 
			\begin{tabular}{cccccccccccccccc}
				\hline
				& 	& \multicolumn{14}{c}{Two-sample tests} \\
				&  & \multicolumn{2}{c}{$J_i$} && \multicolumn{2}{c}{$F$-test}
				&& \multicolumn{2}{c}{KS}
				&& \multicolumn{2}{c}{Ansari}
				&& \multicolumn{2}{c}{Levene}\\
				\cline{3-4}\cline{6-7} \cline{9-10} \cline{12-13} \cline{15-16}\\
				$\widehat{\pi}_0$	& &  Bias & Sd &&	Bias & Sd && Bias & Sd && Bias & Sd && Bias & Sd\\	
				Liang		&&	0.0103&	0.0186	&&	-&-		&&	0.0683	&0.0165	&&	0.0874	&0.0232	&&	-&-		\\
				ST		&&	-0.0022&	0.0676	&&	-0.0888	&0.0546	&&	0.5000&	0.0000	&&	0.4989&	0.0092	&&	-0.0031&	0.0636	\\
				Chen		&&	-0.0044	&0.0189	&&	-&-		&&	0.0697&	0.0165	&&	0.0854	&0.0212	&&	-&-		\\
				SS		&&	0.0057	&0.0240	&&	-0.0720&	0.0230	&&	0.3154&	0.0237	&&	0.1219&	0.0247	&&	0.0260&	0.0235	\\
				Rand		&&	0.0057	&0.0240	&&		-&-	&&	0.0449&	0.0191	&&	0.0874&	0.0239	&&	-&-		\\
				\hline 		
				&  & \multicolumn{2}{c}{$J_i$} && \multicolumn{2}{c}{$F$-test}
				&& \multicolumn{2}{c}{KS}
				&& \multicolumn{2}{c}{Ansari}
				&& \multicolumn{2}{c}{Levene}\\
				\cline{3-4}\cline{6-7} \cline{9-10} \cline{12-13} \cline{15-16}\\
				MCP	& &  FDR & POWER &&	FDR & POWER && FDR & POWER && FDR & POWER && FDR & POWER\\	
				Liang&&	0.0451&	0.8193	&&	-&-		&&	0.0211&	0.1172	&&	0.0323&	0.3615	&&	-&-		\\
				ST&&	0.0471&	0.8253	&&	0.1473&	0.8896	&&	0.0196&	0.1102	&&	0.0210&	0.2105	&&	0.0557	&0.1292	\\
				Chen&&	0.0464&	0.8256	&&	-&-		&&	0.0211&	0.1172	&&	0.0323&	0.3613	&&	-&-		\\
				Real&&	0.0459&	0.8238	&&	0.1315&	0.8747	&&	0.0285&	0.1726	&&	0.0447&	0.4261	&&	0.0553&	0.1225	\\
				SS&&	0.0455&	0.8212	&&	0.1433&	0.8863	&&	0.0210	&0.1162	&&	0.0318&	0.3584	&&	0.0540&	0.1140	\\
				Rand&&	0.0455&	0.8213	&&	-&-		&&	0.0225&	0.1255	&&	0.0323&	0.3613	&&	-&-		\\	
				\hline			
		\end{tabular}}	
	\end{center}
	\caption{Two-sample setting. Location, Scale and Shape differences with ${n}_1={n}_2=8$, $m=1000$, $\delta=0.5$ and $A$ given by (\ref{ec:A13}). The Monte Carlo estimator of the FDR and power are reported for each test and $q$-value method.  The Monte Carlo bias and standard deviation of each $\pi_0$ estimator are also provided.}
	\label{ca:lss_8_1000_0.5_ind}
\end{table}